\documentclass[12pt, draftclsnofoot, onecolumn]{IEEEtran}
\ifCLASSINFOpdf

\else

\fi
\usepackage{mathrsfs}

\usepackage{amsmath}
\usepackage{ntheorem}

\usepackage{makeidx}  
\usepackage{algorithm}
\usepackage{algorithmic}
\usepackage{graphicx}
\usepackage{subfigure}
\usepackage{epstopdf}
\usepackage{bm}
\usepackage{cite}
\usepackage{stfloats}

\usepackage{amssymb}
\usepackage{graphicx}

\usepackage{url}
\newtheorem{proposition}{Proposition}
\usepackage{tabularx,booktabs}
\newcolumntype{C}{>{\centering\arraybackslash}X} 
\usepackage{lipsum}

\usepackage{makecell} 

\usepackage{graphicx}
\usepackage{color}
\newtheorem{thm}{Theorem}
\newtheorem{lem}{Lemma}

\newtheorem{corollary}{Corollary}

\begin{document}

\title{Power Scaling Law Analysis and Phase Shift Optimization of  RIS-aided Massive MIMO Systems with Statistical CSI}
\author{Kangda Zhi, Cunhua Pan, Hong Ren and Kezhi Wang\thanks{(Corresponding author: Cunhua Pan).
		 		
		K. Zhi, C. Pan are with the School of Electronic Engineering and Computer Science at Queen Mary University of London, London E1 4NS, U.K. (e-mail: k.zhi, c.pan@qmul.ac.uk).

H. Ren is with the National Mobile Communications Research Laboratory, Southeast University, Nanjing 210096, China. (hren@seu.edu.cn).

K. Wang is with Department of Computer and Information Sciences, Northumbria University, UK. (e-mail: kezhi.wang@northumbria.ac.uk).}}

\maketitle
\begin{abstract}
This paper considers an uplink reconfigurable intelligent surface (RIS)-aided massive multiple-input multiple-output (MIMO) system with statistical channel state information (CSI). The RIS is deployed to help conventional massive MIMO networks serve the users in the dead zone. We consider the Rician channel model and exploit the long-time statistical CSI to design  the phase shifts of the RIS, while the maximum ratio combination (MRC) technique is applied for the active beamforming at the base station (BS) relying on the instantaneous CSI. Firstly, we derive the closed-form expressions for the uplink achievable rate which holds for arbitrary numbers of base station (BS) antennas. Based on the theoretical expressions, we reveal the power scaling laws, provide the average asymptotic rate when using random phase shifts and discuss the rate performance under some special cases. Then, we consider the  sum-rate maximization and the minimum user rate maximization problems by optimizing the phase shifts at the RIS. However, these two optimization problems are challenging to solve due to the complicated data rate expression. To solve these problems, we propose a novel genetic algorithm (GA) with low complexity but can achieve considerable performance. Finally, extensive simulations are provided to validate the benefits by integrating RIS into conventional massive MIMO systems.  Besides, our simulations demonstrate the feasibility of deploying large-size but low-resolution RIS in massive MIMO systems.
\end{abstract}


\IEEEpeerreviewmaketitle

\section{Introduction}
The massive multiple-input multiple-output (MIMO) technology is an essential technique to provide the extremely high network throughput in current and future communication systems\cite{bjornson2019massive}. However, to achieve such high throughput, hundreds of antennas should be equipped at the base station (BS), which raises the issues of high cost and energy consumption. Besides, to provide seamless coverage in the urban environment with dense obstacles, active relay or small BSs should be densely deployed, which also sharply increases the cost. On the other hand, reconfigurable intelligent surface (RIS), also known as intelligent reflecting surface (IRS), has been proposed as a revolutionary technology to support high data rate while maintaining at a low cost and energy consumption\cite{wu2020intelligent,di2020smart,di2019smart}. Specifically, RIS can constructively reflect the signal from the multi-antenna BS to multiple users which cannot directly communicate with the BS due to blockages, and it can also operate in a full-duplex (FD) mode without self-interference. Therefore, RIS is an efficient and cost-effective solution for the blockage problem of conventional massive MIMO systems.

Due to the above advantages, RIS-aided communication systems have been widely investigated in various scenarios\cite{wu2019intelligent,huang2020DRL,zhang2019capacity,zhou2020multicast,pan2020intelligent,peng2020multiuser,huang2019reconfigu,pan2020multicell,bai2020latency,bai2020resource,li2020reconfigurable,hua2020uav,wang2020joint,zhou2020framework,Basar2020spacekey,Basar2020modulation,hongsheng2020noise,yuxianghao2020robost,hong2020robust,zhang2020intelligent,yuxianghao2020cogitiveradio}. Specifically, for single-cell multi-antenna systems, the authors in \cite{wu2019intelligent} jointly considered the active and passive beamforming optimizations to demonstrate the potential of RIS, while a deep reinforcement learning-based method was proposed in \cite{huang2020DRL}. Zhang \emph{et al.} \cite{zhang2019capacity} characterized the fundamental capacity limit of RIS-aided MIMO systems with the narrowband and broadband transmission. Downlink multigroup multicast communication systems were presented in \cite{zhou2020multicast} and the RIS-aided simultaneous wireless information and power transfer (SWIPT) systems were studied in \cite{pan2020intelligent}. The benefits of using RIS in multi-user FD two-way communication systems were demonstrated in \cite{peng2020multiuser}. Meanwhile, an energy efficiency maximization problem was considered in \cite{huang2019reconfigu}. To investigate the performance of RIS-aided multi-cell MIMO networks, the authors in \cite{pan2020multicell} proposed to deploy an RIS at the cell edge and demonstrated the benefits of the RIS to mitigate the inter-cell interference.  Furthermore, RIS-aided mobile edge computing (MEC) systems were  studied in \cite{bai2020latency}, which showed that significant latency can be reduced by integrating RIS into conventional MEC systems. The authors in \cite{bai2020resource} further investigated the wireless powered orthogonal-frequency-division-multiplexing (OFDM) MEC systems under the assistance of an RIS. Meanwhile, RIS-aided unmanned aerial vehicle (UAV) networks were studied in \cite{li2020reconfigurable,hua2020uav,wang2020joint}. Specifically, the work in \cite{li2020reconfigurable} considered the joint optimization of UAV's trajectory and RIS's phase shifts in a single-user network, and a novel symbiotic UAV-aided multiple RIS radio system was studied in \cite{hua2020uav}. Wang \emph{et al.} \cite{wang2020joint} further investigated the UAV-aided multi-RIS multi-user systems using a deep reinforcement learning approach. Taking into consideration the impact of imperfect cascaded channels, the authors in \cite{zhou2020framework} firstly studied the robust active and passive beamforming optimization problem to minimize the total transmit power. Besides, RIS-aided space shift keying and RIS-aided spatial modulation schemes were investigated in \cite{Basar2020spacekey,Basar2020modulation}. Considering the secure communication scenarios, the authors in\cite{hongsheng2020noise} studied the performance of artificial noise-aided MIMO systems with the aid of an RIS. RIS-aided secure communications with imperfect RIS-eavesdropper channels were considered in \cite{yuxianghao2020robost}, while the authors in \cite{hong2020robust} further investigated the robust transmission design in RIS-aided secure communications with cascaded channel error. Furthermore, RIS-aided MIMO and FD cognitive radio systems were respectively studied in  \cite{zhang2020intelligent} and \cite{yuxianghao2020cogitiveradio}.

However,  all of the above contributions considered to design the phase shifts of the RIS based on instantaneous channel state information (CSI). Those schemes are suitable for the scenarios with a fixed location or low mobility, which enable the BS to carry out the channel estimation, design the optimal RIS phase shifts and adjust the phase shifts of the RIS in each channel coherence time. However, for the scenarios with high mobility and short channel coherence time, it is more practical to design and tune the phase shifts of the RIS relying on statistical CSI. Furthermore, this statistical CSI-based strategy can effectively reduce the feedback overhead required for RIS\cite{di2019smart}, reduce the power consumed by RIS's controller and release the capacity requirement for the RIS's control link. In addition, significant computational complexity can be reduced at the BS since the phase shift matrix is only needed to be updated when the statistical CSI varies, which occurs in a much larger time scale than the instantaneous CSI.

Due to the above benefits,  some researchers have exploited the statistical CSI to design the RIS-aided communication systems \cite{han2019large,jia2020analysis,peng2020analysis,yu2020design,hu2020location,you2020reconfigurable,kammoun2020asymptotic,zhao2020intelligent,chaaban2020opportunistic}. For the single-user systems, Han \emph{et al.} \cite{han2019large} first presented the optimal RIS phase shift design based on the derived ergodic capacity expression under the Rician channel model. The authors in \cite{jia2020analysis} further designed the RIS-aided systems with a serving BS and an interfered BS. For the multi-user case, Peng \emph{et al.} \cite{peng2020analysis} investigated the performance of RIS-aided multi-pair communication systems and verified the effectiveness of applying genetic algorithm (GA) in the optimization of the phase shifts of the RIS. The performance of RIS-aided Internet of Things under correlated Rayleigh channels was evaluated in \cite{yu2020design}. The authors in \cite{hu2020location} proposed a location information-aided multi-RIS system, where a low-complexity BS-RIS maximum-ratio transmission beamforming scheme was proposed. By resorting to random matrix theory, You \emph{et al.}\cite{you2020reconfigurable} considered the energy efficiency maximization problem in MIMO networks under the correlated Rayleigh channel model, and Nadeem  \emph{et al.} \cite{kammoun2020asymptotic} considered the  minimum signal-to-interference-plus-noise ratio (SINR) maximization problem with line-of-sight (LoS) BS-RIS channel matrix. A novel two-timescale beamforming optimization scheme was proposed in \cite{zhao2020intelligent}, where the passive beamforming was first optimized based on statistical CSI and then the active beamforming was designed based on instantaneous CSI. Besides, the authors in \cite{chaaban2020opportunistic} studied the IRS-aided opportunistic beamforming scheme with statistical CSI.

However, based on the statistical CSI, the RIS-aided massive MIMO systems under the Rician channel model have not been investigated. On  one hand, since the RIS is often deployed on the facade of tall buildings, the RIS-related channels may possess the LoS channel components. Therefore,  the more general Rician fading model should be adopted. On the other hand, it is crucial to characterize the interplay between the promising RIS technology and the existing massive MIMO technology, and evaluate the potential of RIS-aided massive MIMO systems. To the best of our knowledge, only \cite{wang2020intelligent} studied the RIS-aided massive MIMO networks. However, in \cite{wang2020intelligent}, the correlated Rayleigh channel was considered and the phase shifts of RIS are simply set as an identity matrix.

Against the above background, in this paper, we theoretically analyze and optimize the uplink RIS-aided massive MIMO systems with the Rician channel model and statistical CSI. Specifically, the low-complexity maximum-ratio combination (MRC) technique is employed for the active beamforming based on the instantaneous CSI, while the phase shifts of the RIS are designed and adjusted by exploiting the statistical CSI. The Rician channel model is applied in this paper to capture the achievable spatial multiplexing gain of RIS-aided massive MIMO systems. We present the closed-form analytical expression for the uplink achievable rate which holds for arbitrary numbers of antennas at the BS. Our main contributions are summarized as follows:
\begin{itemize}
	\item First, we derive the closed-form expression of the uplink achievable rate using the Rician channel model that holds for any finite number of antennas at the BS, and this analytical expression only depends on the locations and angles information and Rician factors. Based on the derived expressions, we reveal the scaling laws with respect to the number of RIS's elements and the number of BSs' antennas. We also evaluate the average asymptotic rate achieved by  random phase shifts.
	
	\item Then, by using the derived expression, we utilize the GA-based method to solve the sum-rate maximization problem and the minimum user rate maximization problem, by taking into consideration the impact of discrete phase shifts.
	
	\item Finally, extensive simulations are carried out to characterize the gains by employing RIS into massive MIMO networks. Our results reveal the trade-off between the increase of spatial multiplexing gain and the decrease of path loss in the RIS-aided massive MIMO systems. Meanwhile, we validate the feasibility of deploying large-size RIS with low-resolution hardware into existing massive MIMO systems.
\end{itemize}

The remainder of this paper is organized as follows. Section \ref{section2} describes the model of uplink RIS-aided massive MIMO systems with Rician channel. Section \ref{section3} derives the closed-form analytical expressions for the uplink achievable rate with arbitrary numbers of BS antennas, and discusses the power scaling laws and some special cases. Section \ref{section4} presents the GA-based method to solve the sum-rate maximization and the minimum user rate maximization problems. Section \ref{section5} provides extensive simulation results to characterize the achievable spatial multiplexing gain and other benefits brought by RIS. Finally, Section \ref{section6} concludes this paper.

\emph{Notations}: The vectors and the matrices are respectively expressed in lowercase blodface and uppercase blodface letters. $\mathbf{A}^H$,  $\mathbf{A}^T$ and $\mathbf{A}^*$ represent the conjugate transpose, transpose and conjugate operators, respectively. $\left|\mathbf{a}\right|$  denotes the modulus of the complex number and  $\left\|\mathbf{a}\right\|$ denotes $l_2$-norm of the vector. $\left[\mathbf{A}\right]_{m,n}$ denotes the $\left(m,n\right)$-th entry of the matrix. $\left[\mathbf{a}\right]_m$ denotes the $m$-th entry of the vector. Re$\left\{\cdot\right\}$ represents the real part. $\mathbb{E}\left\{\cdot\right\}$ and $\mathrm{Tr}\left\{\cdot\right\}$ denote the expectation and trace operator, respectively. $\mathbf{I}_N$ is the identity matrix with $N$ dimension. $\mathbb{C}^{M\times N}$ represents the $M\times N$ complex-valued matrix. Besides, $ x\sim{\cal C}{\cal N}\left( {a,b} \right)$ denotes that random variable  $x$ follows the complex Gaussian distribution with mean $a$ and variance $b$. Operation $\left\lfloor n \right\rfloor $ means rounding $n$ toward the negative infinity and operation $\bmod$ means taking the remainder after division.
 
\section{System Model}\label{section2}
\begin{figure}[h]
\setlength{\abovecaptionskip}{0pt}
\setlength{\belowcaptionskip}{-20pt}
\centering
\includegraphics[width= 0.6\textwidth]{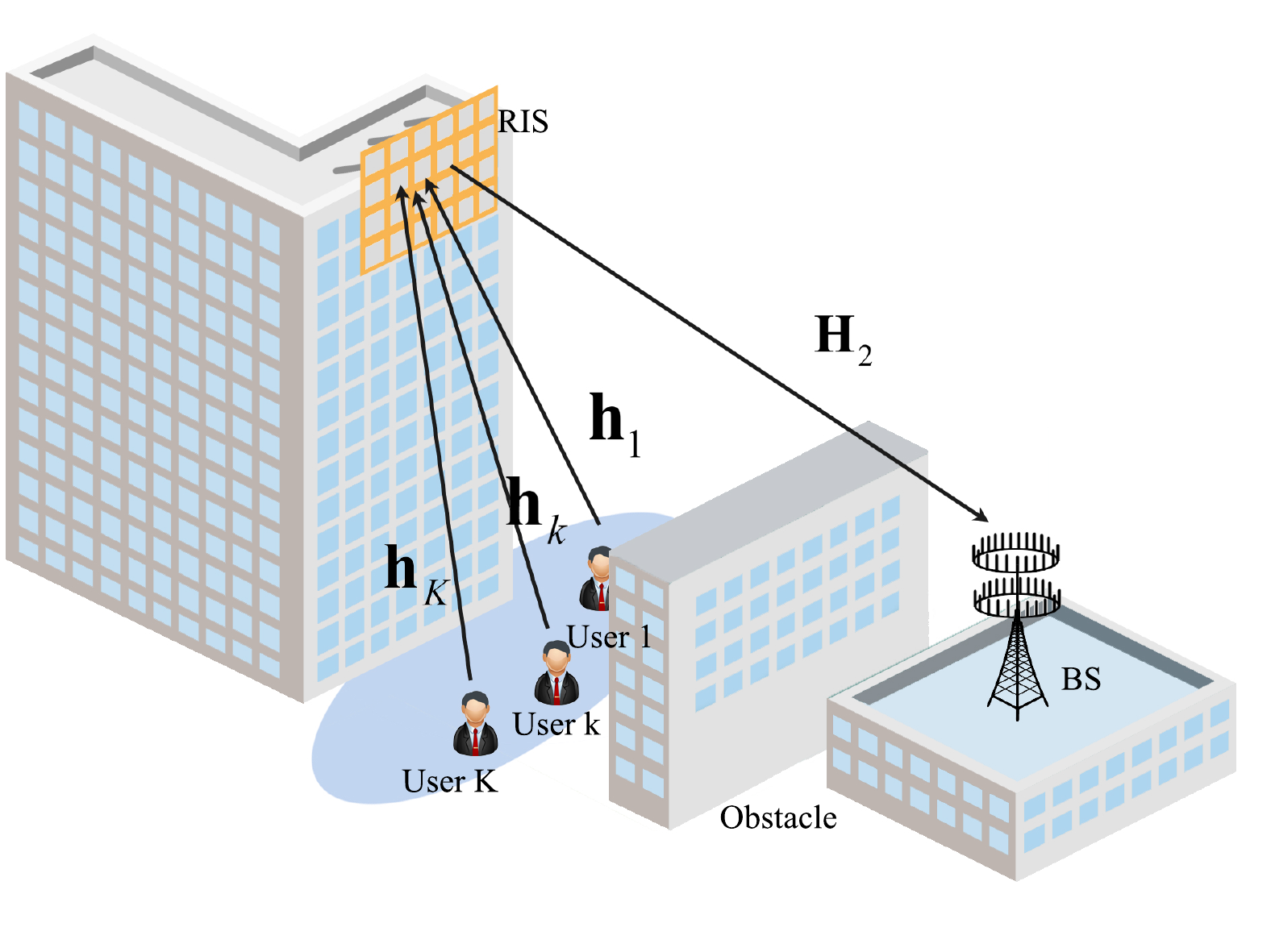}
\DeclareGraphicsExtensions.
\caption{An RIS-assisted uplink massive MIMO communication system.}
\label{figure0}
\end{figure}

Consider a typical uplink RIS-aided communication system with a base station (BS), an RIS and $K$  single-antenna users. The BS and RIS are equipped with $M$  antennas and $N$ reflecting elements, respectively.  The  RIS is connected with the BS with a dedicated transmission link and its phase shifts can be controlled by the BS.

Since the ground communication links can be obstructed by buildings, humans and trees, we assume that the direct links between the BS and users are blocked similar to \cite{kammoun2020asymptotic,hu2020location,you2020reconfigurable}. To assist the communications for users, an RIS is deployed on the building and helps users communicate with the BS, as shown in Fig. \ref{figure0}. Considering the fact that the RIS can be deployed on the wall of tall buildings, it can assist in creating channels dominated by LoS propagation along with a few scatters. Therefore, we adopt the Rician fading model, and the channels between users and the RIS and that between the RIS and the BS can be respectively expressed as:

\begin{align}
&{{\bf {H}}_1} = \left[ {{{\bf h}_1},{{\bf h}_2},...,{{\bf h}_K}} \right],{{\bf h}_k} = \sqrt {{\alpha _k}} \left( {\sqrt {\frac{{{\varepsilon _k}}}{{{\varepsilon _k} + 1}}} {{\bf  \bar h}_k} + \sqrt {\frac{1}{{{\varepsilon _k} + 1}}} {{\bf \tilde h}_k}} \right),\label{rician1}\\
&{{\bf H}_2} = \sqrt \beta  \left( {\sqrt {\frac{{{\delta _{}}}}{{{\delta _{}} + 1}}} {{\bf \bar H}_2} + \sqrt {\frac{1}{{\delta  + 1}}} {{\bf \tilde H}_2}} \right),\label{rician2}
\end{align}
where ${{\bf H}_1} \in {\mathbb{C}^{N \times K}}$,  ${{\bf H}_2} \in {\mathbb{C}^{M \times N}}$, ${\alpha _k}$  and  $\beta $ are the large-scale path loss of the user $k$-RIS link and RIS-BS link, respectively.  ${\varepsilon _k}$ and ${\delta}$ are the Rician factors of the user $k$-RIS link and RIS-BS link, respectively.  ${{\bf \bar h}_k} \in {{\mathbb{C}}^{N \times 1}}$ and ${{\bf \bar H}_2} \in {{\mathbb{C}}^{M \times N}}$  denote the LoS components of the user $k$-RIS link and RIS-BS link.   ${{\bf \tilde h}_k} \in {{\mathbb{C}}^{N \times 1}}$ and ${{\bf \tilde H}_2} \in {{\mathbb{C}}^{M \times N}}$  represent the non-line-of-sight (NLoS) components, whose elements are independently and identically distributed (i.i.d) complex Gaussian random variables following the distribution of $ {\cal C}{\cal N}\left( {0,1} \right)$.

 Assume that the BS and RIS are equipped with uniform square planar array (USPA) with size of $\sqrt M  \times \sqrt M $ and $\sqrt N  \times \sqrt N $, respectively. Therefore,   the LoS components ${{\bf \bar h}_k} \in {{\mathbb{C}}^{N \times 1}}$ and ${{\bf \bar H}_2} \in {{\mathbb{C}}^{M \times N}}$    can be respectively expressed as:
\begin{align}
&{{\bf \bar h}_k} = {{\bf a}_N}\left( {\varphi _{kr}^a,\varphi _{kr}^e} \right),\\
&{{\bf \bar H}_2} = {{\bf a}_M}\left( {\phi _r^a,\phi _r^e} \right){\bf a}_N^H\left( {\varphi _t^a,\varphi _t^e} \right),
\end{align}
with array response vector as
\begin{align}\label{upa}
{{\bf a}_X}\left( {\vartheta _{}^a,\vartheta _{}^e} \right) = {\left[ {1,...,{e^{j2\pi \frac{d}{\lambda }\left( {x\sin \vartheta _{}^a\sin \vartheta _{}^e + y\cos \vartheta _{}^e} \right)}},...,{e^{j2\pi \frac{d}{\lambda }\left( {\left( {\sqrt X  - 1} \right)\sin \vartheta _{}^a\sin \vartheta _{}^e + \left( {\sqrt X  - 1} \right)\cos \vartheta _{}^e} \right)}}} \right]^T},
\end{align}
where ${\rm{0}} \le x,y \le \sqrt X  - 1$, $d$ and $\lambda$ are the element spacing and carrier wavelength,   $\varphi _{kr}^a$  and $\varphi _{kr}^e$   are respectively the azimuth and elevation angles of arrival (AoA) at the RIS from  user  $k$.   $\varphi _{t}^a$  and $\varphi _{t}^e$ respectively denote the azimuth and elevation angles of departure (AoD) from the RIS towards the BS.  $\phi _r^a$ and   $\phi _r^e$ respectively represent the AoA at the BS from the RIS. Note that ${{\bf \bar h}_k}$  and ${{\bf \bar H}_2}$  only rely on the AoA and AoD, which could keep invariant within the considered time period. Besides, we assume that these angles are  known based on some technologies. For example, it can be calculated by the locations obtained from the global position system (GPS). 

With the help of RIS, the received signal at the BS can be written as:
\begin{align}
{\bf y} = {\bf G\mathbf{ P} x + n} =  {\bf {H_2}\Phi {H_1}\mathbf{ P} x + n},
\end{align}
where $\mathbf{ P}={\rm{diag}}\left\{\sqrt{p_1},\sqrt{p_2},...,\sqrt{p_K} \right\}$, $p_k$ is the transmit power of user $k$. ${{ {\bf G} \buildrel \Delta \over = \bf {H_2}\Phi {H_1}}}\in {{\mathbb{C}}^{M \times K}}$ represents the cascaded user-RIS-BS channel.  ${\bf \Phi } = {\rm{diag}}\left\{ {{e^{j{\theta _1}}},{e^{j{\theta _2}}},...,{e^{j{\theta _N}}}} \right\}$ is the reflection matrix of RIS and  ${\theta _n}\in [0,2\pi)$ is the phase shift introduced by the RIS reflector $n$.   ${\bf x} = {\left[ {{x_1},{x_2},...,{x_K}} \right]^T}\in {{\mathbb{C}}^{K \times 1}}$  denotes the information symbols from $K$ users, where $\mathbb{E}\left\{\left|x_{k}\right|^{2}\right\}=1$.   ${\bf n} \sim {\cal C}{\cal N}\left( {0,{\sigma ^2}{\bf I}} \right)$  is the additional white Gaussian noise (AWGN).

Adopting the maximal-ratio-combining (MRC) technique, the received signal at the BS can be written as
\begin{align}
{{\bf r} = {{\bf G}^H}{\bf y} }=  { {{\bf G}^H}{\bf G\mathbf{P}x} + {{\bf G}^H}\bf n},
\end{align}
and the signal of user $k$  can be expressed as
\begin{align}
{r_k} = \sqrt p_k {{\bf g}_k^H} {{\bf g}_k}{x_k} + \sum\limits_{i = 1,i \ne k}^K {\sqrt p_i {{{\bf g}_k^H}} {{\bf g}_i}{x_i}}  + {{{\bf g}_k^H}}{\bf n}.
\end{align}
where  ${{\bf g}_k} \buildrel \Delta \over = {{\bf H}_2}{\bf \Phi }{{\bf h}_k} \in {{\mathbb{C}}^{M \times 1}}$ is the  $k$-th column of matrix  $\bf G$ representing the cascaded user $k$-RIS-BS channel.

Considering the ergodic channel,  the uplink achievable rate of user $k$  can be expressed as
\begin{align}\label{rate_userk}
R_{k}=\mathbb{E}\left\{\log _{2}\left(1+\frac{{p_k}\left\|\mathbf{g}_{k}\right\|^{4}}{ \sum\limits_{i=1, i\neq k}^{K}{p_i}\left|\mathbf{g}_{k}^{H} \mathbf{g}_{i}\right|^{2}+\sigma^{2}\left\|\mathbf{g}_{k}\right\|^{2}}\right)\right\},
\end{align}
and the sum rate is as
\begin{align}
R = \sum\limits_{i = 1}^K {{R_k}}.
\end{align}

\section{Uplink Achievable Rate Analysis}\label{section3}
In this section, we derive the closed-form expression of the achievable rate in the uplink RIS-aided multi-user system. The theoretical results can capture  the impacts of various variables, including the number of antennas at the BS, the number of  reflecting elements at the RIS, the transmit power and  Rician factors. We will also present asymptotic expressions in some special cases.

\subsection{Preliminary Results}
We first give a key Lemma which will be used in further derivations.
\begin{lem}\label{lemma2}
The expectation of  ${{{\left\| {{{\bf g}_k}} \right\|}^2}}$, ${{{\left\| {{{\bf g}_k}} \right\|}^4}}$  and ${{{\left| {{\bf g}_{_k}^H{{\bf g}_i}} \right|}^2}}$ are respectively given by
\begin{align}\label{gk2}
\mathbb{E}\left\{\left\|\mathbf{g}_{k}\right\|^{2}\right\} =\frac{M\beta \alpha_{k}}{(\delta+1)\left(\varepsilon_{k}+1\right)}\left(\delta \varepsilon_{k}\left|f_{k}({\bf \Phi}	)\right|^{2}+\left(\delta+\varepsilon_{k}+1\right) N\right),\qquad\qquad\qquad\qquad\quad
\end{align}
\begin{align}\label{ES4}
\begin{array}{l}
\mathbb{E}\left\{\left\|\mathbf{g}_{k}\right\|^{4}\right\}=M\left(\frac{\beta \alpha_{k}}{(\delta+1)\left(\varepsilon_{k}+1\right)}\right)^{2} \times \\
\left\{M\left(\delta \varepsilon_{k}\right)^{2}\left|f_{k}({\bf \Phi})\right|^{4}+2  \delta \varepsilon_{k}\left|f_{k}({\bf \Phi})\right|^{2}\left(2 M N \delta+M N \varepsilon_{k}+M N+2 M+N \varepsilon_{k}+N+2\right)\right. \\
+M N^{2}\left(2 \delta^{2}+\varepsilon_{k}^{2}+2 \delta \varepsilon_{k}+2 \delta+2 \varepsilon_{k}+1\right)+N^{2}\left(\varepsilon_{k}^{2}+2 \delta \varepsilon_{k}+2 \delta+2 \varepsilon_{k}+1\right) \\
\left.+M N\left( 2\delta + 2 \varepsilon_{k} +1\right)+ N\left( 2\delta+2 \varepsilon_{k} +1\right)\right\},
\end{array}
\end{align}
and
\begin{align}\label{EI2}
\begin{array}{l}
\mathbb{E}\left\{\left|\mathbf{g}_{k}^{H} \mathbf{g}_{i}\right|^{2}\right\}=M\frac{\beta^{2} \alpha_{i} \alpha_{k}}{(\delta+1)^{2}\left(\varepsilon_{i}+1\right)\left(\varepsilon_{k}+1\right)} \times\left\{M \delta^{2} \varepsilon_{k} \varepsilon_{i}\left|f_{k}({\bf \Phi})\right|^{2}\left|f_{i}({\bf \Phi})\right|^{2}\right. \\
+ \delta \varepsilon_{k}\left|f_{k}({\bf \Phi})\right|^{2}\left(\delta M N+N \varepsilon_{i}+N+2 M\right)+ \delta \varepsilon_{i}\left|f_{i}({\bf \Phi})\right|^{2}\left(\delta M N+N \varepsilon_{k}+N+2 M\right) \\
+N^{2}\left(M \delta^{2}+\delta\left(\varepsilon_{i}+\varepsilon_{k}+2\right)+\left(\varepsilon_{k}+1\right)\left(\varepsilon_{i}+1\right)\right)+M N\left(2 \delta+\varepsilon_{i}+\varepsilon_{k}+1\right) \\
\left.+M \varepsilon_{k} \varepsilon_{i}\left|\overline{\mathbf{h}}_{k}^{H} \overline{\mathbf{h}}_{i}\right|^{2}+2 M \delta \varepsilon_{k} \varepsilon_{i} \operatorname{Re}\left\{f_{k}^{H}({\bf \Phi}) f_{i}({\bf \Phi}) \overline{\mathbf{h}}_{i}^{H} \overline{\mathbf{h}}_{k}\right\}\right\},
\end{array}
\end{align}
	where ${f_c}\left( {\bf \Phi}  \right) \in {{\mathbb{C}}^{1 \times 1}}, c\in \left\{k,i\right\}$ is defined as
	\begin{align}\label{fc_Phi}
{f_c}\left( {\bf \Phi}  \right) \buildrel \Delta \over ={\bf a}_N^H\left( {\varphi _t^a,\varphi _t^e} \right){\bf \Phi} \overline{\bf h}_c= \sum\limits_{n=1}^{N} {{e^{j2\pi \frac{d}{\lambda }\left( {x{p_c} + y{q_c}} \right) + j{\theta _n}}}},
\end{align}
with $x = \left\lfloor {\left( {n - 1} \right)/\sqrt N} \right\rfloor $, $y = \left( {n - 1} \right)\bmod \sqrt N $, ${p_c} = \sin \varphi _{cr}^a\sin \varphi _{cr}^e - \sin \varphi _t^a\sin \varphi _t^e$, ${q_c} = \cos \varphi _{cr}^e - \cos \varphi _t^e$.
\end{lem}

\itshape \textbf{Proof:}  \upshape Please refer to Appendix \ref{appB}.  \hfill $\blacksquare$

Note that $ \left|{f_c}\left( {\bf \Phi}  \right) \right|\le N$ and the equality holds when $\theta_n = -2\pi \frac{d}{\lambda }\left( {x{p_c} + y{q_c}} \right), \forall n$. In this paper, we refer to the phase shift solution that maximizes $\left|{f_k}\left( {\bf \Phi}  \right) \right|$  as ``phase aligned to user $k$". In this setting, when $N\rightarrow\infty$, $\left|{f_k}\left( {\bf \Phi}  \right) \right|$ can grow without bound. However,  $\left|{f_i}\left( {\bf \Phi}  \right)\right|, i \neq k $ will be bounded unless user $i$ has nearly the same azimuth and elevation AoA with user $k$. Note that we ignore this rare situation in this section.

Lemma \ref{lemma2} shows that both $ \mathbb{E}\left\{\left\|\mathbf{g}_{k}\right\|^{4}\right\} $ and $ \mathbb{E}\left\{\left|\mathbf{g}_{k}^{H} \mathbf{g}_{i}\right|^{2}\right\} $ are on the order of $\mathcal{O}\left(M^2\right)$. However, their scaling laws with respect to $N$ depends on the value of $\bf\Phi$. For example, when the phase shifts of RIS are aligned to user $k$, $ \mathbb{E}\left\{\left\|\mathbf{g}_{k}\right\|^{4}\right\} $ is on the order of $\mathcal{O}\left(N^4\right) $ whereas $\mathbb{E}\left\{\left|\mathbf{g}_{k}^{H} \mathbf{g}_{i}\right|^{2}\right\} $ is on the order of $\mathcal{O}\left(N^3\right) $.

\subsection{Main Results}

Next, with the above results (\ref{gk2})$\sim$ (\ref{EI2}), we provide the closed-form expression of the uplink achievable rate under the general case with any number of  antennas.
\begin{thm}\label{theorem1}
	In the RIS-aided massive MIMO systems, the uplink achievable rate of user $k$ can be approximated as
	\begin{align}\label{rate}
	R_{k} \approx \log _{2}\left(1+\frac{p_k {E}_{k}^{(signal)}({\bf\Phi})}{ \sum\limits_{i=1, i \neq k}^{K} p_i I_{k i}({\bf\Phi})+\sigma^{2} E_{k}^{(noise)}({\bf\Phi})}\right),
	\end{align}
	where $ E_{k}^{(signal)}({\bf\Phi}) \triangleq \mathbb{E}\left\{\left\|\mathbf{g}_{k}\right\|^{4}\right\}  $, $ I_{ki}({\bf\Phi}) \triangleq \mathbb{E}\left\{\left|\mathbf{g}_{k}^{H} \mathbf{g}_{i}\right|^{2}\right\} $, and $ E_{k}^{(noise)}({\bf\Phi}) \triangleq \mathbb{E}\left\{\left\|\mathbf{g}_{k}\right\|^{2}\right\} $.
\end{thm}

\itshape \textbf{Proof:}  \upshape   It can be readily proved by using Jensen's inequality  as in \cite[Lemma 1]{6816003}.\hfill $\blacksquare$

Rate expression (\ref{rate}) characterizes the impacts of $\bf\Phi$, $M$, $N$, $P$, different kinds of AoA and AoD, path-loss parameters and Rician factors on the data rate performance.  We can see that this theoretical  expression is only determined by locations, AoA and AoD of the BS, the RIS and users, which could keep invariant for a long time. Therefore, designing the phase shifts of RIS based on statistical CSI can significantly reduce the computational complexity and channel estimation overhead
 in practical systems.

\begin{corollary}
	In the RIS-aided single user systems, i.e., without the multi-user interference, the achievable rate of user $k$ is
	\begin{align}
		R_{k} \approx \log _{2}\left(1+\frac{p_k E_{k}^{(signal)}({\bf\Phi})}{ \sigma^{2} E_{k}^{(noise)}({\bf\Phi})}\right),
	\end{align}
	which can achieve the gain of $\mathcal{O}\left(\mathrm{log_2}(MN^2)\right)$.
\end{corollary}

It is well known that this performance gain comes from the active beamforming gain of multi-antenna, passive beamforming gain of RIS and the inherent aperture gain of RIS\cite{wu2019intelligent}. However, when considering the multi-user interference, this performance gain cannot be obtained. We can see that both  $ E_{k}^{(signal)}({\bf\Phi}) $ and  $ I_{ki}({\bf\Phi}) $ in (\ref{rate}) increase on the order of $\mathcal{O}\left(M^2\right)$, therefore it cannot achieve a gain of $\mathcal{O}\left(\mathrm{log_2}(M)\right)$, i.e., the rate in (\ref{rate}) cannot grow without bound when $M\rightarrow \infty$. This is because the channels of different users share the same RIS-BS channel $\mathbf{H}_2$. For example, recalling (\ref{gk1gi1_2}) and (\ref{HHAHH}) in Appendix \ref{appB}, we can see that the common term $\overline{\mathbf{H}}_2$ and $\tilde{\mathbf{H}}_2$ bring the factor $M^2$ to the interference term  $ I_{ki}({\bf\Phi}) $. 

Meanwhile, rate expression (\ref{rate}) shows that the order of magnitude with respect to $N$ is determined by the setting of $\mathbf{ \Phi}$. If we adjust RIS's phase shifts to maximize the desired signal power of user $k$, i.e.,  $ \left|{f_k}\left( {\bf \Phi}  \right) \right|= N$, we can find that the rate $R_k$ is on the order of $\mathcal{O}\left(\mathrm{log_2}(N)\right)$. This means that the sum achievable rate can grow without bound when $N\rightarrow\infty$. Although the rate does not have the order of $\mathcal{O}\left(\mathrm{log_2}(N^2)\right)$ in this simple case, the rate performance could be improved by properly designing the phase shifts of RIS to increase the desired signal as well as mitigating the interference, and RIS's  interference mitigation capability has been validated in the previous contributions \cite{jia2020analysis,pan2020multicell}. This discussion emphasizes the importance of the phase shift design in RIS-aided massive MIMO systems.

To better understand Theorem \ref{theorem1}, we will present the asymptotic results under some special cases. Firstly, we reveal the power-scaling law of the uplink achievable rate in the RIS-aided massive MIMO systems in the following.
\begin{corollary}\label{power_scale_law}
	Assume that the transmit power of each user is scaled with the number of antennas at the BS according to $p_k={E_u}/{M}$, $\forall k$, where ${E_u}$ is fixed. When $M\rightarrow\infty $, we have
\begin{align}\label{Rk_scale}
R_{k} \to \log _{2}\left(1+\frac{E_u  \frac{\beta \alpha_{k}}{(\delta+1)\left(\varepsilon_{k}+1\right)} \mathcal{A}_{k}^{(1)}({\bf\Phi})}{ \sum_{i=1, i \neq k}^{K} E_u   \frac{\beta \alpha_{i}}{(\delta+1)\left(\varepsilon_{i}+1\right)}    {  \mathcal{A}^{(2)}_{k i}}  ({\bf\Phi})+\sigma^{2} \mathcal{A}_{k}^{(3)}({\bf\Phi})}\right),
\end{align}
	where
	\begin{align}
	\mathcal{A}_{k}^{(1)}({\bf\Phi})&=\left(\mathcal{A}_{k}^{(3)}({\bf\Phi})\right)^{2}+2 \delta \varepsilon_{k}\left|f_{k}(\mathbf{\Phi})\right|^{2}(N \delta+2)+N\left(N \delta^{2}+2 \delta+2 \varepsilon_{k}+1\right),\\
	\mathcal{A}_{ki}^{(2)}({\bf\Phi})&=\varepsilon_{k} \varepsilon_{i}\left|\delta f_{k}^{H}(\mathbf{\Phi}) f_{i}(\mathbf{\Phi})+\overline{\mathbf{h}}_{k}^{H} \overline{\mathbf{h}}_{i}\right|^{2} \nonumber\\
	&\;\;\;+\left(\delta^{2} N+2 \delta\right)\left(\varepsilon_{k}\left|f_{k}(\mathbf{\Phi})\right|^{2}+\varepsilon_{i}\left|f_{i}(\mathbf{\Phi})\right|^{2}\right)+N\left(N \delta^{2}+2 \delta+\varepsilon_{i}+\varepsilon_{k}+1\right),\\
	\mathcal{A}^{(3)}_{k}({\bf\Phi}) &=\delta \varepsilon_{k}\left|f_{k}(\mathbf{\Phi})\right|^{2}+\left(\delta+\varepsilon_{k}+1\right) N.
	\end{align}
	
\end{corollary}

\itshape \textbf{Proof:}  \upshape By substituting $p_k=E_u/M,\forall k$ into rate expression (\ref{rate}), when $M\to\infty$, we can ignore the insignificant terms which don't scale with $M$. Then, after some simplifications, we can complete the proof.  \hfill $\blacksquare$

From Corollary \ref{power_scale_law}, we can see that similar to traditional massive MIMO systems, users in RIS-aided systems can cut down their transmit power by a factor $1/M$ while the rate will converge to a non-zero value as $M\to\infty$. However, different from the traditional systems, both the signal, interference and noise terms in rate (\ref{Rk_scale}) are related with $\bf\Phi$. To clearly show the difference, we consider a special case where the RIS is deployed in the environment with pure NLoS channels, i.e., $\delta=\varepsilon_{k}=0,\forall k$. Then, the power scaling law in Corollary \ref{power_scale_law} becomes
\begin{align}\label{sclale_law_rayleigh}
\tilde{R}_{k} \rightarrow \log _{2}\left(1+\frac{E_{u} \beta \alpha_{k}(N+1)}{\sum_{i=1, i \neq k}^{K} E_{u} \beta \alpha_{i}+\sigma^{2}}\right), \text{as } M\to\infty.
\end{align}

By contrast, in traditional non-RIS massive MIMO systems with large-scale path-loss $\gamma_{k}$, when scaling the power by $p_k=E_u/M$, the rate can be written as \cite[Theorem 1]{6816003}:
\begin{align}\label{rate_nonIRS}
{R}^{(w/o)}_{k} \rightarrow \log _{2}\left(1+\frac{E_{u} \gamma_{k} }{\sigma^{2}}\right), \text{as } M\to\infty.
\end{align}

Comparing Eq. (\ref{sclale_law_rayleigh}) with Eq. (\ref{rate_nonIRS}), we can see that the rate can reap significant benefits
by deploying RIS with large number of elements in massive MIMO systems.

\begin{corollary}\label{corollary1}
	If the phase shifts of RIS are aligned to user $k$, the transmit power of user $k$ is scaled down by $p_k = \frac{E_u}{MN^2}$, while the transmit power of other users are scaled down by $p_i = \frac{E_u}{MN}, \forall i \neq k$. When both $M$ and $N$ are large, we have
	\begin{align}
&R_{k} \rightarrow \log _{2}\left(1+\frac{E_u\frac{\varepsilon_{k}}{\left(\varepsilon_{k}+1\right)}}{E_u\sum_{i=1, i \neq k}^{K} \frac{\alpha_{i}}{\left(\varepsilon_{i}+1\right) \alpha_{k}}+\left(1+\frac{1}{\delta}\right) \frac{\sigma^{2}}{\beta \alpha_{k}}}\right),\\
&R_{i}\rightarrow 0, \forall i \neq k,
	\end{align}
\end{corollary}

\itshape \textbf{Proof:}  \upshape Please refer to Appendix \ref{D2}. \hfill $\blacksquare$

Corollary \ref{corollary1} means that with large $M$ and $N$, we can  further cut down user $k$'s transmission power to $E_u/(MN^2)$  while keeping the data rate as a non-zero value. Meanwhile, this rate will be improved if the environment has few scatters, i.e., with a larger $\alpha_{k}$, $\beta$ and $\delta$.

\begin{corollary}\label{corollary2}
	For both the ideal RIS with continuous phase shifts and non-ideal RIS with $b>1$ bits discrete phase shifts, if the phase shift matrix $\bf\Phi$ is randomly adjusted in each time block, when $N\rightarrow\infty $ and $M\rightarrow\infty$, we have
	\begin{align}\label{rate_3}
R_{k} \rightarrow \log _{2}\left(1+\frac{p_k \alpha_{k}\left(2 \delta^{2}+2 \delta+1\right)}{\sum_{i=1, i \neq k}^{K} p_i \alpha_{i} \delta^{2}}\right).
	\end{align}
\end{corollary}

\itshape \textbf{Proof:}  \upshape Please refer to Appendix \ref{appE}.  \hfill $\blacksquare$

Corollary \ref{corollary2} shows that with a large number of antennas at the BS and a large number of reflecting elements at the RIS, the sum achievable rate is still bounded if the phase shifts are randomly adjusted. This conclusion shows the necessity of optimizing the phase shifts of RIS in the RIS-aided massive MIMO systems. Besides, we can see that the data rate in (\ref{rate_3}) decreases when $\delta$ increases, which has a different trend from Corollary \ref{corollary1}. The reason lies in that when the phase shifts are adjusted randomly in each time block, it tends to equally allocate the passive beamforming gain to all the users. However, when $\delta\rightarrow\infty$, the channel with unit rank will be unable to support the multi-user communications.

\begin{corollary}\label{corollary4}
	If $\delta=\varepsilon_{k}=0, \forall k$, i.e., only NLoS paths exist in the environment, we have
	\begin{align}\label{rate_2}
\tilde{R}_{k}\rightarrow\log _{2}\left(1+\frac{p_{k} \beta \alpha_{k}(M N+M+N+1)}{\sum_{i=1, i \neq k}^{K} p_{i} \beta \alpha_{i}(M+N)+\sigma^{2}}\right).
	\end{align}
\end{corollary}

\itshape \textbf{Proof:}  \upshape The proof can be completed by removing the terms with zero values when setting $\delta=\varepsilon_{k}=0, \forall k$.  \hfill $\blacksquare$

Corollary \ref{corollary4} represents the environment where rich scatters exist and the Rician channel degrades to the Rayleigh channel. We can see that with uncorrelated Rayleigh channel, there is no need to design the phase shifts of RIS. Therefore, in the environment with rich scatters, the phase shifts of RIS can be set arbitrarily. Besides, with a large number of antennas or a large number of reflecting elements, the rate in (\ref{rate_2}) will converge to
\begin{align}
\tilde{R}_{k}\rightarrow\log _{2}\left(1+\frac{p_{k} \alpha_{k} (N+1)}{\sum_{i=1, i \neq k}^{K} p_{i} \alpha_{i}}\right), \mathrm{as} \; M\rightarrow\infty,\\
\tilde{R}_{k}\rightarrow\log _{2}\left(1+\frac{p_{k} \alpha_{k}(M+1)}{\sum_{i=1, i \neq k}^{K} p_{i} \alpha_{i}}\right), \mathrm{as} \; N\rightarrow\infty.
\end{align}

Therefore, even the LoS link does not exist, significant performance gain can be achieved by deploying RIS with large numbers of elements in the massive MIMO systems.

\begin{corollary}\label{corollary5}
	When $\delta = \varepsilon_{k} \rightarrow \infty, \forall k$, i.e., only LoS paths exist, we have
	\begin{align}\label{rate_los}
\bar{R}_{k}\rightarrow\log _{2}\left(1+\frac{p_{k} \beta \alpha_{k} M\left|f_{k}({\bf\Phi})\right|^{2}}{  \sum_{i=1, i \neq k}^{K} p_{i} { \beta \alpha_{i} M\left|f_{i}({\bf\Phi})\right|^{2}}+\sigma^{2}}\right).
	\end{align}
	
	By contrast, in the conventional massive MIMO systems without RIS, the rate under LoS channel $\sqrt {\gamma_{k}}\mathbf{\bar{h}}_k^{w/o}$ is
	\begin{align}
\bar{R}_{k}^{w / o}=\log _{2}\left(1+\frac{p_{k} \gamma_{k}  M}{\sum_{i=1, i \neq k}^{K} p_{i} \gamma_{i}  \frac{\left|\left(\overline{\mathbf{h}}_{k}^{w / o}\right)^{H} \overline{\mathbf{h}}_{i}^{w / o}\right|^{2}}{M}+\sigma^{2}}\right).
\end{align}
\end{corollary}

\itshape \textbf{Proof:}  \upshape Please refer to Appendix \ref{D3}.  \hfill $\blacksquare$

Corollary \ref{corollary5} clearly presents the difference between RIS-aided massive MIMO systems and conventional non-RIS  massive MIMO systems. We can see that in  conventional uplink massive MIMO systems without fast fading, when the number of antennas is large, the multi-user interference term will be zero compared with the useful signal power. However, this property does not hold in the RIS-aided massive MIMO systems with a low-complexity MRC scheme. However, this rate degradation can be compensated by properly designing phase shifts $\mathbf{ \Phi}$. For example,  when the phase shifts are aligned to user $k$, the sum inter-user interference suffered by user $k$ will become negligible compared with the desired signal received by user $k$. This observation emphasizes the importance of the optimization of $\bf\Phi$.

\begin{corollary}\label{corollary6}
	In the RIS-aided massive MIMO systems, if the RIS has discrete phase shifts with $b$ bits resolution, the sum achievable rate can still achieve a gain of $\mathcal{O}\left(\mathrm{log_2}(N)\right)$.
\end{corollary}

\itshape \textbf{Proof:}  \upshape Please refer to Appendix \ref{appF}.  \hfill $\blacksquare$

Corollary \ref{corollary6} states that the phase noise of RIS does not impact the scaling laws, and the rate can still grow without bound when $N\rightarrow\infty$. Corollary \ref{corollary6} indicates that the negative effect brought by RIS's low-resolution elements can be easily compensated by increasing the size of RIS. Therefore, this result demonstrates the feasibility of deploying  low-resolution RIS with a large number of reflecting elements in the massive MIMO systems.

\section{Phase Shifts Design}\label{section4}
In this section, we will design the phase shifts of RIS based on the long-term statistical CSI, which could effectively reduce the training overhead and the frequency of updating phase shifts at the RIS. For the RIS-aided massive MIMO systems, we respectively formulate two optimization problems with different objective functions, and both continuous and discrete phase shifts of RIS are considered. The sum user rate-oriented optimization problem which can characterize the system capacity limitation is formulated as
\begin{subequations}\label{p1}
	\begin{equation}\label{objective1}
	\max\limits_{\mathbf{\Phi}}  \;\; \sum\limits_{k=1}^{K} R_{k},\qquad\qquad\qquad\qquad\qquad\qquad\qquad
	 \end{equation}
	\begin{equation}\text {s.t. } \;\;\theta_{n} \in[0,2 \pi), \forall n, \text{or}\qquad\quad\qquad\qquad\qquad\label{constraint1}\end{equation}
	\begin{equation}\qquad\qquad{\theta}_n \in\left\{0, \frac{2 \pi}{2^{b}}, 2 \times \frac{2 \pi}{2^{b}}, \ldots,\left(2^{b}-1\right) \frac{2 \pi}{2^{b}}\right\}, \forall n,\label{constraint2}\end{equation}
\end{subequations}
where $R_k$ is given by (\ref{rate}).  Note that constraint (\ref{constraint1}) corresponds to the continuous phase shift case, while constraint (\ref{constraint2}) corresponds to the discrete phase shift case with  $b$ bits precision.

Next, the minimum user rate-oriented optimization problem which could guarantee fairness and characterize networks spatial multiplexing is formulated as follows
\begin{subequations}\label{p2}
	\begin{equation}\label{objective2}
	\max\limits_\mathbf{\Phi}  \;\; \min\limits_{k} \;\;R_{k},\qquad\;\end{equation}
	\begin{equation}\text {s.t. } \;\;(\text{\ref{constraint1}}) \; \text{or} \; \text{(\ref{constraint2})}.\nonumber\end{equation}
\end{subequations}

\subsection{Special Cases}
To begin with, we will discuss phase shifts design in some special cases.
\begin{proposition}
For problem (\ref{p1}) and (\ref{p2}), if $N=1$, any phase shift satisfying (\ref{constraint1}) or (\ref{constraint2}) is optimal.
\end{proposition}

\itshape \textbf{Proof:}  \upshape Recalling (\ref{fc_Phi}), if $N=1$ we have $x=y=0$ and $ {f_c}\left( {\bf \Phi}  \right)=e^ { j{\theta _{1}}}$. Therefore, any $\theta_{1}$ will have the same results  of $ \left|{f_c}\left( {\bf \Phi}  \right)\right| = 1$ and $f_{k}^{H}\left( {\bf \Phi}  \right)f_{i}\left( {\bf \Phi}  \right)= 1$.\hfill $\blacksquare$

\begin{proposition}
For problem (\ref{p1}) and (\ref{p2}), 	if $\delta=0$ or $\varepsilon_{k}=0, \forall k$, any phase shift satisfying (\ref{constraint1}) or (\ref{constraint2}) is optimal.
\end{proposition}

\itshape \textbf{Proof:}  \upshape Recalling (\ref{rate}), if $\delta=0$ or $\varepsilon_{k}=0, \forall k$, all terms related to $\bf\Phi$  become zero. \hfill $\blacksquare$

This result indicates that if the environment between the BS and the RIS or that between the RIS and all users has rich scatters, there is no need to design the phase shifts of RIS based on the statistical CSI.

\begin{proposition}\label{propostion3}
For problem (\ref{p1}),	if $p_i=0, \forall i \neq k$, aligning $\bf\Phi$ to user $k$ is optimal.
\end{proposition}

\itshape \textbf{Proof:}  \upshape  If $p_i=0, \forall i \neq k$, it becomes RIS-aided single-user systems. In this case, we can directly apply the results in \cite{han2019large}.  \hfill $\blacksquare$

Proposition \ref{propostion3} also indicates that if a user $k$ locates very close to the RIS, aligning the phase shifts of RIS to this user will almost yield an optimal sum rate.

\subsection{General Case}
Next, we consider the optimization problems (\ref{p1}) and (\ref{p2}) in the general case. Since the expression of the rate has a complicated form and the active and passive beamforming are closely coupled, it is hard to obtain a globally optimal solution in general. Therefore, we propose a GA-based method to solve the two optimization problems.

\begin{figure}[h]
	\setlength{\abovecaptionskip}{0pt}
	\setlength{\belowcaptionskip}{-20pt}
	\centering
	\includegraphics[width= 1\textwidth]{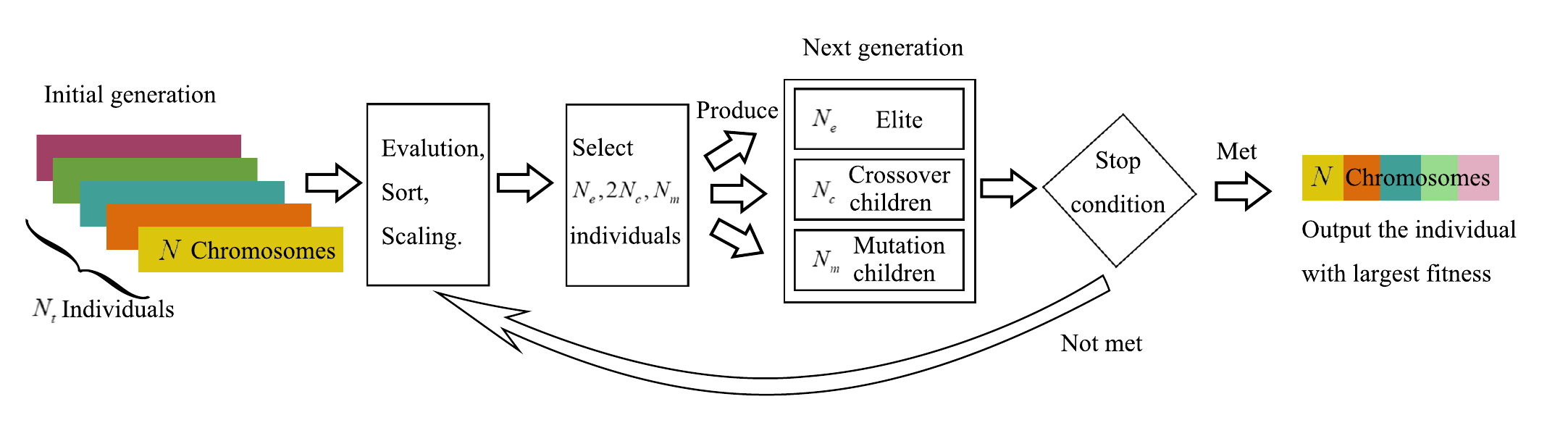}
	\DeclareGraphicsExtensions.
	\caption{The outline of GA-based method.}
	\label{figure2}
\end{figure}

GA simulates the evolution of a population in the nature\cite{matlab2003ga,mitchell1998introduction}, and its main steps are summarized in Fig. \ref{figure2}. Next, we  describe the implementation details of the GA-based optimization method for both the problem (\ref{p1}) and (\ref{p2}).
\subsubsection{Initial population}
GA is initialized by generating a population with $N_t$ individuals. Each individual contains $N$ chromosomes, and the $n$-th chromosome corresponds to RIS's phase shift $\theta_{n}$. For continuous phase shifts, we randomly generate the initial chromosomes of  individuals in the population in $\left[0,2\pi\right)$.  For discrete phase shifts, we randomly generate the initial chromosomes of  individuals from the set in (\ref{constraint2}). Then, we tend to evolve this initial population to the next generation following the steps below.
\subsubsection{Fitness evaluation and scaling}
We first evaluate the fitness of each individual in the current population. The fitness evaluation function is the objective function in (\ref{objective1}) or (\ref{objective2}), respectively. This operation means that  an individual with better fitness in the population corresponds to a better solution for optimization problems (\ref{p1}) or (\ref{p2}). Next, we need to scale the raw fitness value of individuals based on their rank in the population. We sort the raw fitness of individuals and compute their scaled fitness as follows
\begin{align}
\begin{aligned}
&f_{i}=\frac{1}{\sqrt{\operatorname{rank}_{i}}}, \operatorname{rank}_{i} \in\left[1, \ldots, i, \ldots, N_{t}\right],\\
&f_{i}^{\text {scaled}}=2 N_{c} \frac{f_{i}}{\sum_{i=1}^{N_{\text {t }}} f_{i}},
\end{aligned}
\end{align}
where $\operatorname{rank}_{i}$ is the index of raw fitness of individual $i$ after descending sort, $f_{i}^{\text {scaled}}$ is the scaled fitness of individual $i$, $N_c$ is a parameter used in the selection operation. This scaling operation can restrict the effect of individuals with large fitness which may reproduce their chromosomes too frequently and cause prematurity. After the adjustment of raw fitness values, raw fitness values are converted to a more suitable range and we can carry out the selection operation better.
\subsubsection{Selection}
Here we will select some individuals from current population, and some of them are selected as elites, some of them are chosen as parents which could generate offspring. First, $N_e$ individuals with larger $f_{i}^{\text {scaled}}$ are selected as elites, and they will be directly passed to the next generation. Then we will select $2N_c$ parents based on stochastic universal sampling, which has a fast and accurate selection mechanism. To perform stochastic universal sampling, we first form a roulette wheel which has $2N_c$ slots and the size of  slot $i$ is proportional to $f_{i}^{\text {scaled}}$ as follows 
\begin{align}
\operatorname{slot}_{i}=\frac{f_{i}^{\text {scaled}}}{2 N_{c}},
\end{align}
where we have $ \sum_{i=1}^{N_{t}} \operatorname{slot}_{i}=1 $. Therefore, each slot corresponds to an individual. Then we rotate the roulette wheel $2N_c$ times, each time forwarding with an equal step $\frac{1}{2N_c}$. After each time rotation, we find where the wheel  pointer falls and select the corresponding individual as a parent. After $2N_c$ times rotation, we can select $2N_c$ parents which will be used for crossover operation. Note that  one individual may appear multiple times in this $2N_c$ combination, and its appearance probability is proportional to its scaled fitness value. Finally, the remaining $N_m = N_t - N_e -N_c$ individuals will be used for mutation operation.
\subsubsection{Crossover}
We will use  previously selected $2N_c$ parents to perform crossover and generate $N_c$ offspring. Crossover operation can extract the best chromosome from different parents and recombine them into potentially superior offspring. When $N\le 2$, we will use the single point crossover method. Otherwise,  two points crossover method is adopted in this paper. The pseudo codes of crossover operation are shown in Algorithm \ref{algorithm1}.
\begin{algorithm}
	\caption{Crossover Algorithm}
	\begin{algorithmic}[1]\label{algorithm1}
		\STATE Set $c_1=1$, $c_2=2$;
		\IF{$ N> 2 $}
		\FOR {$i=1:N_c$}
		\STATE Select the $c_1$-th and the $c_2$-th parents in the $2N_c$ combination;
		\STATE  Generate different integers $i_1$ and $i_2$ randomly from $\left[1,N-1\right]$;
		\IF{$i_1>i_2$}
		\STATE Swap $i_1$ and $i_2$;
		\STATE Swap parents $c_1$ and $c_2$;
		\ENDIF
		\STATE Generate the $i$-th offspring by $\left[ \text{parent } c_1 (1:i_1) , \text{parent } c_2 (i_1+1,i_2) ,\text{parent } c_1 (i_2+1,N)  \right]$;
		\STATE $c_1=c_1+2$, $c_2=c_2+2$;
		\ENDFOR
		\ELSE
		\STATE Generate a random integer $i_1$ and perform single point crossover;
		\ENDIF
	\end{algorithmic}
\end{algorithm}
\subsubsection{Mutation}
$N_m$ parents will experience mutation operation with probability $p_m$ and produce $N_m$ offspring. Mutation operation can increase the diversity of the population and bring the likelihood that offspring with better fitness is generated. We use the uniform mutation method and its pseudo codes are shown in Algorithm \ref{algorithm2}.
\begin{algorithm}
	\caption{Mutation Algorithm}
	\begin{algorithmic}[1]\label{algorithm2}
		\FOR{$i=1:N_m$}
		\FOR{$ n=1:N $}
		\IF{$\text{rand}\left(1\right)<p_m$}
		\IF{RIS has continous phase shifts}	
		\STATE the $n$-th chromosome $\theta_{n}$ of parent $i$ mutates to $2\pi\times\text{rand}(1)$;
		\ELSE
		\STATE the $n$-th chromosome ${\theta}_{n}$ of parent $i$ mutates to a value randomly selected from the set in (\ref{constraint2});
		\ENDIF
		\ENDIF		
		\ENDFOR
		\ENDFOR
	\end{algorithmic}
\end{algorithm}

After the above operations, we  combine the $N_e$ elite children, $N_c$ children coming from crossover and $N_m$ children coming from mutation to form the next generation population. The GA will stop if the number of generations is larger than $N_{max}$ or the change of the average fitness value is less than $\xi $.

\section{Numerical Results}\label{section5}
In this section, we  validate our analysis and demonstrate the benefits brought by deploying RIS into massive MIMO systems. Our simulation parameters are set as in \cite{wu2019intelligent,pan2020multicell}. We assume the locations of the BS and the RIS are $(0,0,25)$ and $(5,100,30)$, respectively. Similar to \cite{wu2019intelligent}, we assume users are located on a half circle centered at $(5,100)$ with radius of $5$m and height of $1.6$m. The AoA and AoD of BS, RIS and users are generated randomly from $[0,2\pi]$\cite{pan2020intelligent,pan2020multicell} and these angles will be fixed after initial generation. Unless otherwise stated, our simulation parameters are set as follows: element spacing of $d=\frac{\lambda}{2}$, number of users of $K=4$, number of reflecting elements of $N=16$, number of antennas of  $M=64$, transmit power of  $p_k = 30$ dBm, $\forall k$, noise power of $\sigma^2=-104$ dBm and Rician factor of $\delta=1$, $\varepsilon_{k}=10, \forall k$. Large-scale path-loss is calculated as $\alpha_k=\frac{1}{1000 {d_{k}}^{\alpha_{kUR}}}, \forall k$ and $\beta=\frac{1}{1000 d_{0}^{\beta_{RB}}}$ \cite{pan2020multicell} where $d_{k}$ and $d_{0}$ are respectively the distances of user $k$-RIS and RIS-BS, and the path-loss exponents are $\alpha_{kUR}=\beta_{RB}=2.8, \forall k$ \cite{wu2019intelligent}. GA population $ N_t $ is 200, elite number is $N_e=10$, number of crossover parents is $N_c=152$, number of mutation parents is $N_m=38$. The following  simulation results are obtained by averaging over $10000$ random channel generations.

\begin{figure}
	\setlength{\abovecaptionskip}{0pt}
	\setlength{\belowcaptionskip}{-20pt}
	\centering
	\includegraphics[width=5in]{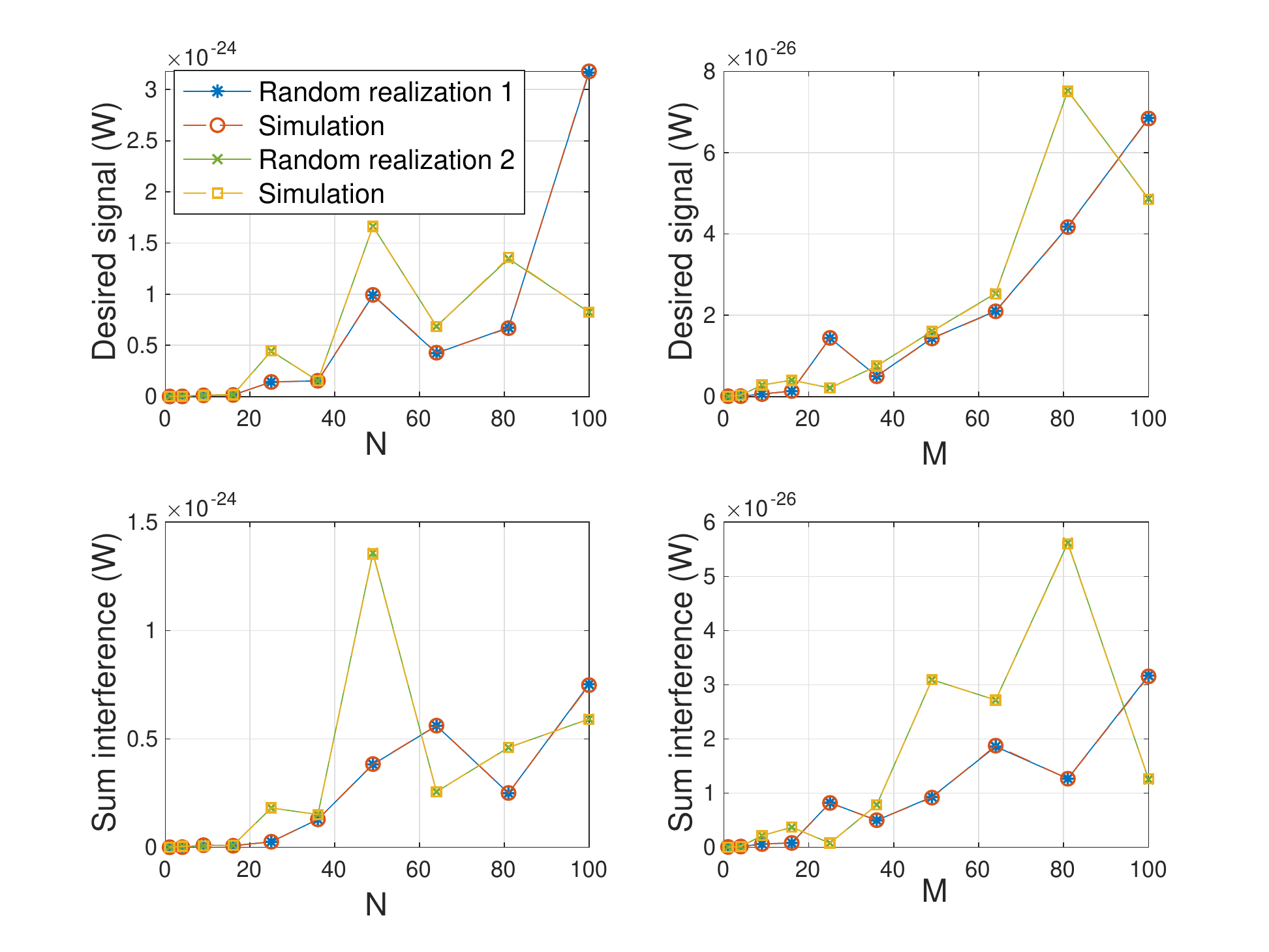}
	\DeclareGraphicsExtensions.
	\caption{Desired signal power and sum interference power of user $1$ under  random RIS phase shifts.}
	\label{figure3}
\end{figure}
To begin with, we validate the correctness of our key derivation in Lemma \ref{lemma2}. In Fig. \ref{figure3}, we show the desired signal $\mathbb{E}\left\{\left\|\mathbf{g}_{1}\right\|^{4}\right\}$ and sum inter-user interference $\sum_{i=2}^{4} \mathbb{E}\left\{\left|\mathbf{g}_{1}^{H} \mathbf{g}_{i}\right|^{2}\right\} $ for user $1$ under two  independent random realizations of $\mathbf{\Phi}$. Fig. \ref{figure3} shows that our derived expressions perfectly match the Monte Carlo simulation, which verifies the accuracy of our results.

Next, we evaluate the impact of various system parameters on the data rate of the RIS-aided massive MIMO system. To this end, two kinds of optimal phase shifts $\mathbf{\Phi^*}$ are obtained by respectively solving optimization problem (\ref{p1}) and problem (\ref{p2}), and the obtained $\mathbf{\Phi^*}$ will be used to  calculate two different performance metrics, i.e., the sum user rate $\sum_{k=1}^{K} R_k\left(\mathbf{\Phi}^*\right)$ and the minimum user rate $\min\limits_{k} R_k\left(\mathbf{\Phi^*}\right)$. We refer to the sum user rate calculated by $\mathbf{\Phi}^*$ obtained from problem (\ref{p1}) as ``sum rate by max-sum'', refer to the minimum user rate calculated by  $\mathbf{\Phi}^*$ obtained from problem (\ref{p1}) as ``min rate by max-sum'', refer to the sum  rate calculated by  $\mathbf{\Phi}^*$ obtained from problem (\ref{p2}) as ``sum rate by max-min'' and refer to the minimum user rate calculated by  $\mathbf{\Phi}^*$ obtained from problem (\ref{p2}) as ``min rate by max-min'', respectively. Besides, we will calculate the sum rate and minimum user rate under random RIS phase shifts setting by averaging over 1000 random phase shifts generations.
\subsection{Trade-off between path-loss and spatial multiplexing}
\begin{figure}
	\setlength{\abovecaptionskip}{0pt}
	\setlength{\belowcaptionskip}{-20pt}
	\centering
	\includegraphics[width=5in]{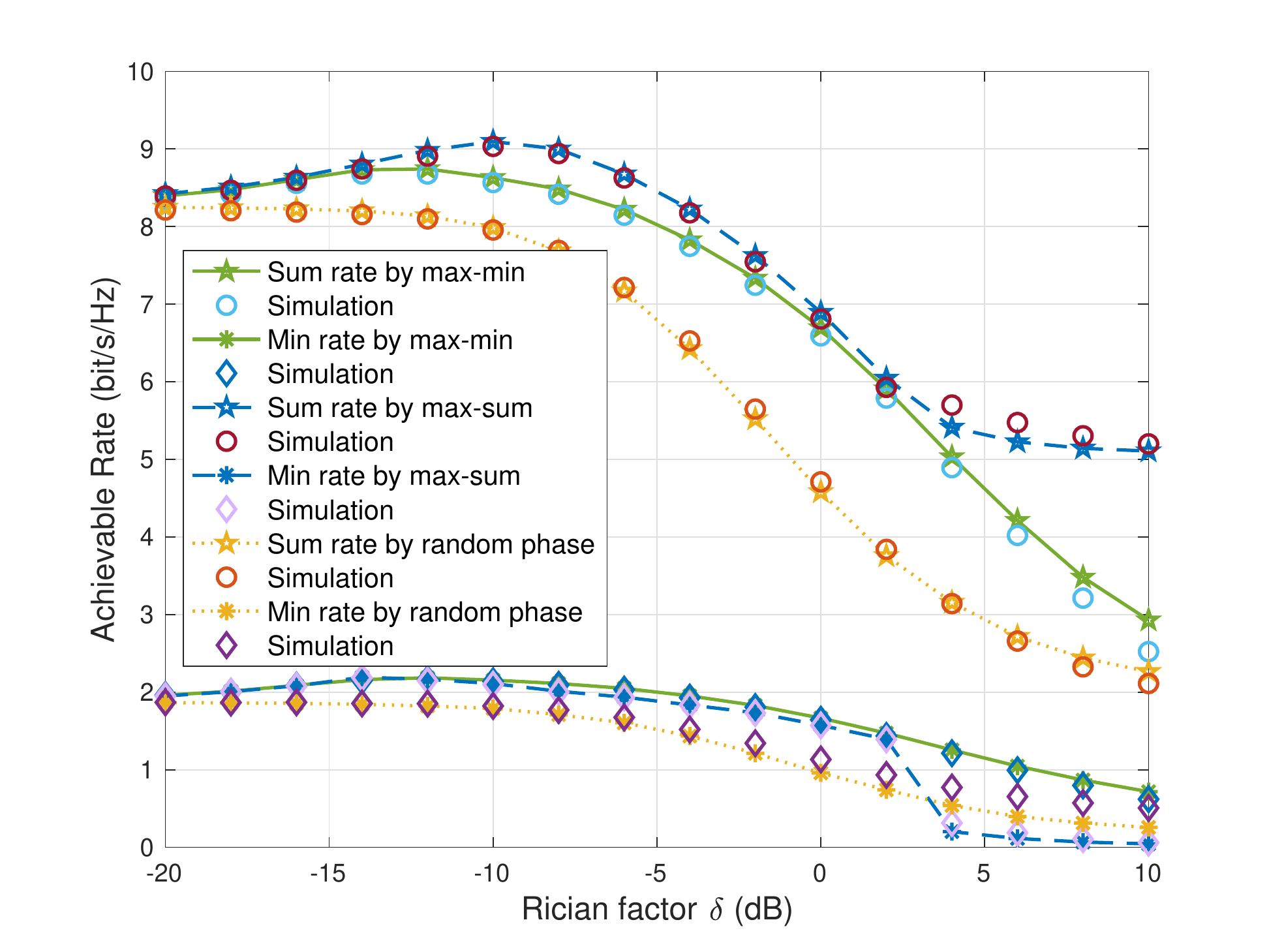}
	\DeclareGraphicsExtensions.
	\caption{Sum rate and minimum user rate vesus the Rician factor of RIS-BS channel. }
	\label{figure4}
\end{figure}
Fig. \ref{figure4} shows the four different kinds of rate versus the Rician factor $\delta$ of RIS-BS channel $\mathbf{H}_2$.  Results show that our approximated analytical rate expression (\ref{rate}) matches well with the  simulation result, which verifies the correctness of the derived results.  We can see that when $\delta$ is small, both the sum rate maximization (\ref{p1}) and minimum rate maximization (\ref{p2}) lead to a similarly good  performance (both in terms of sum rate and minimum rate). This means that in the rich scattering environment, one can simultaneously achieve a large system capacity while guaranteeing  user fairness.  However, when  $\delta$ increases, it becomes impossible to balance the system capacity and  fairness. If we maximize the sum rate, the minimum user rate will approach zero. Conversely, if we want to maintain the minimum rate, the  sum rate will be severely degraded, which nearly equals the rate  achieved by random phases. This result is totally different from the RIS-aided single-user system with statistical CSI\cite{han2019large,jia2020analysis}, whose rate performance will be improved by increasing the Rician factor. The reason lies in that when $\delta$ increases,  channel LoS components will become more dominant, which increases the channel correlation between different users, as well as increases the inter-user interference and reduces the  spatial multiplexing gain. Specifically, when $\delta\rightarrow\infty$, the rank of the cascaded channel $\mathbf{G}$ will approach one, and the system will be incapable of supporting the communication of multiple users. 

\begin{figure}
	\setlength{\abovecaptionskip}{0pt}
	\setlength{\belowcaptionskip}{-20pt}
	\centering
	\includegraphics[width=5in]{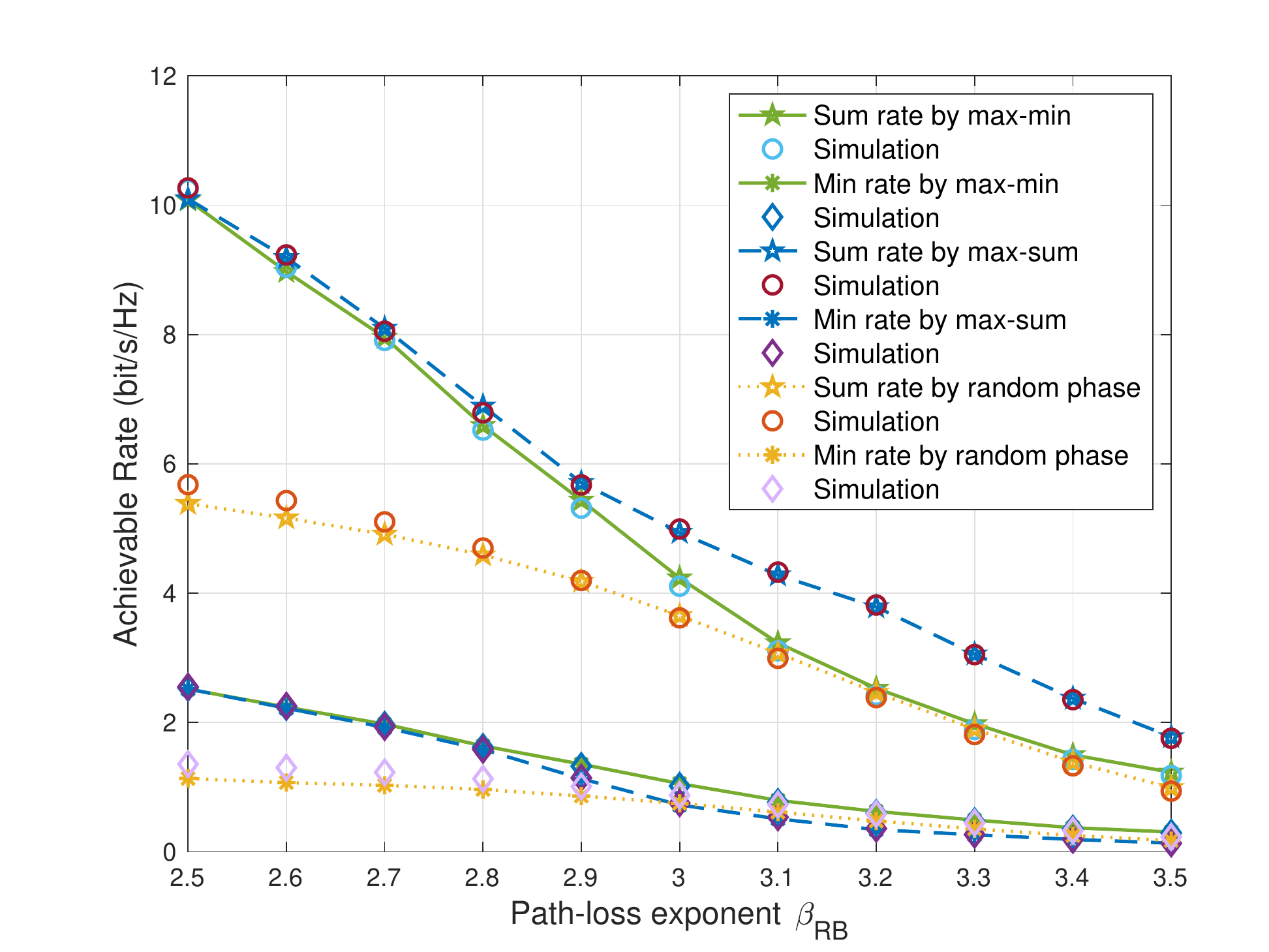}
	\DeclareGraphicsExtensions.
	\caption{Sum rate and minimum rate versus the path-loss exponent $\beta_{RB}$ of RIS-BS channel.}
	\label{figure5}
\end{figure}

According to the above discussion, we know it is better to deploy the RIS in the environment with relatively rich scatters to support multi-user communications. However, to ensure the rich scatters, the distance between the BS and RIS should be increased, yielding an increased path loss and a larger path-loss exponent. Therefore, we  present Fig. \ref{figure5} to show the impacts of  RIS-BS channel path-loss exponent $\beta_{RB}$.
Firstly, we can see that when $\beta_{RB}$ is small, both the max-sum problem (\ref{p1}) and max-min problem (\ref{p2}) can achieve similarly good performance. Secondly, we can see that as $\beta_{RB}$ keeps increasing, if we want to maintain fairness, the rate performance (sum-rate and min-rate) will decrease and eventually approach the rate achieved by random phases. These observations indicate that if we want to simultaneously achieve high system throughput and guarantee fairness, the path-loss exponent should be as small as possible, which corresponds to short distances and high value of Rician factors. Therefore, there exists a trade-off between the achievable spatial multiplexing gain and the unwanted channel path-loss.

\subsection{The interplay between RIS and massive MIMO}
In this subsection, we aim to answer the question about what benefits are brought by deploying RIS in massive MIMO systems. Note that to guarantee  fairness, only the minimum user rate maximization (\ref{p2}) is considered in the following simulation.

\begin{figure}
	\setlength{\abovecaptionskip}{0pt}
	\setlength{\belowcaptionskip}{-20pt}
	\centering
	\includegraphics[width=5in]{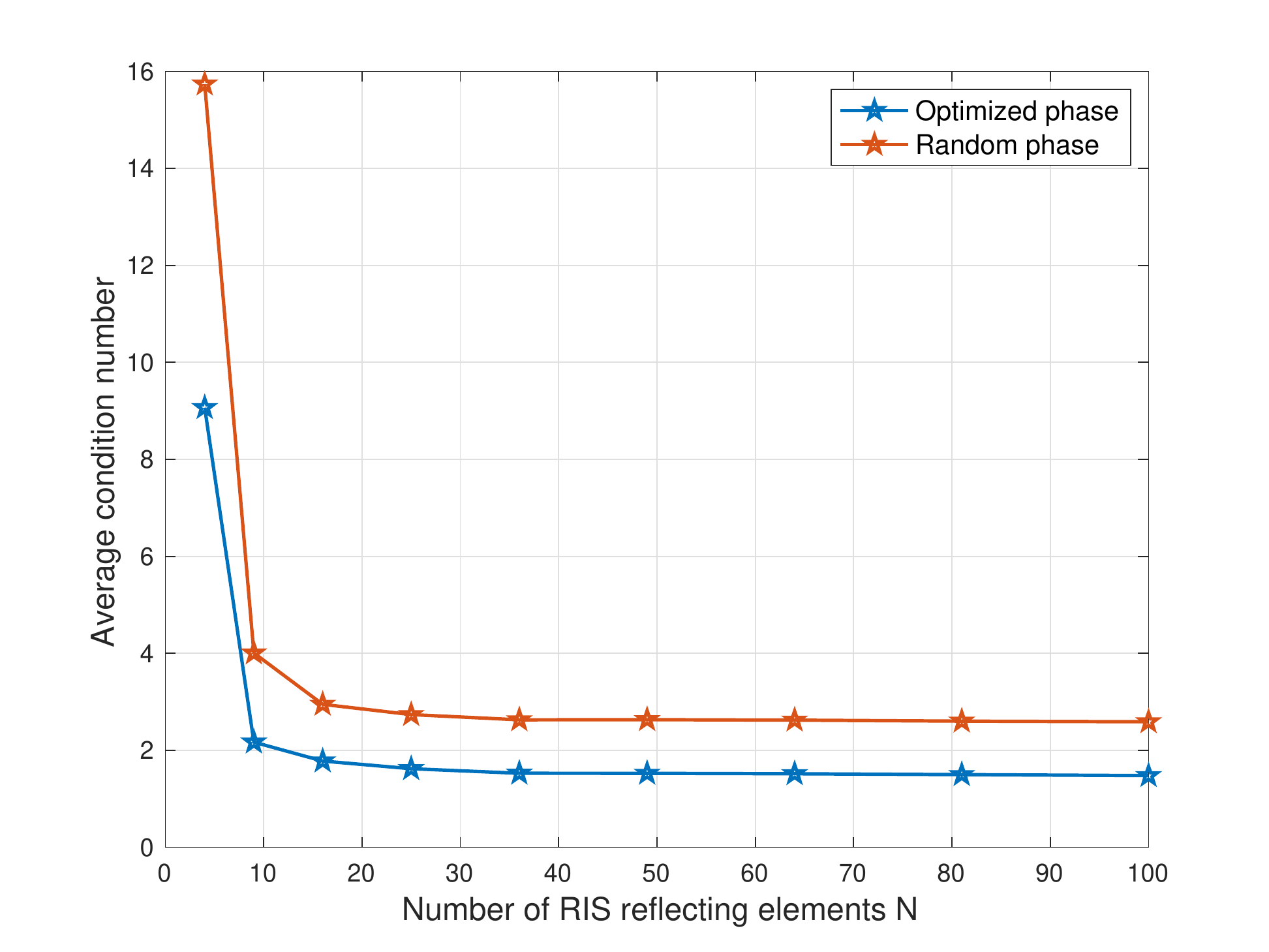}
	\DeclareGraphicsExtensions.
	\caption{Average condition numbers of the cascaded channel $\mathbf{G}$.}
	\label{figure6}
\end{figure}
Fig. \ref{figure6} shows the standard condition number (i.e., the ratio between the largest to the smallest eigenvalue \cite{matthaiou2010condition}) of the cascaded channel $\mathbf{G}$ versus the number of RIS elements $N$, and the result is obtained from Monte Carlo simulation. It is well known that channel matrix with a lower condition number can achieve better performance in the high signal-to-noise ratio (SNR) regime\cite{tse2005fundamentals}, and the channel matrix with condition number $1$ is referred to as ``well-conditioned''. Fig. \ref{figure6} shows that the condition number of the cascaded channel decreases quickly as $N$ increases. Besides, after the optimization of RIS's phase shifts, we can see that the channel will  become nearly well-conditioned. This finding indicates that RIS can reshape the channel in massive MIMO  systems, reduce the disparity among the channel singular values and achieve a higher capacity in the high SNR regime.

\begin{figure}
	\setlength{\abovecaptionskip}{0pt}
	\setlength{\belowcaptionskip}{-20pt}
	\centering
	\includegraphics[width=5in]{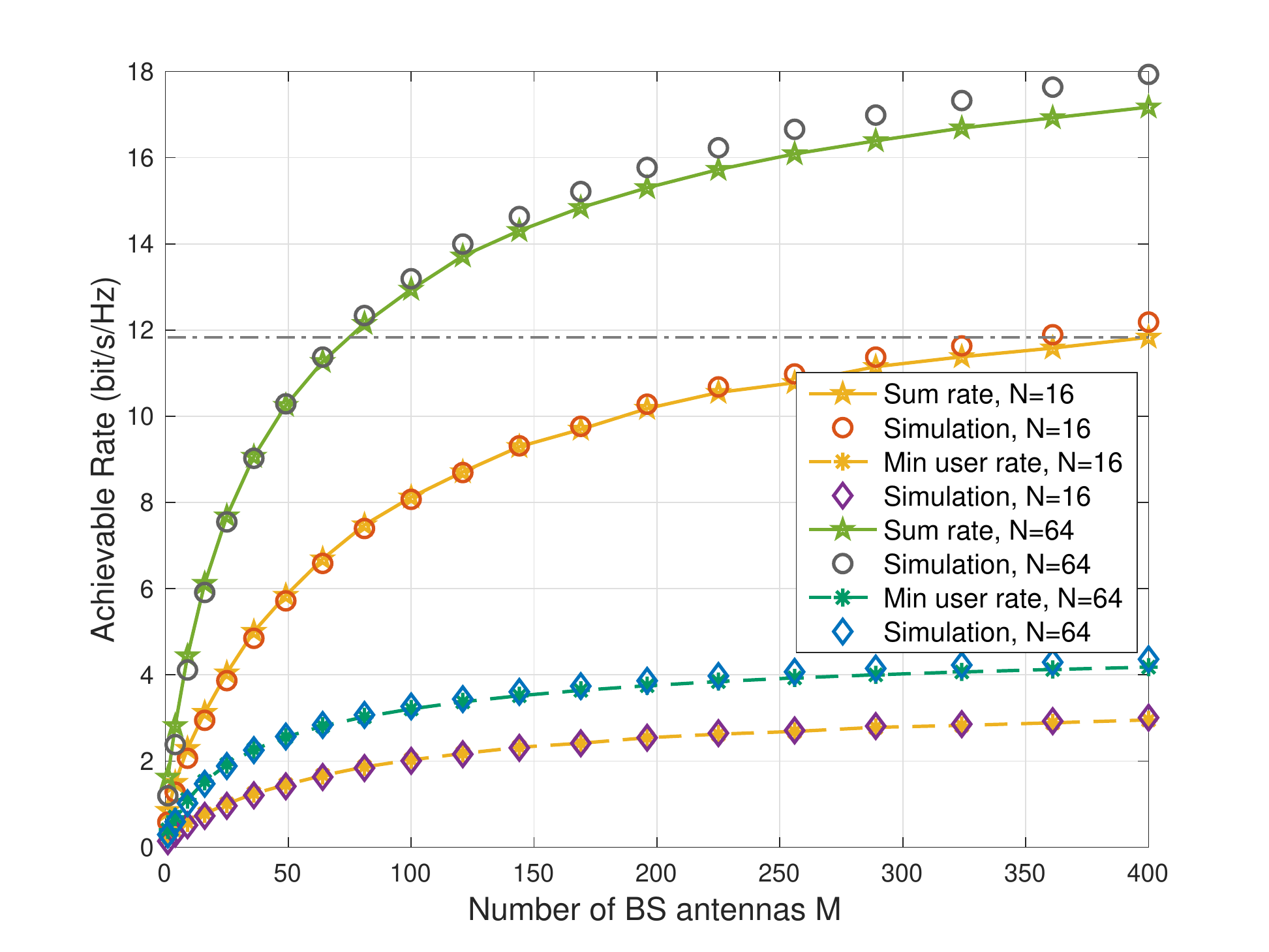}
	\DeclareGraphicsExtensions.
	\caption{Sum rate and minimum user rate versus the number of BS antennas $M$.}
	\label{figure7}
\end{figure}
Fig. \ref{figure7} shows the data rate performance  of RIS-aided massive MIMO systems with the simple MRC technique. We can see that although the inter-user interference makes the minimum rate and sum rate approach saturation when $M\rightarrow\infty$, it still has some promising features. Firstly, by increasing the number of RIS elements $N$, the data rate can be significantly improved, which demonstrates the benefits of integrating RIS into massive MIMO networks. By contrast, in the conventional massive MIMO networks without RIS, the number of the antennas should be extremely large to serve excessive number of users. However, the increase of the number of active antennas requires a large-sized  array, high power consumption and high hardware cost. By observing Fig. \ref{figure7}, we can find that thanks to RIS's passive beamforming gain, only a moderate number of antennas are enough to bring promising throughput. For example, $100$ antennas with $64$ RIS elements can outperform $400$ antennas with $16$ RIS elements. Therefore, RIS-aided massive MIMO systems are promising to be applied in future communication systems with much reduced hardware cost and power consumption, while still maintaining the network capacity requirement.

\begin{figure}
	\setlength{\abovecaptionskip}{0pt}
	\setlength{\belowcaptionskip}{-20pt}
	\centering
	\includegraphics[width=5in]{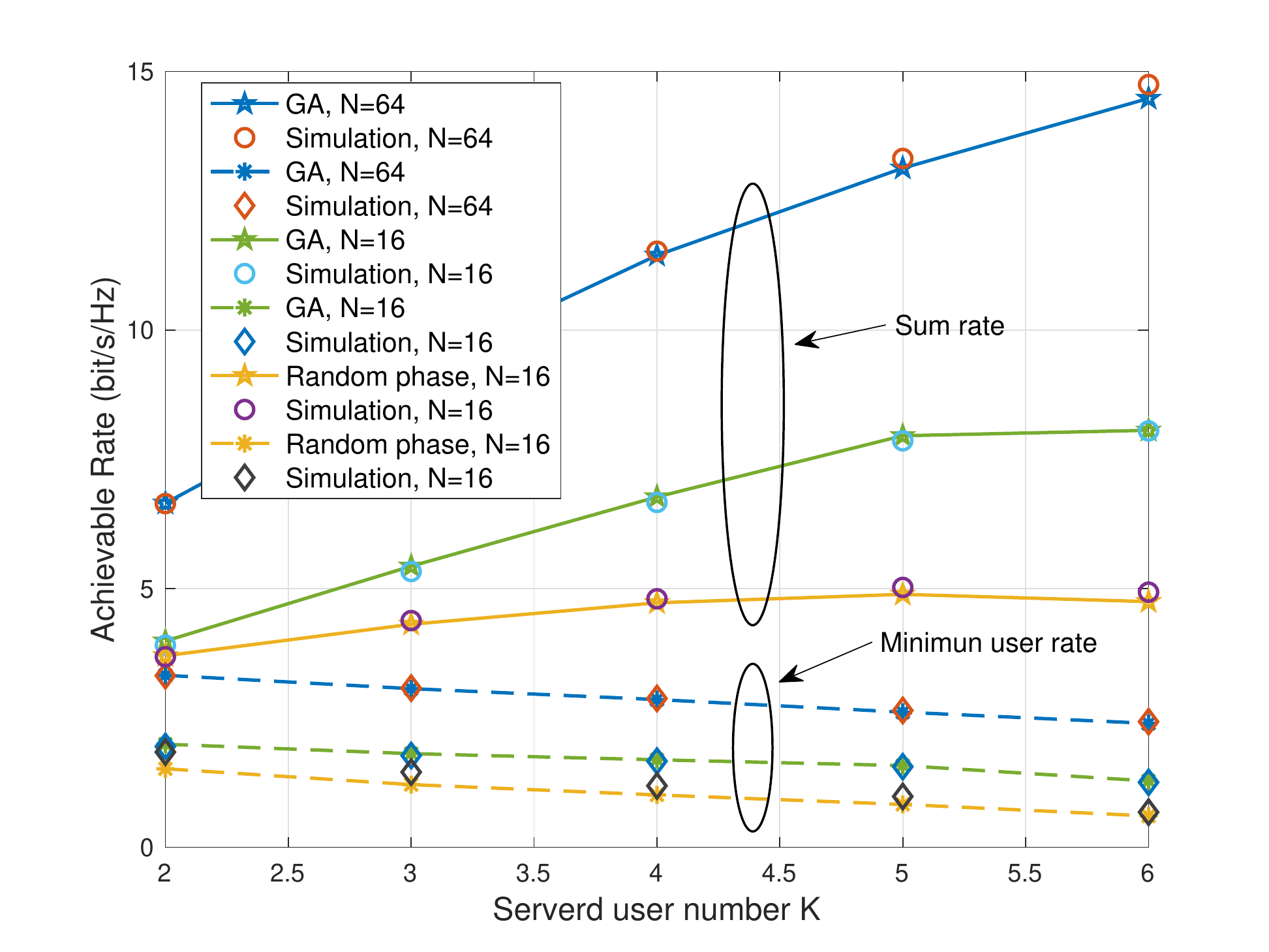}
	\DeclareGraphicsExtensions.
	\caption{Achievable rate versus the number of served users $K$.}
	\label{figure9}
\end{figure}
Fig. \ref{figure9} further examines the capability of supporting multiple users in RIS-aided massive MIMO systems. Here we increase the number of users which are located on the same circle centered at the IRS with a radius of $5$m, and six users' angles are randomly generated. We can see that the minimum user rate decreases with the increase of the number of users, but the sum rate increases with $K$. This result is reasonable since we consider the minimum user rate maximization problem. Although the minimum user rate decreases by increasing the number of users served simultaneously, it can be significantly promoted by increasing the number of RIS's elements and carefully designing RIS's phase shifts.

\begin{figure}
	\setlength{\abovecaptionskip}{0pt}
	\setlength{\belowcaptionskip}{-20pt}
	\centering
	\includegraphics[width=5in]{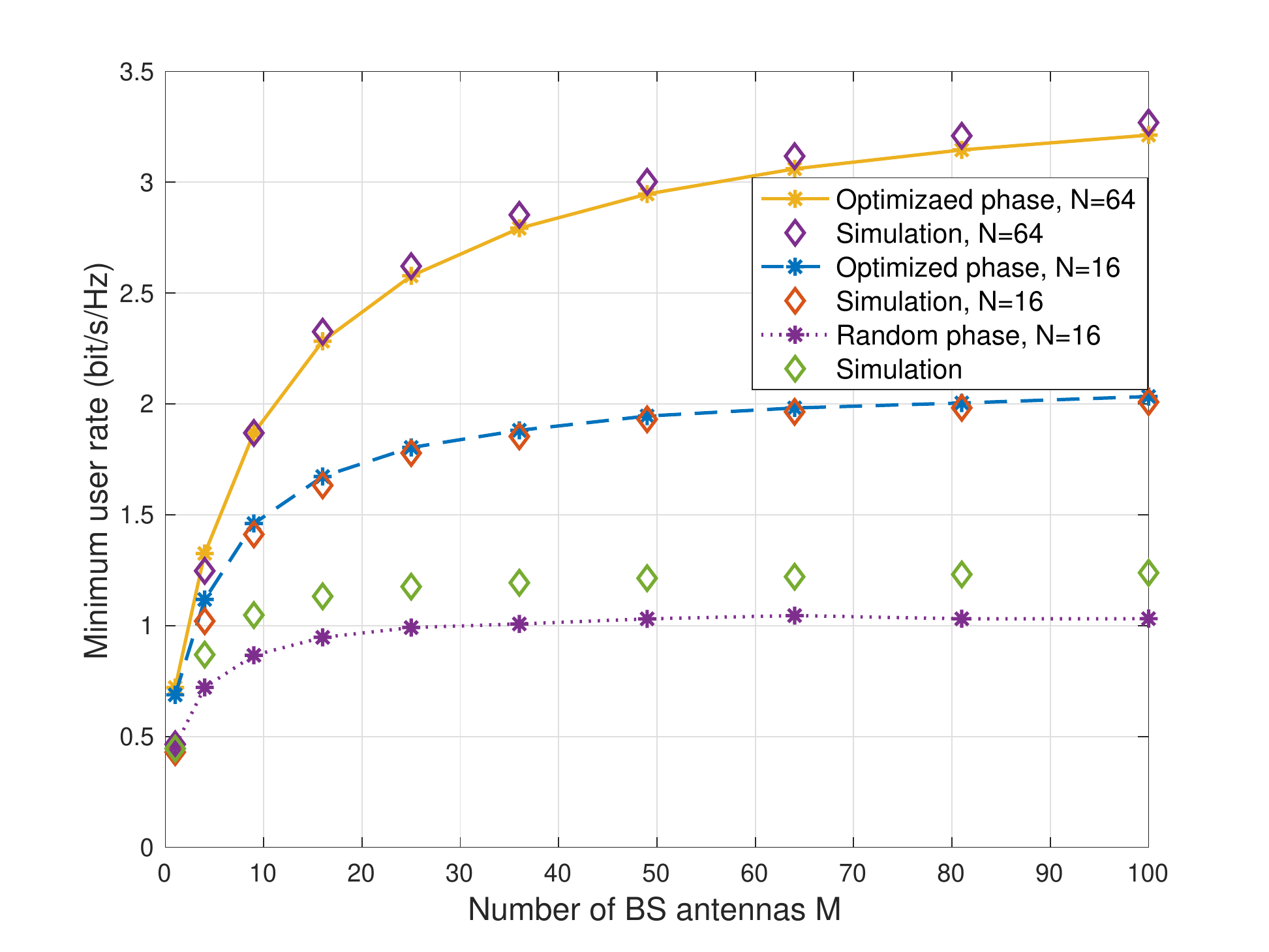}
	\DeclareGraphicsExtensions.
	\caption{Minimum user rate versus the number of BS antennas, with scaled transmission power $p_k=100/M, \forall k$.}
	\label{figure8}
\end{figure}

In  Fig. \ref{figure8},  we examine the power scaling laws in the RIS-aided massive MIMO systems, where the transmit power of each user is scaled down as  $p_k=100/M, \forall k$.
It has been proved that the massive MIMO technique can help users decrease their uplink transmitting power while maintaining the data rate performance \cite{ngo2013energy}. Besides, in the RIS-aided massive MIMO systems, the transmit power of each user can be further cut down by carefully designing the phase shifts of RIS relying on statistical CSI. Meanwhile, the increase of RIS's size also has a positive impact on saving  power consumption.

\subsection{The impacts brought by RIS limited precision}
\begin{figure}
	\setlength{\abovecaptionskip}{0pt}
	\setlength{\belowcaptionskip}{-20pt}
	\centering
	\includegraphics[width=5in]{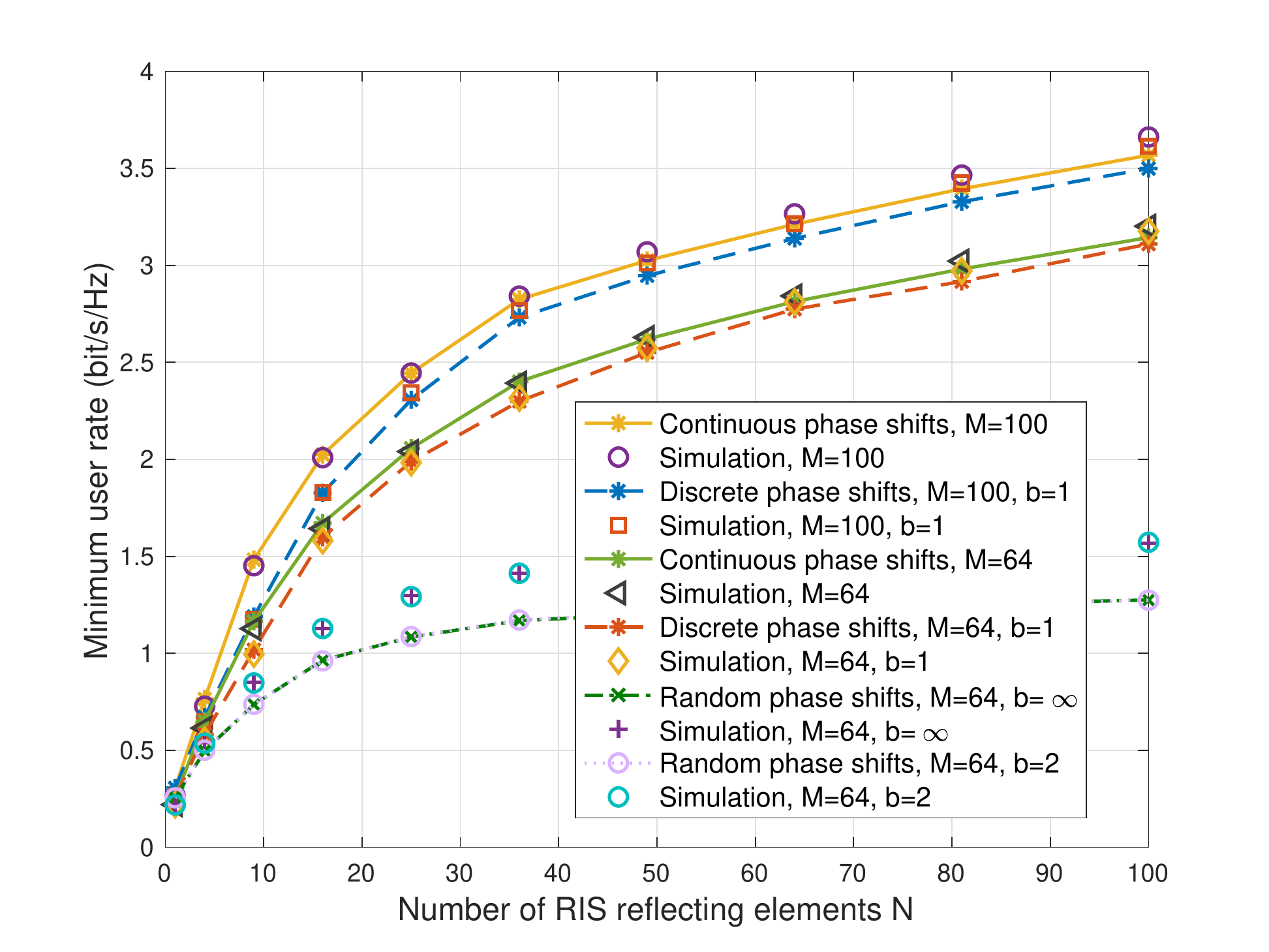}
	\DeclareGraphicsExtensions.
	\caption{Minimum user rate under continuous and discrete phase shifts.}
	\label{figure10}
\end{figure}

Finally, in Fig. \ref{figure10}, we assess the performance degradation brought by RIS's discrete phase shifts in massive MIMO systems.
Firstly, we can see that both random continuous phase shifts and random discrete phase shifts lead to the same rate performance, which is consistent with our derivation in Corollary \ref{corollary2}. Secondly, we can see that in the RIS-aided massive MIMO systems, the degradation due to low-resolution reflecting elements is marginal which does not enlarge when $N$ increases. Hence, it will not be an implementation bottleneck in practical systems. Meanwhile, the degradation can be easily compensated by increasing $N$, and the degradation does not enlarge when increasing the number of antennas at BS.  We conjecture that the reason for this phenomenon lies in that the robustness of data rate against the low-resolution of individual reflecting elements is increased by means of large $N$ and large $M$. Therefore, it is feasible to deploy RIS with low hardware cost but large size in massive MIMO systems.

\section{Conclusion}\label{section6}
This paper has analyzed and optimized the uplink achievable rate performance in the uplink RIS-aided massive MIMO systems with statistical CSI. We have applied the RIS to provide an additional  communication link to the user in the dead zone of conventional massive MIMO systems. We have designed  the phase shifts of the RIS based on statistical CSI, which could reduce the implementation complexity and the signaling overhead. To this end, first, we have derived the closed-form expressions for the uplink achievable rate which hold for any finite numbers of BS antennas. We have then investigated the power scaling laws, analyzed the rate under some special cases and presented the average asymptotic rate achieved by the random phase shift setting. Then, we have studied the optimal phase shifts in some special cases and used the GA-based method to solve the sum-rate maximization and the minimum user rate maximization problems in the general case. Finally, we have provided the numerical results to validate the potential of integrating RIS into existing massive MIMO systems. Our results have revealed the trade-off between the achievable spatial multiplexing gain and unwanted path-loss. Besides, we have demonstrated that it is promising to use RIS with low-resolution hardware to enhance the coverage in massive MIMO systems.

\begin{appendices}

%
%

	\section{}\label{appB}
	To begin with, we present some definition and properties which will be utilized in the following derivation.
	
	According to the definition of Rician channels in (\ref{rician1}) and (\ref{rician2}), we can rewrite the cascaded channels ${\bf g}_k$ for user $k$ and ${\bf g}_i$ for user $i$ as follows
	\begin{align}
	&\mathbf{g}_{k}=\!\mathbf{H}_{2} 	{\bf \Phi} \mathbf{h}_{k}=\!\sqrt{\frac{\beta \alpha_{k}}{(\delta+1)\left(\varepsilon_{k}+1\right)}}(\underbrace{\sqrt{\delta \varepsilon_{k}} \overline{\mathbf{H}}_{2} 	{\bf \Phi} \overline{\mathbf{h}}_{k}}_{\mathbf{g}_{k}^{1}}+\underbrace{\sqrt{\delta} \overline{\mathbf{H}}_{2} 	{\bf \Phi} \tilde{\mathbf{h}}_{k}}_{\mathbf{g}_{k}^{2}}+\underbrace{\sqrt{\varepsilon_{k}} \tilde{\mathbf{H}}_{2} 	{\bf \Phi} \overline{\mathbf{h}}_{k}}_{\mathbf{g}_{k}^{3}}+\underbrace{\tilde{\mathbf{H}}_{2} 	{\bf \Phi} \tilde{\mathbf{h}}_{k}}_{\mathbf{g}_k^{4}}),\label{gk}\\
	&\mathbf{g}_{i}=\mathbf{H}_{2} 	{\bf \Phi} \mathbf{h}_{i}=\sqrt{\frac{\beta \alpha_{i}}{(\delta+1)\left(\varepsilon_{i}+1\right)}}(\underbrace{\sqrt{\delta \varepsilon_{i}} \overline{\mathbf{H}}_{2} 	{\bf \Phi} \overline{\mathbf{h}}_{i}}_{\mathbf{g}_{i}^{1}}+\underbrace{\sqrt{\delta} \overline{\mathbf{H}}_{2} 	{\bf \Phi} \tilde{\mathbf{h}}_{i}}_{\mathbf{g}_{i}^{2}}+\underbrace{\sqrt{\varepsilon_{i}} \tilde{\mathbf{H}}_{2} 	{\bf \Phi} \overline{\mathbf{h}}_{i}}_{\mathbf{g}_{i}^{3}}+\underbrace{\tilde{\mathbf{H}}_{2} 	{\bf \Phi} \tilde{\mathbf{h}}_{i}}_{\mathbf{g}_{i}^{4}}).\label{gi}
	\end{align}
	
	Note that $ \tilde{\mathbf{H}}_{2}$, $\tilde{\mathbf{h}}_{k}$ and $\tilde{\mathbf{h}}_{i}$ are independent with each other, and $ \tilde{\mathbf{H}}_{2}$, $\tilde{\mathbf{h}}_{k}$ and $\tilde{\mathbf{h}}_{i}$ are composed of independent and identically distributed random variables following $\mathcal{CN}\left(0,1\right)$. Therefore, for arbitrary $m$ and $n$, we have
	\begin{align}\label{property1}
	\begin{array}{l}
	\mathbb{E}\left\{\left[\tilde{\mathbf{H}}_{2}\right]_{m n}\right\}=\mathbb{E}\left\{\tilde{\mathrm{h}}_{k m}\right\}=\mathbb{E}\left\{\tilde{\mathrm{h}}_{i m}\right\}=0, \\
		\mathbb{E}\left\{\left[\tilde{\mathbf{H}}_{2}\right]_{m n} \tilde{\mathrm{h}}_{k m} \tilde{\mathrm{h}}_{i m}\right\}=\mathbb{E}\left\{\left[\tilde{\mathbf{H}}_{2}\right]_{m n} \right\}  \mathbb{E}\left\{ \tilde{\mathrm{h}}_{k m}\right\}  \mathbb{E}\left\{ \tilde{\mathrm{h}}_{i m}\right\}=0, \\
	\mathbb{E}\left\{\tilde{\mathrm{h}}_{k m}\tilde{\mathrm{h}}^*_{k n}\right\}=\mathbb{E}\left\{\tilde{\mathrm{h}}_{k m}\right\}\mathbb{E}\left\{\tilde{\mathrm{h}}^*_{k n}\right\}=0, \forall m\neq n\\
	\mathbb{E}\left\{\left|\tilde{\mathrm{h}}_{k m}\right|^{2}\right\}=\mathbb{E}\left\{\left|\tilde{\mathrm{h}}_{i m}\right|^{2}\right\}=\mathbb{E}\left\{\left|\left[\tilde{\mathbf{H}}_{2}\right]_{m n}\right|^{2}\right\}=1,
	\end{array}
	\end{align}
	where $\left[{\mathbf{H}}\right]_{m n}$ denotes the $(m,n)$-th entry of matrix ${\mathbf{H}}$ and $\left[\mathbf{h}_{c}\right]_{m} \triangleq {\rm{h}}_{c m}$ represents the $m$-th element of column vector ${\bf h}_c$.
	
	Next, we will derive $\mathbb{E}\left\{\left\|\mathbf{g}_{k}\right\|^{2}\right\}$, $\mathbb{E}\left\{\left\|\mathbf{g}_{k}\right\|^{4}\right\}$ and $\mathbb{E}\left\{\left|\mathbf{g}_{k}^{H} \mathbf{g}_{i}\right|^{2}\right\}$, respectively.

	 \subsection{Derivation of $\mathbb{E}\left\{\left\|\mathbf{g}_{k}\right\|^{2}\right\}$}
	 Using the definition in (\ref{gk}), $\mathbb{E}\left\{\left\|\mathbf{g}_{k}\right\|^{2}\right\}$ can be written as
	 \begin{align}
	 \mathbb{E}\left\{\left\|\mathbf{g}_{k}\right\|^{2}\right\} = \mathbb{E}\left\{\mathbf{g}^H_{k}\mathbf{g}_{k}   \right\}  =\frac{\beta \alpha_{k}}{(\delta+1)\left(\varepsilon_{k}+1\right)} \mathbb{E}\left\{\sum_{\omega=1}^{4}\left(\mathbf{g}_{k}^{\omega}\right)^{H} \sum_{\psi=1}^{4} \mathbf{g}_{k}^{\psi}\right\}.
	 \end{align}
	 
	 Based on (\ref{property1}), we have
	 \begin{align}
	 \begin{array}{l}
	 \mathbb{E}\left\{\left(\mathbf{g}_{k}^{\omega}\right)^{H} \mathbf{g}_{k}^{\psi}\right\}=0, \forall \omega \neq \psi. 
	 \end{array}
	 \end{align}
	 
	 Therefore, we have
	 \begin{align}\label{lemma1kk}
	 \begin{array}{l}
	 \mathbb{E}\left\{\mathbf{g}_{k}^{H} \mathbf{g}_{k}\right\}\\
	 =\frac{\beta \alpha_{k}}{(\delta+1)\left(\varepsilon_{k}+1\right)} \mathbb{E}\left\{\sum_{\omega=1}^{4}\left(\mathbf{g}_{k}^{\omega}\right)^{H}  \mathbf{g}_{k}^{\omega}\right\} \\
	 =\frac{\beta \alpha_{k}}{(\delta+1)\left(\varepsilon_{k}+1\right)}\left(\delta \varepsilon_{k}\left\|\overline{\mathbf{H}}_{2} {\bf \Phi}	 \overline{\mathbf{h}}_{k}\right\|^{2}+\delta \mathbb{E}\left\{\left\|\overline{\mathbf{H}}_{2} {\bf \Phi}	 \tilde{\mathbf{h}}_{k}\right\|^{2}\right\}+\varepsilon_{k} \mathbb{E}\left\{\left\|\tilde{\mathbf{H}}_{2} {\bf \Phi}	 \overline{\mathbf{h}}_{k}\right\|^{2}\right\}+\mathbb{E}\left\{\left\|\tilde{\mathbf{H}}_{2} {\bf \Phi}	 \tilde{\mathbf{h}}_{k}\right\|^{2}\right\}\right) \\
	 {\mathop  = \limits^{\left( a \right)} }\frac{\beta \alpha_{k}}{(\delta+1)\left(\varepsilon_{k}+1\right)}\left(\delta \varepsilon_{k} M\left|f_{k}({\bf \Phi}	)\right|^{2}+\delta M N+\varepsilon_{k} M N+M N\right) \\
	 =M\frac{\beta \alpha_{k}}{(\delta+1)\left(\varepsilon_{k}+1\right)}\left(\delta \varepsilon_{k}\left|f_{k}({\bf \Phi}	)\right|^{2}+\left(\delta+\varepsilon_{k}+1\right) N\right),
	 \end{array}
	 \end{align}
	 where $(a)$ utilizes the following results
	 \begin{align}
	 \begin{array}{l}
	 \left\|\overline{\mathbf{H}}_{2} {\bf \Phi}	 \overline{\mathbf{h}}_{k}\right\|^{2}=\left\|\mathbf{a}_{M}\left(\phi_{r}^{a}, \phi_{r}^{e}\right)\right\|^{2}\left\|\mathbf{a}_{N}^{H}\left(\varphi_{t}^{a}, \varphi_{t}^{e}\right) {\bf \Phi}	 \overline{\mathbf{h}}_{k}\right\|^{2}=M\left|f_{k}({\bf \Phi}	)\right|^{2}, \\
	 \mathbb{E}\left\{\tilde{\mathbf{h}}_{k} \tilde{\mathbf{h}}_{k}^{H}\right\}=\mathbf{I}_{N}, {\bf\Phi}{\bf\Phi}^H={\bf I}_N,\\
	 \mathbb{E}\left\{\tilde{\mathbf{h}}_{k}^{H} \tilde{\mathbf{h}}_{k}\right\}=\overline{\mathbf{h}}_{k}^{H} \overline{\mathbf{h}}_{k}=N, \\
	 \mathbb{E}\left\{\tilde{\mathbf{H}}_{2}^{H} \tilde{\mathbf{H}}_{2}\right\}=M \mathbf{I}_{N},\\
	 \mathbb{E}\left\{ \tilde{\mathbf{H}}_{2}\tilde{\mathbf{H}}_{2}^{H}\right\}=N \mathbf{I}_{M}.
	 \end{array}
	 \end{align}

	 \subsection{Derivation of $\mathbb{E}\left\{\left\|\mathbf{g}_{k}\right\|^{4}\right\}$}
	 We can divide $\mathbb{E}\left\{\left\|\mathbf{g}_{k}\right\|^{4}\right\}$ into the following two parts
	 \begin{align}\label{gk4}
\begin{array}{l}
\mathbb{E}\left\{\left\|\mathbf{g}_{k}\right\|^{4}\right\}
\\=\mathbb{E}\left\{\left(\sum\limits_{m=1}^{M}\left|{\rm{g}}_{k m}\right|^{2}\right)^{2}\right\} \\
=\sum\limits_{m=1}^{M} \mathbb{E}\left\{\left|{\rm{g}}_{k m}\right|^{4}\right\}+2 \sum\limits_{m=1}^{M-1} \sum\limits_{h=m+1}^{M} \mathbb{E}\left\{\left|{{\rm{g}}}_{k m}\right|^{2}\left|{\rm{g}}_{k h}\right|^{2}\right\},
\end{array}
	 \end{align}
	 where $ {\rm{g}}_{k m} $ is the $m$-th entry of $\mathbf{g}_{k}$.
	

	Next, we will calculate $\mathbb{E}\left\{\left|{\rm{g}}_{k m}\right|^{4}\right\}$ and $ \mathbb{E}\left\{\left|{\rm{g}}_{k m}\right|^{2}\left|{\rm{g}}_{k h}\right|^{2}\right\}$, respectively.
	
\subsubsection{Calculate $\mathbb{E}\left\{\left|{\mathrm{g}}_{k m}\right|^{4}\right\}$ }
	Recalling (1) $\sim$ (4), we can rewrite $\mathrm{g}_{km}$ in the following form
	\begin{align}\label{gkm}
\begin{aligned}
&\mathrm{g}_{k m}=\sqrt{\frac{\beta \alpha_{k}}{(\delta+1)\left(\varepsilon_{k}+1\right)}} \times\left(\underbrace{\sqrt{\delta \varepsilon_{k}} \mathrm{a}_{Mm}\left(\phi_{r}^{a}, \phi_{r}^{e}\right) f_{k}({\bf \Phi}	)}_{\mathrm{g}_{k m}^{1}}+\underbrace{\sqrt{\delta} \mathrm{a}_{M m}\left(\phi_{r}^{a}, \phi_{r}^{e}\right) \sum_{n=1}^{N} \mathrm{a}_{N n}^{*}\left(\varphi_{t}^{a}, \varphi_{t}^{e}\right) e^{j \theta_{n}} \tilde{\mathrm{h}}_{k n}}_{\mathrm{g}_{k m}^{2}}\right.\\
&\qquad+\left.\underbrace{\sqrt{\varepsilon_{k}} \sum_{n=1}^{N}\left[\tilde{\mathbf{H}}_{2}\right]_{m n} e^{j \theta_{n}} \mathrm{a}_{N n}\left(\varphi_{k r}^{a}, \varphi_{k r}^{e}\right)}_{{\rm{g}}_{k m}^{3}}+\underbrace{\sum_{n=1}^{N}\left[\tilde{\mathbf{H}}_{2}\right]_{m n} e^{j \theta_{n}} \tilde{\mathrm{h}}_{k n}}_{\mathrm{g}_{k m}^{4}}\right),
\end{aligned}
	\end{align}
	where $\mathrm{a}_{X i}\left( {\vartheta _{}^a,\vartheta _{}^e} \right) $ is the $i$-th element of ${\bf a}_{X}\left( {\vartheta _{}^a,\vartheta _{}^e} \right) $.

	Therefore, $\mathbb{E}\left\{\left|{\mathrm{g}}_{k m}\right|^{4}\right\}$ can be calculated as follows
	\begin{align}\label{gkm4_overall}
\begin{array}{l}
\mathbb{E}\left\{\left|\mathrm{g}_{k m}\right|^{4}\right\}
\\=\left(\frac{\beta \alpha_{k}}{(\delta+1)\left(\varepsilon_{k}+1\right)}\right)^{2} \mathbb{E}\left\{\left|\mathrm{g}_{k m}^{1}+\mathrm{g}_{k m}^{2}+\mathrm{g}_{k m}^{3}+\mathrm{g}_{k m}^{4}\right|^{4}\right\} \\
{\mathop  = \limits^{\left( b \right)} }\left(\frac{\beta \alpha_{k}}{(\delta+1)\left(\varepsilon_{k}+1\right)}\right)^{2}   \left(  \mathbb{E}\left\{\sum\limits_{\omega=1}^{4}\left|\mathrm{g}_{k m}^{\omega}\right|^{4}\right\}+ 2 \mathbb{E}\left\{\sum\limits_{\omega=1}^{3} \sum\limits_{\psi=\omega+1}^{4}\left|\mathrm{g}_{k m}^{\omega}\right|^{2}\left|\mathrm{g}_{k m}^{\psi}\right|^{2}\right\}\right.\\
\qquad\qquad\qquad\qquad+\left.4 \mathbb{E}\left\{\sum\limits_{\omega=1}^{3} \sum\limits_{\psi=\omega+1}^{4}\left(\operatorname{Re}\left\{\left(\mathrm{g}_{k m}^{\omega}\right)^{*} \mathrm{g}_{k m}^{\psi}\right\}\right)^{2}\right\}\right.\\
\qquad\qquad\qquad\qquad+8 \mathbb{E}\left\{\operatorname{Re}\left\{
\left(\mathrm{g}_{k m}^{1}\right)^*\mathrm{g}_{k m}^{2}\right\} \operatorname{Re}\left\{\left(\mathrm{g}_{k m}^{3}\right)^*\mathrm{g}_{k m}^{4}\right\}\right\} \\
\qquad\qquad\qquad\qquad+8 \mathbb{E}\left\{\operatorname{Re}\left\{\left(\mathrm{g}_{k m}^{1}\right)^*\mathrm{g}_{k m}^{3}\right\} \operatorname{Re}\left\{\left(\mathrm{g}_{k m}^{2}\right)^*\mathrm{g}_{k m}^{4}\right\}\right\}\Bigg)
\end{array}
	\end{align}
	where $  (b)  $ is obtained by removing the zero terms. Since each element in $ \tilde{\mathbf{H}}_{2}$ and $\tilde{\mathbf{h}}_{k}$ is composed of independent real and imaginary parts following $\mathcal{N}\left(0, \frac{1}{2}\right)$, we can filter the zero items based on the property that the $k$-order raw moment $\mathbb{E}\left\{s^{k}\right\}=0$, when $k$ is odd and $s$ is a normal distribution variable with zero mean\cite{winkelbauer2012moments}. 
	
	Next, we will calculate the above terms in (\ref{gkm4_overall}) one by one.
	
	Firstly, we calculate $\mathbb{E}\left\{\left|\mathrm{g}_{k m}^{\omega}\right|^{4}\right\},1\leq\omega\leq4$. When $\omega=1$ we have
	\begin{align}\label{gkm14}
\mathbb{E}\left\{\left|{\rm{g}}_{k m}^{1}\right|^{4}\right\}=\left|{\rm{g}}_{k m}^{1}\right|^{4}=\left(\delta \varepsilon_{k}\left|f_{k}({\bf\Phi})\right|^{2}\right)^{2}.
	\end{align}
	
	 When $\omega=2$, we have
	 \begin{align}\label{gkm2_4}
	 \begin{aligned}
	 \mathbb{E}\left\{\left|\mathrm{g}_{k m}^{2}\right|^{4}\right\}&=\mathbb{E}\left\{\left|\sqrt{\delta} \mathrm{a}_{M m}\left(\phi_{r}^{a}, \phi_{r}^{e}\right) \sum_{n=1}^{N} \mathrm{a}_{N n}^{*}\left(\varphi_{t}^{a}, \varphi_{t}^{e}\right) e^{j \theta_{n}} \tilde{\mathrm{h}}_{k n}\right|^{4}\right\}\\
	 &=\delta^{2} \mathbb{E}\left\{\left|\sum_{n=1}^{N} \mathrm{a}_{N n}^{*}\left(\varphi_{t}^{a}, \varphi_{t}^{e}\right) e^{j \theta_{n}} \tilde{\mathrm{h}}_{k n}\right|^{4}\right\}=\delta^{2} \mathbb{E}\left\{\left(\left|\sum_{n=1}^{N} \mathrm{a}_{N n}^{*}\left(\varphi_{t}^{a}, \varphi_{t}^{e}\right) e^{j \theta_{n}} \tilde{\mathrm{h}}_{k n}\right|^{2}\right)^{2}\right\}\\
	 &=\delta^{2} \mathbb{E}\left\{\left(\sum_{n=1}^{N}\left|\mathrm{a}_{N n}^{*}\left(\varphi_{t}^{a}, \varphi_{t}^{e}\right) e^{j \theta_{n}} \tilde{\mathrm{h}}_{k n}\right|^{2} \right.\right. \\
	 &\qquad\qquad\left.\left.+  2 \sum_{n_{1}=1}^{N-1} \sum_{n_{2}=n_{1}+1}^{N} \operatorname{Re}\left\{\mathrm{a}_{N n_{1}}^{*}\left(\varphi_{t}^{a}, \varphi_{t}^{e}\right) e^{j \theta_{n_1}} \tilde{\mathrm{h}}_{k n_{1}} \tilde{\mathrm{h}}_{k n_{2}}^{*} e^{-j \theta_{n_{2}}} \mathrm{a}_{N n_{2}}\left(\varphi_{t}^{a}, \varphi_{t}^{e}\right)\right\}\right)^{2}\right\}\\
	 &{\mathop  = \limits^{\left( c \right)} } \,\delta^{2}\mathbb{E}\left\{\left(\sum_{n=1}^{N}\left|\mathrm{a}_{N n}^{*}\left(\varphi_{t}^{a}, \varphi_{t}^{e}\right) e^{j \theta_{n}} \tilde{\mathrm{h}}_{k n}\right|^{2}\right)^{2}\right\}\\
	 &\quad+4\delta^{2} \mathbb{E}\left\{\left(\sum_{n_{1}=1}^{N-1} \sum_{n_{2}=n_{1}+1}^{N} \operatorname{Re}\left\{\mathrm{a}_{N n_{1}}^{*}\left(\varphi_{t}^{a}, \varphi_{t}^{e}\right) e^{j \theta_{n_1}} \tilde{\mathrm{h}}_{k n_{1}} \tilde{\mathrm{h}}_{k n_{2}}^{*} e^{-j \theta_{n_{2}}} \mathrm{a}_{N n_{2}}\left(\varphi_{t}^{a}, \varphi_{t}^{e}\right)\right\}\right)^{2}\right\}\\
	 &{\mathop  = \limits^{\left( d \right)} }\delta^{2}\sum_{n=1}^{N} \mathbb{E}\left\{\left|\mathrm{a}_{N n}^{*}\left(\varphi_{t}^{a}, \varphi_{t}^{e}\right) e^{j \theta_{n}} \tilde{\mathrm{h}}_{k n}\right|^{4}\right\}\\
	 &\quad +2\delta^{2} \sum_{n_{1}=1}^{N-1} \sum_{n_{2}=n_{1}+1}^{N} \mathbb{E}\left\{\left|\mathrm{a}_{N n_{1}}^{*}\left(\varphi_{t}^{a}, \varphi_{t}^{e}\right) e^{j \theta_{n_1}} \tilde{\mathrm{h}}_{k n_{1}}\right|^{2}\right\} \mathbb{E}\left\{\left|\mathrm{a}_{N n_{2}}^{*}\left(\varphi_{t}^{a}, \varphi_{t}^{e}\right) e^{j \theta_{n_2}} \tilde{\mathrm{h}}_{k n_{2}}\right|^{2}\right\}\\
	 &\quad+4\delta^{2} \sum_{n_{1}=1}^{N-1} \sum_{n_{2}=n_{1}+1}^{N} \mathbb{E}\left\{\left(\operatorname{Re}\left\{\mathrm{a}_{N n_{1}}^{*}\left(\varphi_{t}^{a}, \varphi_{t}^{e}\right) e^{j \theta_{n_1}} \tilde{\mathrm{h}}_{k n_{1}} \tilde{\mathrm{h}}_{k n_{2}}^{*} e^{-j \theta_{n_{2}}} \mathrm{a}_{N n_{2}}\left(\varphi_{t}^{a}, \varphi_{t}^{e}\right)\right\}\right)^{2}\right\},
	 \end{aligned}
	 \end{align}
	where $ (c) $ and $ (d) $ are obtained by removing the zero expectation terms in binomial expansion.
	
	Assume $ \tilde{\mathrm{h}}_{k n}=s+j t $, where $s\sim\mathcal{N}\left(0,1/2\right)$ and $t\sim\mathcal{N}\left(0,1/2\right)$. Then, we have
	\begin{align}
\mathbb{E}\left\{\left|\tilde{\mathrm{h}}_{k n}\right|^{4}\right\}=\mathbb{E}\left\{|s+j t|^{4}\right\}=\mathbb{E}\left\{s^{4}+t^{4}+2 s^{2} t^{2}\right\}=2,
	\end{align}
according to the fact that $\mathbb{E}\left\{s^{4}\right\}=\mathbb{E}\left\{t^{4}\right\}=3/4$ and $\mathbb{E}\left\{s^{2}\right\}=\mathbb{E}\left\{t^{2}\right\}=1/2$ .

	Thus, we can derive
	\begin{align}\label{gkm2_4_1}
\begin{array}{l}
\sum\limits_{n=1}^{N} \mathbb{E}\left\{\left|\mathrm{a}_{N n}^{*}\left(\varphi_{t}^{a}, \varphi_{t}^{e}\right) e^{j \theta_{n}} \tilde{\mathrm{h}}_{k n}\right|^{4}\right\} =\sum\limits_{n=1}^{N} \mathbb{E}\left\{\left|\tilde{\mathrm{h}}_{k n}\right|^{4}\right\} =2 N.
\end{array}
	\end{align}
	
	Likewise, we have
	\begin{align}\label{gkm2_4_2}
\begin{array}{l}
2 \sum\limits_{n_{1}=1}^{N-1} \sum\limits_{n_{2}=n_{1}+1}^{N} \mathbb{E}\left\{\left|\mathrm{a}_{N n_{1}}^{*}\left(\varphi_{t}^{a}, \varphi_{t}^{e}\right) e^{j \theta_{n_1}} \tilde{\mathrm{h}}_{k n_{1}}\right|^{2}\right\} \mathbb{E}\left\{\left|\mathrm{a}_{N n_{2}}^{*}\left(\varphi_{t}^{a}, \varphi_{t}^{e}\right) e^{j \theta_{n_2}} \tilde{\mathrm{h}}_{k n_{2}}\right|^{2}\right\} \\
=2 \sum\limits_{n_{1}=1}^{N-1} \sum\limits_{n_{2}=n_{1}+1}^{N} \mathbb{E}\left\{\left|\tilde{\mathrm{h}}_{k n_{1}}\right|^{2}\right\} \mathbb{E}\left\{\left|\tilde{\mathrm{h}}_{k n_{2}}\right|^{2}\right\} \\
=N(N-1).
\end{array}
	\end{align}
	
	Assume that $\mathrm{a}_{N n_{1}}^{*}\left(\varphi_{t}^{a}, \varphi_{t}^{e}\right) e^{j \theta_{n_1}} e^{-j \theta_{n_{2}}} \mathrm{a}_{N n_{2}}\left(\varphi_{t}^{a}, \varphi_{t}^{e}\right)=\sigma_{n}^{c}+j \sigma_{n}^{s}$, where $ \left(\sigma_{n}^{c}\right)^{2}+\left(\sigma_{n}^{s}\right)^{2}=1 $. Besides, assume that  $ \tilde{\mathrm{h}}_{k n_{1}}=s_{k n_{1}}+j t_{k n_{1}} $ and $ \tilde{\mathrm{h}}_{k n_{2}}=s_{k n_{2}}+j t_{k n_{2}} $, then we have
	\begin{align}\label{gkm2_4_3}
\begin{array}{l}
4 \sum\limits_{n_{1}=1}^{N-1} \sum\limits_{n_{2}=n_{1}+1}^{N} \mathbb{E}\left\{\left(\operatorname{Re}\left\{\mathrm{a}_{N n_{1}}^{*}\left(\varphi_{t}^{a}, \varphi_{t}^{e}\right) e^{j \theta_{n_1}} \tilde{\mathrm{h}}_{k n_{1}} \tilde{\mathrm{h}}_{k n_{2}}^{*} e^{-j \theta_{n_2}} \mathrm{a}_{N n_{2}}\left(\varphi_{t}^{a}, \varphi_{t}^{e}\right)\right\}\right)^{2}\right\} \\
=4 \sum\limits_{n_{1}=1}^{N-1} \sum\limits_{n_{2}=n_{1}+1}^{N} \mathbb{E}\left\{\left(\sigma_{n}^{c} s_{k n_{1}} s_{k n_{2}}-\sigma_{n}^{s} t_{k n_{1}} s_{k n_{2}}+\sigma_{n}^{c} t_{k n_{1}} t_{k n_{2}}+\sigma_{n}^{s} s_{k n_{1}} t_{k n_{2}}\right)^{2}\right\} \\
=4 \sum\limits_{n_{1}=1}^{N-1} \sum\limits_{n_{2}=n_{1}+1}^{N} \left(\mathbb{E}\left\{\left(\sigma_{n}^{c} s_{k n_{1}} s_{k n_{2}}\right)^{2}\right\}+\mathbb{E}\left\{\left(\sigma_{n}^{s} t_{k n_{1}} s_{k n_{2}}\right)^{2}\right\}\right.\\
\left.\qquad\quad\qquad\qquad+\mathbb{E}\left\{\left(\sigma_{n}^{c} t_{k n_{1}} t_{k n_{2}}\right)^{2}\right\}+\mathbb{E}\left\{\left(\sigma_{n}^{s} s_{k n_{1}} t_{k n_{2}}\right)^{2}\right\} \right)\\
=4 \sum\limits_{n_{1}=1}^{N-1} \sum\limits_{n_{2}=n_{1}+1}^{N}\left(\left(\sigma_{n}^{c}\right)^{2}+\left(\sigma_{n}^{s}\right)^{2}\right) \frac{1}{4} \times 2 \\
=N(N-1).
\end{array}
	\end{align}

	Substituting (\ref{gkm2_4_1}), (\ref{gkm2_4_2}) and (\ref{gkm2_4_3}) into (\ref{gkm2_4}), we complete the calculation of $\mathbb{E}\left\{\left|{\rm{g}}_{k m}^{2}\right|^{4}\right\}$ as follows
	\begin{align}\label{gkm24}
\mathbb{E}\left\{\left|\mathrm{g}_{k m}^{2}\right|^{4}\right\}=\delta^{2}(2 N+2 N(N-1))=2 \delta^{2} N^{2}.
	\end{align}
	
	When $\omega=3, 4$, similarly, we have
		\begin{align}\label{gkm34}
	\mathbb{E}\left\{\left|\mathrm{g}_{k m}^{3}\right|^{4}\right\}={\varepsilon^{2}_k}(2 N+2 N(N-1))=2 {\varepsilon^{2}_k} N^{2},
	\end{align}
	and
	\begin{align}\label{gkm44}
	\mathbb{E}\left\{\left|\mathrm{g}_{k m}^{4}\right|^{4}\right\}=4 N+2 N(N-1)=2 N(N+1).
	\end{align}
	
	Secondly, we focus on $ \sum\limits_{\omega=1}^{3} \sum\limits_{\psi=\omega+1}^{4}    \mathbb{E}\left\{  \left|\mathrm{g}_{k m}^{\omega}\right|^{2}\left|\mathrm{g}_{k m}^{\psi}\right|^{2}\right\}$. 
	
	When $\omega=1$, we can derive
	\begin{align}
\begin{array}{l}
\mathbb{E}\left\{\left|\mathrm{g}_{k m}^{1}\right|^{2}\left|\mathrm{g}_{k m}^{2}\right|^{2}\right\}=\left|\mathrm{g}_{k m}^{1}\right|^{2} \mathbb{E}\left\{\left|\mathrm{g}_{k m}^{2}\right|^{2}\right\}=\delta^{2} \varepsilon_{k}\left|f_{k}(\mathbf{\Phi})\right|^{2} N, \\
\mathbb{E}\left\{\left|\mathrm{g}_{k m}^{1}\right|^{2}\left|\mathrm{g}_{k m}^{3}\right|^{2}\right\}=\left|\mathrm{g}_{k m}^{1}\right|^{2} \mathbb{E}\left\{\left|\mathrm{g}_{k m}^{3}\right|^{2}\right\}=\delta \varepsilon_{k}^{2}\left|f_{k}(\mathbf{\Phi})\right|^{2} N, \\
\mathbb{E}\left\{\left|\mathrm{g}_{k m}^{1}\right|^{2}\left|\mathrm{g}_{k m}^{4}\right|^{2}\right\}=\left|\mathrm{g}_{k m}^{1}\right|^{2} \mathbb{E}\left\{\left|\mathrm{g}_{k m}^{4}\right|^{2}\right\}=\delta \varepsilon_{k}\left|f_{k}({\bf \Phi})\right|^{2} N.
\end{array}
	\end{align}
	
	When $\omega=2$, by utilizing the property of independence and removing the terms with zero expectation, we have
	\begin{align}
	\begin{array}{l}
	\mathbb{E}\left\{\left|\mathrm{g}_{k m}^{2}\right|^{2}\left|\mathrm{g}_{k m}^{3}\right|^{2}\right\} \\
	=\delta \varepsilon_{k} \mathbb{E}\left\{\left|\sum_{n=1}^{N} \mathrm{a}_{N n}^{*}\left(\varphi_{t}^{a}, \varphi_{t}^{e}\right) e^{j \theta_{n}} \tilde{\mathrm{h}}_{k n}\right|^{2}\left|\sum_{n=1}^{N}\left[\tilde{\mathbf{H}}_{2}\right]_{m n} e^{j \theta_{n}} \mathrm{a}_{N n}\left(\varphi_{k r}^{a}, \varphi_{k r}^{e}\right)\right|^{2}\right\} \\
	=\delta \varepsilon_{k} \mathbb{E}\left\{\sum_{n=1}^{N}\left|\mathrm{a}_{N n}^{*}\left(\varphi_{t}^{a}, \varphi_{t}^{e}\right) e^{j \theta_{n}} \tilde{\mathrm{h}}_{k n}\right|^{2} \sum_{n=1}^{N}\left|\left[\tilde{\mathbf{H}}_{2}\right]_{m n} e^{j \theta_{n}} \mathrm{a}_{N n}\left(\varphi_{k r}^{a}, \varphi_{k r}^{e}\right)\right|^{2}\right\} \\
	=\delta \varepsilon_{k} \mathbb{E}\left\{\sum_{n=1}^{N}\left|\tilde{\mathrm{h}}_{k n}\right|^{2} \sum_{n=1}^{N}\left|\left[\tilde{\mathbf{H}}_{2}\right]_{m n}\right|^{2}\right\} \\
	=\delta \varepsilon_{k} \sum_{n=1}^{N} \mathbb{E}\left\{\left|\tilde{\mathrm{h}}_{k n}\right|^{2}\right\} \sum_{n=1}^{N} \mathbb{E}\left\{\left.\left[\tilde{\mathbf{H}}_{2}\right]_{m n}\right|^{2}\right\} \\
	=\delta \varepsilon_{k} N^{2},
	\end{array}
	\end{align}
	and
	\begin{align}
	\begin{array}{l}
	\mathbb{E}\left\{\left|\mathrm{g}_{k m}^{2}\right|^{2}\left|\mathrm{g}_{k m}^{4}\right|^{2}\right\} \\
			=\delta \mathbb{E}\left\{\left|\sum_{n=1}^{N} \mathrm{a}_{N n}^{*}\left(\varphi_{t}^{a}, \varphi_{t}^{e}\right) e^{j \theta_{n}} \tilde{\mathrm{h}}_{k n}\right|^{2} \left| \sum_{n=1}^{N}\left[\tilde{\mathbf{H}}_{2}\right]_{m n} e^{j \theta_{n} }\tilde{\mathrm{h}}_{k n}\right|^{2}\right\} \\
	=\delta \mathbb{E}\left\{\sum_{n=1}^{N}\left|\tilde{\mathrm{h}}_{k n}\right|^{2} \sum_{n=1}^{N}\left|\left[\tilde{\mathbf{H}}_{2}\right]_{m n}\right|^{2}\left|\tilde{\mathrm{h}}_{k n}\right|^{2}\right\} \\
	=\delta \mathbb{E}\left\{\sum_{n_{1}=1}^{N} \sum_{n_{2}=1, n_{2} \neq n_{1}}^{N}\left|\tilde{\mathrm{h}}_{k n_{1}}\right|^{2}\left|\tilde{\mathrm{h}}_{k n_{2}}\right|^{2}\left|\left[\tilde{\mathbf{H}}_{2}\right]_{m n_{2}}\right|^{2}+\sum_{n_{1}=1}^{N}\left|\tilde{\mathrm{h}}_{k n_{1}}\right|^{4}\left|\left[\tilde{\mathbf{H}}_{2}\right]_{m n_{1}}\right|^2\right\} \\
	=\delta N(N+1).
	\end{array}
	\end{align}
	
	When $\omega=3$, similarly, we have
	\begin{align}
\mathbb{E}\left\{\left|\mathrm{g}_{k m}^{3}\right|^{2}\left|\mathrm{g}_{k m}^{4}\right|^{2}\right\} =\varepsilon_{k} N(N+1).
	\end{align} 
	
	Thirdly, we calculate $ \sum\limits_{\omega=1}^{3} \sum\limits_{\psi=\omega+1}^{4}    \mathbb{E}\left\{\left(\operatorname{Re}\left\{\left(\mathrm{g}_{k m}^{\omega}\right)^{*} \mathrm{g}_{k m}^{\psi}\right\}\right)^{2}\right\}$. Using the similar methods in (\ref{gkm2_4_3}), we can extract the real parts and then calculate the expectation of their square. Then we can obtain the following results after some straightforward simplifications:
	\begin{align}\label{E_Re_2}
	&\mathbb{E}\left\{\left(\operatorname{Re}\left\{\left(\mathrm{g}_{k m}^{1}\right)^{*} \mathrm{g}_{k m}^{2}\right\}\right)^{2}\right\}=\frac{\delta^{2} \varepsilon_{k}}{2} N\left|f_{k}({\bf \Phi})\right|^{2},\quad\mathbb{E}\left\{\left(\operatorname{Re}\left\{\left(\mathrm{g}_{k m}^{1}\right)^{*} \mathrm{g}_{k m}^{3}\right\}\right)^{2}\right\}=\frac{\delta {\varepsilon_{k}^{2}}  }{2} N\left|f_{k}({\bf \Phi})\right|^{2},\nonumber\\
	&\mathbb{E}\left\{\left(\operatorname{Re}\left\{\left(\mathrm{g}_{k m}^{1}\right)^{*} \mathrm{g}_{k m}^{4}\right\}\right)^{2}\right\}=\frac{\delta {\varepsilon_{k} }  }{2} N\left|f_{k}({\bf \Phi})\right|^{2},\;\,\quad\mathbb{E}\left\{\left(\operatorname{Re}\left\{\left(\mathrm{g}_{k m}^{2}\right)^{*} \mathrm{g}_{k m}^{3}\right\}\right)^{2}\right\}=\frac{\delta {\varepsilon_{k} }  }{2} N^2,\nonumber\\	
	&\mathbb{E}\left\{\left(\operatorname{Re}\left\{\left(\mathrm{g}_{k m}^{2}\right)^{*} \mathrm{g}_{k m}^{4}\right\}\right)^{2}\right\}={\frac{\delta  }{2} N\left(N+1\right)},\quad\quad\;\;\mathbb{E}\left\{\left(\operatorname{Re}\left\{\left(\mathrm{g}_{k m}^{3}\right)^{*} \mathrm{g}_{k m}^{4}\right\}\right)^{2}\right\}={\frac{ {\varepsilon_{k} }  }{2} N\left(N+1\right)}.
	\end{align}
	
	The remaining two terms in (\ref{gkm4_overall}) can also be derived similarly, which is presented as follows
	\begin{align}\label{8fold_term1}
	 \mathbb{E}\left\{\operatorname{Re}\left\{\left(\mathrm{g}_{k m}^{1}\right)^*\mathrm{g}_{k m}^{2}\right\} \operatorname{Re}\left\{\left(\mathrm{g}_{k m}^{3}\right) ^* \mathrm{g}_{k m}^{4}\right\}\right\} = \frac{1}{2}\delta\varepsilon_{k} \left|f_{k}({\bf \Phi})\right|^{2},\\\label{8fold_term2}
	\mathbb{E}\left\{\operatorname{Re}\left\{\left(\mathrm{g}_{k m}^{1}\right)^* \mathrm{g}_{k m}^{3}\right\} \operatorname{Re}\left\{\left(\mathrm{g}_{k m}^{2}\right)^*\mathrm{g}_{k m}^{4}\right\}\right\} = \frac{1}{2}\delta\varepsilon_{k} \left|f_{k}({\bf \Phi})\right|^{2}.
	\end{align}

Substituting the above intermediate results (\ref{gkm14}) and (\ref{gkm24}) $\sim$ (\ref{8fold_term2}) into (\ref{gkm4_overall}), we complete the calculation of $\mathbb{E}\left\{\left|{\mathrm{g}}_{k m}\right|^{4}\right\}$, which is not related with its subscript $m$. Here we omit its detailed expression since it is straightforward.

\subsubsection{Calculate $ \mathbb{E}\left\{\left|{\rm{g}}_{k m}\right|^{2}\left|{\rm{g}}_{k h}\right|^{2}\right\}$}
Similar to (\ref{gkm}), we can express $\mathrm{g}_{kh}$ as follows
	\begin{align}\label{gkh}
\begin{aligned}
&\mathrm{g}_{k h}=\sqrt{\frac{\beta \alpha_{k}}{(\delta+1)\left(\varepsilon_{k}+1\right)}} \times\left(\underbrace{\sqrt{\delta \varepsilon_{k}} \mathrm{a}_{Mh}\left(\phi_{r}^{a}, \phi_{r}^{e}\right) f_{k}({\bf \Phi}	)}_{\mathrm{g}_{k h}^{1}}+\underbrace{\sqrt{\delta} \mathrm{a}_{Mh}\left(\phi_{r}^{a}, \phi_{r}^{e}\right) \sum_{n=1}^{N} \mathrm{a}_{N n}^{*}\left(\varphi_{t}^{a}, \varphi_{t}^{e}\right) e^{j \theta_{n}} \tilde{\mathrm{h}}_{k n}}_{\mathrm{g}_{k h}^{2}}\right.\\
&\qquad+\left.\underbrace{\sqrt{\varepsilon_{k}} \sum_{n=1}^{N}\left[\tilde{\mathbf{H}}_{2}\right]_{h n} e^{j \theta_{n}} \mathrm{a}_{N n}\left(\varphi_{k r}^{a}, \varphi_{k r}^{e}\right)}_{{\rm{g}}_{k h}^{3}}+\underbrace{\sum_{n=1}^{N}\left[\tilde{\mathbf{H}}_{2}\right]_{h n} e^{j \theta_{n}} \tilde{\mathrm{h}}_{k n}}_{\mathrm{g}_{k h}^{4}}\right),
\end{aligned}
\end{align}

Note that $\left[\tilde{\mathbf{H}}_{2}\right]_{m n}$ is independent to $\left[\tilde{\mathbf{H}}_{2}\right]_{h n}$ and both of them have zero mean. We can extract the terms with non-zero expectation after the binomial expansion as follows
\begin{align}\label{gkm2gkh2}
\begin{aligned}
\mathbb{E}\left\{\left|\mathrm{g}_{k m}\right|^{2}\left|\mathrm{g}_{k h}\right|^{2}\right\}&=\left(\frac{\beta \alpha_{k}}{(\delta+1)\left(\varepsilon_{k}+1\right)}\right)^{2} \mathbb{E}\left\{\left|\sum\limits_{\omega=1}^{4} \mathrm{g}_{k m}^{\omega}\right|^{2}\left|\sum\limits_{\psi=1}^{4} \mathrm{g}_{k h}^{\psi}\right|^{2}\right\} \\
&=\left(\frac{\beta \alpha_{k}}{(\delta+1)\left(\varepsilon_{k}+1\right)}\right)^{2} \times\left(\sum\limits_{\omega=1}^{4} \sum\limits_{\psi=1}^{4}\mathbb{E}\left\{\left|\mathrm{g}_{k m}^{\omega}\right|^2\left|\mathrm{g}_{k h}^{\psi}\right|^{2}\right\} \right. \\
&\quad+4 \mathbb{E}\left\{\operatorname{Re}\left\{\mathrm{g}_{k m}^{1}\left(\mathrm{g}_{k m}^{2}\right)^{*}\right\} \operatorname{Re}\left\{\mathrm{g}_{k h}^{1}\left(\mathrm{g}_{k h}^{2}\right)^{*}\right\}\right\} \\
&\quad+4 \mathbb{E}\left\{\operatorname{Re}\left\{\mathrm{g}_{k m}^{1}\left(\mathrm{g}_{k m}^{2}\right)^{*}\right\} \operatorname{Re}\left\{\mathrm{g}_{k h}^{3}\left(\mathrm{g}_{k h}^{4}\right)^{*}\right\}\right\} \\
&\quad+4 \mathbb{E}\left\{\operatorname{Re}\left\{\mathrm{g}_{k m}^{3}\left(\mathrm{g}_{k m}^{4}\right)^{*}\right\} \operatorname{Re}\left\{\mathrm{g}_{k h}^{1}\left(\mathrm{g}_{k h}^{2}\right)^{*}\right\}\right\} \\
&\quad+\left.4 \mathbb{E}\left\{\operatorname{Re}\left\{\mathrm{g}_{k m}^{3}\left(\mathrm{g}_{k m}^{4}\right)^{*}\right\} \operatorname{Re}\left\{\mathrm{g}_{k h}^{3}\left(\mathrm{g}_{k h}^{4}\right)^{*}\right\}\right\}\right).
\end{aligned}
\end{align}

Next, we will calculate the above terms in (\ref{gkm2gkh2}) one by one.

First, we focus on $ \mathbb{E}\left\{\left|\mathrm{g}_{k m}^{\omega}\right|^2\left|\mathrm{g}_{k h}^{\psi}\right|^{2}\right\}, 1\leq\omega,\psi\leq4 $. These  terms can be derived following the similar process in the calculation of $  \mathbb{E}\left\{\left|\mathrm{g}_{k m}^{\omega}\right|^{2}\left|\mathrm{g}_{k m}^{\psi}\right|^{2}\right\} $. Therefore, we can directly obtain the following results
\begin{align}\label{gkm12gkh12}
&\mathbb{E}\left\{\left|\mathrm{g}_{k m}^{1}\right|^{2}\left|\mathrm{g}_{k h}^{1}\right|^{2}\right\}=\left(\delta \varepsilon_{k}\right)^{2}\left|f_{k}(\mathbf{\Phi})\right|^{4}, &\mathbb{E}\left\{\left|\mathrm{g}_{k m}^{1}\right|^{2}\left|\mathrm{g}_{k h}^{2}\right|^{2}\right\}=\delta^{2} \varepsilon_{k} N\left|f_{k}(\mathbf{\Phi})\right|^{2}, \\
&\mathbb{E}\left\{\left|\mathrm{g}_{k m}^{1}\right|^{2}\left|\mathrm{g}_{k h}^{3}\right|^{2}\right\}=\delta \varepsilon_{k}^{2} N\left|f_{k}(\mathbf{\Phi})\right|^{2},&\mathbb{E}\left\{\left|\mathrm{g}_{k m}^{1}\right|^{2}\left|\mathrm{g}_{k h}^{4}\right|^{2}\right\}=\delta \varepsilon_{k} N\left|f_{k}(\mathbf{\Phi})\right|^{2},\\
&\mathbb{E}\left\{\left|\mathrm{g}_{k m}^{2}\right|^{2}\left|\mathrm{g}_{k h}^{1}\right|^{2}\right\}=\delta^{2} \varepsilon_{k} N\left|f_{k}(\mathbf{\Phi})\right|^{2}, &\mathbb{E}\left\{\left|\mathrm{g}_{k m}^{2}\right|^{2}\left|\mathrm{g}_{k h}^{2}\right|^{2}\right\}={2\delta^{2} N^{2}}, \\
&\mathbb{E}\left\{\left|\mathrm{g}_{k m}^{2}\right|^{2}\left|\mathrm{g}_{k h}^{3}\right|^{2}\right\}=\delta \varepsilon_{k} N^{2}, &\mathbb{E}\left\{\left|\mathrm{g}_{k m}^{2}\right|^{2}\left|\mathrm{g}_{k h}^{4}\right|^{2}\right\}={\delta (N^{2}+N)},\\
	&\mathbb{E}\left\{\left|\mathrm{g}_{k m}^{3}\right|^{2}\left|\mathrm{g}_{k h}^{1}\right|^{2}\right\}=\delta \varepsilon_{k}^{2} N\left|f_{k}(\mathbf{\Phi})\right|^{2}, &\mathbb{E}\left\{\left|\mathrm{g}_{k m}^{3}\right|^{2}\left|\mathrm{g}_{k h}^{2}\right|^{2}\right\}=\delta \varepsilon_{k} N^{2},\\
&\mathbb{E}\left\{\left|\mathrm{g}_{k m}^{3}\right|^{2}\left|\mathrm{g}_{k h}^{3}\right|^{2}\right\}=\varepsilon_{k}^{2} N^{2}, &\mathbb{E}\left\{\left|\mathrm{g}_{k m}^{3}\right|^{2}\left|\mathrm{g}_{k h}^{4}\right|^{2}\right\}=\varepsilon_{k} N^{2},\\
	&\mathbb{E}\left\{\left|\mathrm{g}_{k m}^{4}\right|^{2}\left|\mathrm{g}_{k h}^{1}\right|^{2}\right\}=\delta \varepsilon_{k} N\left|f_{k}(\mathbf{\Phi})\right|^{2}, &\mathbb{E}\left\{\left|\mathrm{g}_{k m}^{4}\right|^{2}\left|\mathrm{g}_{k h}^{2}\right|^{2}\right\}={\delta (N^{2}+N)   }, \\
&\mathbb{E}\left\{\left|\mathrm{g}_{k m}^{4}\right|^{2}\left|\mathrm{g}_{k h}^{3}\right|^{2}\right\}=\varepsilon_{k} N^{2}, &\mathbb{E}\left\{\left|\mathrm{g}_{k m}^{4}\right|^{2}\left|\mathrm{g}_{k h}^{4}\right|^{2}\right\}{=N^{2}+N}.
\end{align}

	Next, we will derive the remaining four parts in (\ref{gkm2gkh2}).
	To begin with, the first one is
	\begin{align}\label{Re_gkm1gkm2_Re_gkh1gkh2}
	\mathbb{E}\left\{\operatorname{Re}\left\{\mathrm{g}_{k m}^{1}\left(\mathrm{g}_{k m}^{2}\right)^{*}\right\} \operatorname{Re}\left\{\mathrm{g}_{k h}^{1}\left(\mathrm{g}_{k h}^{2}\right)^{*}\right\}\right\} =\mathbb{E}\left\{\left(\operatorname{Re}\left\{\mathrm{g}_{k m}^{1}\left(\mathrm{g}_{k m}^{2}\right)^{*}\right\}\right)^{2}\right\} =\frac{\delta^{2} \varepsilon_{k}}{2} N\left|f_{k}({\bf\Phi})\right|^{2}.
	\end{align}
	
	The second one is
	\begin{align}
	\begin{array}{l}
	\mathbb{E}\left\{\operatorname{Re}\left\{\mathrm{g}_{k m}^{1}\left(\mathrm{g}_{k m}^{2}\right)^{*}\right\} \operatorname{Re}\left\{\mathrm{g}_{k h}^{3}\left(\mathrm{g}_{k h}^{4}\right)^{*}\right\}\right\} \\
	=\delta \varepsilon_{k}\mathbb{E}\left\{ \operatorname{Re}\left\{f_{k}({\bf\Phi}) \sum\limits_{n=1}^{N} \mathrm{a}_{N n}\left(\varphi_{t}^{a}, \varphi_{t}^{e}\right) e^{-j \theta_{n}} \tilde{\mathrm{h}}_{k n}^{*}\right\}\right. \\
	\left.\qquad\qquad \times \operatorname{Re}\left\{\sum\limits_{n=1}^{N}\left[\tilde{\mathbf{H}}_{2}\right]_{h n} e^{j \theta_{n}} \mathrm{a}_{N n}\left(\varphi_{k r}^{a}, \varphi_{k r}^{e}\right) \sum\limits_{n=1}^{N}\left[\tilde{\mathbf{H}}_{2}\right]_{h n}^{*} e^{-j \theta_{n}} \tilde{\mathrm{h}}_{k n}^{*}\right\}\right\} \\
	=\delta \varepsilon_{k}\mathbb{E}\left\{ \operatorname{Re}\left\{\sum\limits_{n=1}^{N}\left(f_{k}({\bf\Phi}) \mathrm{a}_{N n}\left(\varphi_{t}^{a}, \varphi_{t}^{e}\right) e^{-j \theta_{n}}\right) \tilde{\mathrm{h}}_{k n}^{*}\right\}\right. \\
	\left.\qquad\qquad\times \operatorname{Re}\left\{\sum\limits_{n=1}^{N}\left|\left[\tilde{\mathbf{H}}_{2}\right]_{h n}\right|^{2} \mathrm{a}_{N n}\left(\varphi_{k r}^{a}, \varphi_{k r}^{e}\right) \tilde{\mathrm{h}}_{k n}^{*}\right\}\right\}.
	\end{array}
	\end{align}
	
	Assume that
	\begin{align}
	\begin{array}{l}
	f_{k}({\bf\Phi}) \mathrm{a}_{N n}\left(\varphi_{t}^{a}, \varphi_{t}^{e}\right) e^{-j \theta_{n}}=\sigma_{c}^{t n}+j \sigma_{s}^{t n}, \\
	\tilde{\mathrm{h}}_{k n}=s^{n}+j t^{n}, \\
	\mathrm{a}_{N n}\left(\varphi_{k r}^{a}, \varphi_{k r}^{e}\right)=\sigma_{c}^{k r n}+j \sigma_{s}^{k r n},
	\end{array}
	\end{align}
	and after some algebraic simplifications, we can obtain
	\begin{align}\label{Re_gkm1gkm2_Re_gkh3gkh4}
	\begin{array}{l}
\mathbb{E}\left\{\operatorname{Re}\left\{\mathrm{g}_{k m}^{1}\left(\mathrm{g}_{k m}^{2}\right)^{*}\right\} \operatorname{Re}\left\{\mathrm{g}_{k h}^{3}\left(\mathrm{g}_{k h}^{4}\right)^{*}\right\}\right\}\\
=\delta \varepsilon_{k} \mathbb{E}\left\{\sum\limits_{n=1}^{N} \sigma_{c}^{t n} \sigma_{c}^{k r n}\left(s^{n}\right)^{2}+\sigma_{s}^{t n} \sigma_{s}^{k r n}\left(t^{n}\right)^{2}\right\} \\
=\frac{\delta \varepsilon_{k}}{2} \sum\limits_{n=1}^{N}\left(\sigma_{c}^{t n} \sigma_{c}^{k r n}+\sigma_{s}^{t n} \sigma_{s}^{k r n}\right) =\frac{\delta \varepsilon_{k}}{2} \sum\limits_{n=1}^{N} \operatorname{Re}\left\{\left(\sigma_{c}^{t n}+j \sigma_{s}^{t n}\right)\left(\sigma_{c}^{k r n}-j \sigma_{s}^{k r n}\right)\right\}\\
=\frac{\delta \varepsilon_{k}}{2} \sum\limits_{n=1}^{N} \operatorname{Re}\left\{f_{k}({\bf \Phi}) \mathrm{a}_{N n}\left(\varphi_{t}^{a}, \varphi_{t}^{e}\right) e^{-j \theta_{n}} \mathrm{a}_{N n}^{*}\left(\varphi_{k r}^{a}, \varphi_{k r}^{e}\right)\right\} \\
=\frac{\delta \varepsilon_{k}}{2} \operatorname{Re}\left\{f_{k}(\mathbf{\Phi}) \mathrm{a}_{N}^{H}\left(\varphi_{k r}^{a}, \varphi_{k r}^{e}\right) \mathbf{\Phi}^{H} \mathrm{a}_{N}\left(\varphi_{t}^{a}, \varphi_{t}^{e}\right)\right\} \\
=\frac{\delta \varepsilon_{k}}{2} \operatorname{Re}\left\{f_{k}(\mathbf{\Phi}) f_{k}^{H}(\mathbf{\Phi})\right\}\\
=\frac{\delta \varepsilon_{k}}{2}\left|f_{k}({\bf \Phi})\right|^{2}.
\end{array}
	\end{align}
	
	Then, we can easily find that
	\begin{align}\label{Re_gkm3gkm4_Re_gkh1gkh2}
\begin{array}{l}
\mathbb{E}\left\{\operatorname{Re}\left\{\mathrm{g}_{k m}^{3}\left(\mathrm{g}_{k m}^{4}\right)^{*}\right\} \operatorname{Re}\left\{\mathrm{g}_{k h}^{1}\left(\mathrm{g}_{k h}^{2}\right)^{*}\right\}\right\} \\
=\mathbb{E}\left\{\operatorname{Re}\left\{\mathrm{g}_{k m}^{1}\left(\mathrm{g}_{k m}^{2}\right)^{*}\right\} \operatorname{Re}\left\{\mathrm{g}_{k h}^{3}\left(\mathrm{g}_{k h}^{4}\right)^{*}\right\}\right\} \\
=\frac{\delta \varepsilon_{k}}{2}\left|f_{k}({\bf \Phi})\right|^{2}.
\end{array}
	\end{align}
	
	The last one can be derived as follows
	\begin{align}\label{Re_gkm3gkm4_Re_gkh3gkh4}
\begin{array}{l}
\mathbb{E}\left\{\operatorname{Re}\left\{\mathrm{g}_{k m}^{3}\left(\mathrm{g}_{k m}^{4}\right)^{*}\right\} \operatorname{Re}\left\{\mathrm{g}_{k h}^{3}\left(\mathrm{g}_{k h}^{4}\right)^{*}\right\}\right\} \\
=\varepsilon_{k} \mathbb{E}\left\{\operatorname{Re}\left\{\sum_{n=1}^{N}\left|\left[\tilde{\mathbf{H}}_{2}\right]_{m n}\right|^{2} \mathrm{a}_{N n}\left(\varphi_{k r}^{a}, \varphi_{k r}^{e}\right) \tilde{\mathrm{h}}_{k n}^{*}\right\}  \operatorname{Re}\left\{\sum_{n=1}^{N}\left|\left[\tilde{\mathbf{H}}_{2}\right]_{h n}\right|^{2} \mathrm{a}_{N n}\left(\varphi_{k r}^{a}, \varphi_{k r}^{e}\right) \tilde{\mathrm{h}}_{k n}^{*}\right\}\right\}\\
=\varepsilon_{k} \mathbb{E}\left\{\left(\sum_{n=1}^{N}\left|\left[\tilde{\mathbf{H}}_{2}\right]_{m n}\right|^{2} \operatorname{Re}\left\{\mathrm{a}_{N n}\left(\varphi_{k r}^{a}, \varphi_{k r}^{e}\right) \tilde{\mathrm{h}}_{k n}^{*}\right\}\right)\left(\sum_{n=1}^{N}\left|\left[\tilde{\mathbf{H}}_{2}\right]_{h n}\right|^{2} \operatorname{Re}\left\{\mathrm{a}_{N n}\left(\varphi_{k r}^{a}, \varphi_{k r}^{e}\right) \tilde{\mathrm{h}}_{k n}^{*}\right\}\right)\right\} \\
=\varepsilon_{k} \mathbb{E}\left\{\sum_{n=1}^{N}\left|\left[\tilde{\mathbf{H}}_{2}\right]_{m n}\right|^{2}\left|\left[\tilde{\mathbf{H}}_{2}\right]_{h n}\right|^{2}\left(\operatorname{Re}\left\{\mathrm{a}_{N n}\left(\varphi_{k r}^{a}, \varphi_{k r}^{e}\right) \tilde{\mathrm{h}}_{k n}^{*}\right\}\right)^{2}\right\} \\
=\varepsilon_{k} \sum_{n=1}^{N}  \mathbb{E}\left\{\left(\operatorname{Re}\left\{\mathrm{a}_{N n}\left(\varphi_{k r}^{a}, \varphi_{k r}^{e}\right) \tilde{\mathrm{h}}_{k n}^{*}\right\}\right)^{2}\right\} \\
=\frac{1}{2}\varepsilon_{k} N.
\end{array}
	\end{align}
	
	Substituting (\ref{gkm12gkh12}) $ \sim $ (\ref{Re_gkm1gkm2_Re_gkh1gkh2}) and (\ref{Re_gkm1gkm2_Re_gkh3gkh4}) $\sim$ (\ref{Re_gkm3gkm4_Re_gkh3gkh4}) into (\ref{gkm2gkh2}), we can obtain the expression of $ \mathbb{E}\left\{\left|{\rm{g}}_{k m}\right|^{2}\left|{\rm{g}}_{k h}\right|^{2}\right\}$, which is not related with its subscript $m$ and $h$. Since we have obtained the expressions of $\mathbb{E}\left\{\left|{\mathrm{g}}_{k m}\right|^{4}\right\}$ and $ \mathbb{E}\left\{\left|{\rm{g}}_{k m}\right|^{2}\left|{\rm{g}}_{k h}\right|^{2}\right\}$ , we can directly obtain $\mathbb{E}\left\{\left\|\mathbf{g}_{k}\right\|^{4}\right\}$ by using
	\begin{align}
\mathbb{E}\left\{\left\|\mathbf{g}_{k}\right\|^{4}\right\}=M \mathbb{E}\left\{\left|\mathrm{g}_{k m}\right|^{4}\right\}+M(M-1) \mathbb{E}\left\{\left|\mathrm{g}_{k m}\right|^{2}\left|\mathrm{g}_{k h}\right|^{2}\right\}.
	\end{align}

	\subsection{Derivation of $\mathbb{E}\left\{\left|\mathbf{g}_{k}^{H} \mathbf{g}_{i}\right|^{2}\right\}$}

	Before the proof, we first provide an important property as follows
	\begin{align}\label{HAH=0}
\mathbb{E}\left\{\operatorname{Re}\left\{\tilde{\mathbf{H}}_{2} \mathbf{A} \tilde{\mathbf{H}}_{2}\right\}\right\}=\mathbf{0},
	\end{align}
	where $ \mathbf{A} \in \mathbb{C}^{N \times M} $ is an arbitrary deterministic matrix. This conclusion can be readily proved by firstly considering the case of one dimension and then generalizing it to high dimensions by mathematical induction.
	
	Note that since the communication of different users goes through the same RIS-BS channel $\mathbf{ H}_2$, $ \mathbf{g}_k $ is no longer independent to $\mathbf{g}_i $, which is different from the scenario without RIS. Recalling (\ref{gk}) and (\ref{gi}), when calculating $\mathbb{E}\left\{\left|\mathbf{g}_{k}^{H}\mathbf{g}_{i}\right|^{2}\right\}$, we can ignore the terms with zero expectation based on (\ref{property1}) and (\ref{HAH=0}), and then we have
	 \begin{align}\label{gkgi_2}
	 \begin{array}{l}
	 \mathbb{E}\left\{\left|\mathbf{g}_{k}^{H} \mathbf{g}_{i}\right|^{2}\right\}\\
	 =\frac{\beta^{2} \alpha_{k} \alpha_{i}}{(\delta+1)^{2}\left(\varepsilon_{k}+1\right)\left(\varepsilon_{i}+1\right)} \mathbb{E}\left\{\left|\sum\limits_{\omega=1}^{4} \sum\limits_{\psi=1}^{4}\left(\mathbf{g}_{k}^{\omega}\right)^{H} \mathbf{g}_{i}^{\psi}\right|^2\right\} \\
	 =\frac{\beta^{2} \alpha_{k} \alpha_{i}}{(\delta+1)^{2}\left(\varepsilon_{k}+1\right)\left(\varepsilon_{i}+1\right)} \times\left(\mathbb{E}\left\{\sum\limits_{\omega=1}^{4} \sum\limits_{\psi=1}^{4}\left|\left(\mathbf{g}_{k}^{\omega}\right)^{H} \mathbf{g}_{i}^{\psi}\right|^{2}\right\}\right. \\
	 +2 \mathbb{E}\left\{ \operatorname{Re}\left\{\left(\mathbf{g}_{k}^{1}\right)^{H} \mathbf{g}_{i}^{1}\left(\mathbf{g}_{i}^{3}\right)^{H} \mathbf{g}_{k}^{3}\right\}\right\}+2\mathbb{E}\left\{  \operatorname{Re}\left\{\left(\mathbf{g}_{k}^{1}\right)^{H} \mathbf{g}_{i}^{2}\left(\mathbf{g}_{i}^{4}\right)^{H} \mathbf{g}_{k}^{3}\right\} \right\} \\
	 \left.+2\mathbb{E}\left\{  \operatorname{Re}\left\{\left(\mathbf{g}_{k}^{2}\right)^{H} \mathbf{g}_{i}^{1}\left(\mathbf{g}_{i}^{3}\right)^{H} \mathbf{g}_{k}^{4}\right\} \right\}+2\mathbb{E}\left\{  \operatorname{Re}\left\{\left(\mathbf{g}_{k}^{2}\right)^{H} \mathbf{g}_{i}^{2}\left(\mathbf{g}_{i}^{4}\right)^{H} \mathbf{g}_{k}^{4}\right\} \right\}\right).
	 \end{array}
	 \end{align}

	Then we will calculate the above terms in (\ref{gkgi_2}) one by one.
	
	First, we focus on $\mathbb{E}\left\{\left|\left(\mathbf{g}_{k}^{\omega}\right)^{H} \mathbf{g}_{i}^{\psi}\right|^{2}\right\}, 1\leq\omega,\psi\leq4$. 
	
	When $\omega=1$, we have
	\begin{align}\label{gk1gi1_2}
\begin{aligned}
&\mathbb{E}\left\{\left|\left(\mathbf{g}_{k}^{1}\right)^{H} \mathbf{g}_{i}^{1}\right|^{2}\right\}= \left|\sqrt{\delta \varepsilon_{k}} \sqrt{\delta \varepsilon_{i}} \overline{\mathbf{h}}_{k}^{H} \mathbf{\Phi}^{H} \overline{\mathbf{H}}_{2}^{H} \overline{\mathbf{H}}_{2} {\bf \Phi} \overline{\mathbf{h}}_{i}\right|^{2} =\delta^{2} \varepsilon_{k} \varepsilon_{i} M^{2}\left|f_{k}(\mathbf{\Phi})\right|^{2}\left|f_{i}(\mathbf{\Phi})\right|^{2},\\
&\mathbb{E}\left\{\left|\left(\mathbf{g}_{k}^{1}\right)^{H} \mathbf{g}_{i}^{2}\right|^{2}\right\} \\
&=\delta^{2} \varepsilon_{k} \mathbb{E}\left\{\left|\overline{\mathbf{h}}_{k}^{H} \mathbf{\Phi}^{H} \overline{\mathbf{H}}_{2}^{H} \overline{\mathbf{H}}_{2} {\bf \Phi} \tilde{\mathbf{h}}_{i}\right|^{2}\right\} =\delta^{2} \varepsilon_{k} M^{2}\left|f_{k}({\bf \Phi})\right|^{2} \mathbb{E}\left\{\left|\mathbf{a}_{N}^{H}\left(\varphi_{t}^{a}, \varphi_{t}^{e}\right) {\bf \Phi} \tilde{\mathbf{h}}_{i}\right|^{2}\right\} \\
&=\delta^{2} \varepsilon_{k} M^{2} N\left|f_{k}({\bf \Phi})\right|^{2},\\
	&\mathbb{E}\left\{\left|\left(\mathbf{g}_{k}^{1}\right)^{H} \mathbf{g}_{i}^{3}\right|^{2}\right\} \\
	&=\delta \varepsilon_{k} \varepsilon_{i}\left|f_{k}({\bf \Phi})\right|^{2} \overline{\mathbf{h}}_{i}^{H} {\bf \Phi}^{H} \mathbb{E}\left\{\tilde{\mathbf{H}}_{2}^{H} \mathbf{a}_{M}\left(\phi_{r}^{a}, \phi_{r}^{e}\right) \mathbf{a}_{M}^{H}\left(\phi_{r}^{a}, \phi_{r}^{e}\right) \tilde{\mathbf{H}}_{2}\right\} {\bf \Phi} \overline{\mathbf{h}}_{i} \\
&	=\delta \varepsilon_{k} \varepsilon_{i}\left|f_{k}({\bf \Phi})\right|^{2} \overline{\mathbf{h}}_{i}^{H} {\bf \Phi}^{H} M \mathbf{I}_{N} {\bf \Phi} \overline{\mathbf{h}}_{i} \\
&	=\delta \varepsilon_{k} \varepsilon_{i}\left|f_{k}({\bf \Phi})\right|^{2} M N,
	\end{aligned}
	\end{align}
	and
	\begin{align}
	\begin{aligned}
		&\mathbb{E}\left\{\left|\left(\mathbf{g}_{k}^{1}\right)^{H} \mathbf{g}_{i}^{4}\right|^{2}\right\} \\
&=\delta \varepsilon_{k}\left|f_{k}({\bf \Phi})\right|^{2} \mathbf{a}_{M}^{H}\left(\phi_{r}^{a}, \phi_{r}^{e}\right) \mathbb{E}\left\{\tilde{\mathbf{H}}_{2} {\bf \Phi} \tilde{\mathbf{h}}_{i} \tilde{\mathbf{h}}_{i}^{H} {\bf {\bf \Phi}}^{H} \tilde{\mathbf{H}}_{2}^{H}\right\} \mathbf{a}_{M}\left(\phi_{r}^{a}, \phi_{r}^{e}\right) \\
&=\delta \varepsilon_{k}\left|f_{k}({\bf \Phi})\right|^{2} M N.
	\end{aligned}
	\end{align}
	
	Similarly, when $\omega=2$, we have
	\begin{align}
	\mathbb{E}\left\{\left|\left(\mathbf{g}_{k}^{2}\right)^{H} \mathbf{g}_{i}^{1}\right|^{2}\right\}=\delta^{2} \varepsilon_{i}\left|f_{i}({\bf \Phi})\right|^{2} M^{2} N.
	\end{align}
	
	Next we have
	\begin{align}
	\begin{aligned}
	&\mathbb{E}\left\{\left|\left(\mathbf{g}_{k}^{2}\right)^{H} \mathbf{g}_{i}^{2}\right|^{2}\right\} \\
	&=M^{2} \delta^{2} \mathbb{E}\left\{\tilde{\mathbf{h}}_{k}^{H} \mathbf{\Phi}^{H} \mathbf{a}_{N}\left(\varphi_{t}^{a}, \varphi_{t}^{e}\right) \mathbf{a}_{N}^{H}\left(\varphi_{t}^{a}, \varphi_{t}^{e}\right) {\bf \Phi} \tilde{\mathbf{h}}_{i} \tilde{\mathbf{h}}_{i}^{H} \mathbf{\Phi}^{H} \mathbf{a}_{N}\left(\varphi_{t}^{a}, \varphi_{t}^{e}\right) \mathbf{a}_{N}^{H}\left(\varphi_{t}^{a}, \varphi_{t}^{e}\right) {\bf \Phi} \tilde{\mathbf{h}}_{k}\right\} \\
	&{\mathop  = \limits^{\left( e \right)} }M^{2} \delta^{2} \mathbb{E}\left\{\tilde{\mathbf{h}}_{k}^{H} \mathbf{\Phi}^{H} \mathbf{a}_{N}\left(\varphi_{t}^{a}, \varphi_{t}^{e}\right) \mathbf{a}_{N}^{H}\left(\varphi_{t}^{a}, \varphi_{t}^{e}\right) {\bf \Phi} \mathbb{E}\left\{\tilde{\mathbf{h}}_{i} \tilde{\mathbf{h}}_{i}^{H}\right\} {\bf \Phi}^{H} \mathbf{a}_{N}\left(\varphi_{t}^{a}, \varphi_{t}^{e}\right) \mathbf{a}_{N}^{H}\left(\varphi_{t}^{a}, \varphi_{t}^{e}\right) {\bf \Phi} \tilde{\mathbf{h}}_{k}\right\} \\
	&=\delta^{2} M^{2} N^{2},
	\end{aligned}
	\end{align}
	where $ (e) $ is due to the independence between $\tilde{\mathbf{h}}_k$ and $\tilde{\mathbf{h}}_i$.
	
	Similarly, we have
	\begin{align}
	&\mathbb{E}\left\{\left|\left(\mathbf{g}_{k}^{2}\right)^{H} \mathbf{g}_{i}^{3}\right|^{2}\right\}=\delta \varepsilon_{i} M N^{2},\\
&\mathbb{E}\left\{\left|\left(\mathbf{g}_{k}^{2}\right)^{H} \mathbf{g}_{i}^{4}\right|^{2}\right\}=\delta M N^{2}.
	\end{align}
	
	When $\omega=3$, we can readily obtain the first two terms as follows
	\begin{align}\label{gk3gi1&2}
\begin{array}{l}
\mathbb{E}\left\{\left|\left(\mathbf{g}_{k}^{3}\right)^{H} \mathbf{g}_{i}^{1}\right|^{2}\right\}=\delta \varepsilon_{i} \varepsilon_{k}\left|f_{i}({\bf \Phi})\right|^{2} M N, \\
\mathbb{E}\left\{\left|\left(\mathbf{g}_{k}^{3}\right)^{H} \mathbf{g}_{i}^{2}\right|^{2}\right\}=\delta \varepsilon_{k} M N^{2}.
\end{array}
	\end{align}
	
	The third term can be derived as follows
	\begin{align}\label{gk3gi3_2}
\begin{array}{l}
\mathbb{E}\left\{\left|\left(\mathbf{g}_{k}^{3}\right)^{H} \mathbf{g}_{i}^{3}\right|^{2}\right\}=\varepsilon_{k} \varepsilon_{i} \mathbb{E}\left\{\left|\overline{\mathbf{h}}_{k}^{H} {\bf \Phi}^{H} \tilde{\mathbf{H}}_{2}^{H} \tilde{\mathbf{H}}_{2} {\bf \Phi} \overline{\mathbf{h}}_{i}\right|^{2}\right\} \\
=\varepsilon_{k} \varepsilon_{i} \overline{\mathbf{h}}_{k}^{H} {\bf \Phi}^{H} \mathbb{E}\left\{\tilde{\mathbf{H}}_{2}^{H} \tilde{\mathbf{H}}_{2} {\bf \Phi} \overline{\mathbf{h}}_{i} \overline{\mathbf{h}}_{i}^{H} {\bf \Phi}^{H} \tilde{\mathbf{H}}_{2}^{H} \tilde{\mathbf{H}}_{2}\right\} {\bf \Phi} \overline{\mathbf{h}}_{k}.
\end{array}
	\end{align}
	
	Assume that $\tilde{\mathbf{H}}_{2}=\left[\mathbf{J}_{1}, \ldots, \mathbf{J}_{i}, \ldots, \mathbf{J}_{N}\right]  $ and $ \left[{\bf \Phi} \overline{\mathbf{h}}_i\overline{\mathbf{h}}_{i}^{H} {\bf \Phi}^{H}\right]_{mn}=\alpha_{m n} $, we can rewrite the $\left(n1,n2\right)$-th entry of $\tilde{\mathbf{H}}_{2}^{H} \tilde{\mathbf{H}}_{2} {\bf \Phi} \overline{\mathbf{h}}_{i} \overline{\mathbf{h}}_{i}^{H} {\bf \Phi}^{H} \tilde{\mathbf{H}}_{2}^{H} \tilde{\mathbf{H}}_{2}$ as follows
	\begin{align}\label{HHHH_n1n2}
\begin{array}{l}
{\left[\tilde{\mathbf{H}}_{2}^{H} \tilde{\mathbf{H}}_{2} {\bf \Phi} \overline{\mathbf{h}}_{i} \overline{\mathbf{h}}_{i}^{H} {\bf \Phi}^{H} \tilde{\mathbf{H}}_{2}^{H} \tilde{\mathbf{H}}_{2}\right]_{n 1, n 2}} =\sum\limits_{h=1}^{N} \sum\limits_{m=1}^{N} \mathbf{J}_{n 1}^{H} \mathbf{J}_{m} \alpha_{m h} \mathbf{J}_{h}^{H} \mathbf{J}_{n 2},
\end{array}
	\end{align}
which can be calculated by discussing the values of $h$ and $m$ under different situations. After some algebraic simplifications, we can obtain the following results
	\begin{align}\label{HHAHH}
\begin{array}{l}
\mathbb{E}\left\{\left[\tilde{\mathbf{H}}_{2}^{H} \tilde{\mathbf{H}}_{2} {\bf \Phi} \overline{\mathbf{h}}_i \overline{\mathbf{h}}_{i}^{H} {\bf \Phi}^{H} \tilde{\mathbf{H}}_{2}^{H} \tilde{\mathbf{H}}_{2}\right]_{n 1, n 2}\right\}=\alpha_{n 1 n 2} M^{2}, \\
\mathbb{E}\left\{\left[\tilde{\mathbf{H}}_{2}^{H} \tilde{\mathbf{H}}_{2} {\bf \Phi} \overline{\mathbf{h}}_i \overline{\mathbf{h}}_{i}^{H} {\bf \Phi}^{H} \tilde{\mathbf{H}}_{2}^{H} \tilde{\mathbf{H}}_{2}\right]_{n 1, n 1}\right\}=M(M+N), \\
\mathbb{E}\left\{\tilde{\mathbf{H}}_{2}^{H} \tilde{\mathbf{H}}_{2} {\bf \Phi} \overline{\mathbf{h}}_{i} \overline{\mathbf{h}}_{i}^{H} {\bf \Phi}^{H} \tilde{\mathbf{H}}_{2}^{H} \tilde{\mathbf{H}}_{2}\right\}=M^{2} {\bf \Phi} \overline{\mathbf{h}}_i \overline{\mathbf{h}}_{i}^{H} {\bf \Phi}^{H}+M N \mathbf{I}_{N}.
\end{array}
	\end{align}
	
	Substituting (\ref{HHAHH}) into (\ref{gk3gi3_2}), we have
	\begin{align}\label{gk3gi3_2_finial}
	\mathbb{E}\left\{\left|\left(\mathbf{g}_{k}^{3}\right)^{H} \mathbf{g}_{i}^{3}\right|^{2}\right\}=\varepsilon_{k} \varepsilon_{i} M\left(N^{2}+M\left|\overline{\mathbf{h}}_{k}^{H} \overline{\mathbf{h}}_{i}\right|^{2}\right).
	\end{align}
	
	Using (\ref{HHAHH}), we know that $\mathbb{E}\left\{\tilde{\mathbf{H}}_{2}^{H} \tilde{\mathbf{H}}_{2} \mathbf{I}_{N} \tilde{\mathbf{H}}_{2}^{H} \tilde{\mathbf{H}}_{2}\right\}=M\left(M+ N\right) \mathbf{I}_{N}$. Therefore, we can obtain the fourth term as follows
	\begin{align}
	\begin{array}{l}
	\mathbb{E}\left\{\left|\left(\mathbf{g}_{k}^{3}\right)^{H} \mathbf{g}_{i}^{4}\right|^{2}\right\} \\
	=\varepsilon_{k} \overline{\mathbf{h}}_{k}^{H} {\bf \Phi}^{H} \mathbb{E}\left\{\tilde{\mathbf{H}}_{2}^{H} \tilde{\mathbf{H}}_{2} \tilde{\mathbf{H}}_{2}^{H} \tilde{\mathbf{H}}_{2}\right\} {\bf \Phi} \overline{\mathbf{h}}_{k} \\
	=\varepsilon_{k} M N(M+N).
	\end{array}
	\end{align}
	
	When $\omega=4$, similarly, we have
	\begin{align}
	\begin{array}{l}
	\mathbb{E}\left\{\left|\left(\mathbf{g}_{k}^{4}\right)^{H} \mathbf{g}_{i}^{1}\right|^{2}\right\}=\delta \varepsilon_{i}\left|f_{i}(\mathbf{\Phi})\right|^{2} M N, \\
	\mathbb{E}\left\{\left|\left(\mathbf{g}_{k}^{4}\right)^{H} \mathbf{g}_{i}^{2}\right|^{2}\right\}=\delta M N^{2}, \\
	\mathbb{E}\left\{\left|\left(\mathbf{g}_{k}^{4}\right)^{H} \mathbf{g}_{i}^{3}\right|^{2}\right\}=\varepsilon_{i} M N(M+N), \\
	\mathbb{E}\left\{\left|\left(\mathbf{g}_{k}^{4}\right)^{H} \mathbf{g}_{i}^{4}\right|^{2}\right\}=M N(M+N).
	\end{array}
	\end{align}
	
	Similar to the above derivation, the remaining four parts in (\ref{gkgi_2}) can be readily derived as follows
	\begin{align}
\begin{array}{l}
\mathbb{E}\left\{\operatorname{Re}\left\{\left(\mathbf{g}_{k}^{1}\right)^{H} \mathbf{g}_{i}^{1}\left(\mathbf{g}_{i}^{3}\right)^{H} \mathbf{g}_{k}^{3}\right\}\right\} \\
=\delta \varepsilon_{k} \varepsilon_{i} \operatorname{Re}\left\{\overline{\mathbf{h}}_{k}^{H} {\bf \Phi}^{H} \overline{\mathbf{H}}_{2}^{H} \overline{\mathbf{H}}_{2} {\bf \Phi} \overline{\mathbf{h}}_{i} \overline{\mathbf{h}}_{i}^{H} \mathbf{\Phi}^{H} \mathbb{E}\left\{\tilde{\mathbf{H}}_{2}^{H} \tilde{\mathbf{H}}_{2}\right\} {\bf \Phi} \overline{\mathbf{h}}_{k}\right\} \\
=\delta \varepsilon_{k} \varepsilon_{i} M^{2} \operatorname{Re}\left\{f_{k}^{H}({\bf \Phi}) f_{i}({\bf \Phi}) \overline{\mathbf{h}}_{i}^{H} \overline{\mathbf{h}}_{k}\right\},
\end{array}
	\end{align}
	\begin{align}
	\begin{array}{l}
	\mathbb{E}\left\{\operatorname{Re}\left\{\left(\mathbf{g}_{k}^{1}\right)^{H} \mathbf{g}_{i}^{2}\left(\mathbf{g}_{i}^{4}\right)^{H} \mathbf{g}_{k}^{3}\right\}\right\} \\
	=\delta \varepsilon_{k} \operatorname{Re}\left\{\overline{\mathbf{h}}_{k}^{H} {\bf \Phi}^{H} \overline{\mathbf{H}}_{2}^{H} \overline{\mathbf{H}}_{2} {\bf \Phi} \mathbb{E}\left\{\tilde{\mathbf{h}}_{i} \tilde{\mathbf{h}}_{i}^{H} {\bf \Phi}^{H} \tilde{\mathbf{H}}_{2}^{H} \tilde{\mathbf{H}}_{2}\right\} {\bf \Phi} \overline{\mathbf{h}}_{k}\right\} \\
	=\delta \varepsilon_{k} M^{2}\left|f_{k}({\bf \Phi})\right|^{2},
	\end{array}
	\end{align}
	\begin{align}
	\begin{array}{l}
	\mathbb{E}\left\{\operatorname{Re}\left\{\left(\mathbf{g}_{k}^{2}\right)^{H} \mathbf{g}_{i}^{1}\left(\mathbf{g}_{i}^{3}\right)^{H} \mathbf{g}_{k}^{4}\right\}\right\} \\
	=\delta \varepsilon_{i} \mathbb{E}\left\{\operatorname{Re}\left\{\tilde{\mathbf{h}}_{k}^{H} {\bf \Phi}^{H} \overline{\mathbf{H}}_{2}^{H} \overline{\mathbf{H}}_{2} {\bf \Phi} \overline{\mathbf{h}}_{i} \overline{\mathbf{h}}_{i}^{H} {\bf \Phi}^{H} \mathbb{E}\left\{\tilde{\mathbf{H}}_{2}^{H} \tilde{\mathbf{H}}_{2}\right\} {\bf \Phi} \tilde{\mathbf{h}}_{k}\right\}\right\} \\
	=\delta \varepsilon_{i}M \operatorname{Re} \left\{  \mathrm{Tr}\left\{      \overline{\mathbf{H}}_{2} {\bf \Phi} \overline{\mathbf{h}}_{i} \overline{\mathbf{h}}_{i}^{H} \mathbb{E}\left\{{\tilde{\mathbf{h}}_{k}  \tilde{\mathbf{h}}_{k}^{H}} \right\} {\bf \Phi}^{H}  \overline{\mathbf{H}}_{2}^{H}  \right\}\right\} \\	
	=\delta \varepsilon_{i}M \operatorname{Re} \left\{  \mathrm{Tr}\left\{   \left|f_{i}({\bf\Phi})\right|^{2} \mathbf{a}_{M}\left(\phi_{r}^{a}, \phi_{r}^{e}\right) \mathbf{a}_{M}^{H}\left(\phi_{r}^{a}, \phi_{r}^{e}\right)     \right\}\right\} \\		
	=\delta \varepsilon_{i} M^{2}\left|f_{i}({\bf \Phi})\right|^{2},
	\end{array}
	\end{align}
	and
	\begin{align}\label{Rek2i2i4k4}
	\begin{array}{l}
	\mathbb{E}\left\{\operatorname{Re}\left\{\left(\mathbf{g}_{k}^{2}\right)^{H} \mathbf{g}_{i}^{2}\left(\mathbf{g}_{i}^{4}\right)^{H} \mathbf{g}_{k}^{4}\right\}\right\} \\
	=\delta \mathbb{E}\left\{\operatorname{Re}\left\{\tilde{\mathbf{h}}_{k}^{H} {\bf \Phi}^{H} \overline{\mathbf{H}}_{2}^{H} \overline{\mathbf{H}}_{2} {\bf \Phi} \mathbb{E}\left\{\tilde{\mathbf{h}}_{i} \tilde{\mathbf{h}}_{i}^{H} {\bf \Phi}^{H} \tilde{\mathbf{H}}_{2}^{H} \tilde{\mathbf{H}}_{2}\right\} {\bf \Phi} \tilde{\mathbf{h}}_{k}\right\}\right\} \\
	=\delta M \operatorname{Re}\left\{\operatorname{Tr}\left\{\overline{\mathbf{H}}_{2}^{H} \overline{\mathbf{H}}_{2} {\bf\Phi} \mathbb{E}\left\{\tilde{\mathbf{h}_{k}} \tilde{\mathbf{h}}_{k}^{H} \right\} {\bf\Phi}^{H}\right\}\right\} \\
	=\delta M \operatorname{Re}\left\{\operatorname{Tr}\left\{\overline{\mathbf{H}}_{2}^{H} \overline{\mathbf{H}}_{2}\right\}\right\}\\
	=\delta M^{2} N.
	\end{array}
	\end{align}
	
	Substituting (\ref{gk1gi1_2}) $\sim$ (\ref{gk3gi1&2}) and (\ref{gk3gi3_2_finial}) $\sim$ (\ref{Rek2i2i4k4})  into (\ref{gkgi_2}), we can complete the proof of Lemma \ref{lemma2} after some trivial simplifications.

\section{}\label{D2}
When the phase shifts of RIS are aligned to user $k$, we have $f_k\left(\mathbf{\Phi}\right)=N$ but $\left|f_i\left(\mathbf{\Phi}\right)\right|$ is bounded when $N\rightarrow \infty$. Therefore, when $N\rightarrow\infty$ and $M\rightarrow\infty$, we can obtain the order of magnitude as follows
\begin{align}
&\mathbb{E}\left\{\left\|\mathbf{g}_{k}\right\|^{4}\right\}= \mathcal{O} \left(M^2N^4\right),\\
 &\mathbb{E}\left\{\left\|\mathbf{g}_{i}\right\|^{4}\right\}= \mathcal{O} \left(M^2N^2\right),\\
& \mathbb{E}\left\{\left|\mathbf{g}_{k}^{H} \mathbf{g}_{i}\right|^{2}\right\} =\mathcal{O} \left(M^2N^3\right),\\
 &\mathbb{E}\left\{\left\|\mathbf{g}_{k}\right\|^{2}\right\}= \mathcal{O} \left(MN^2\right).
\end{align}

Therefore, when $M\rightarrow\infty$ and $N\rightarrow\infty$, user $k$'s rate $R_k$ can maintain a non-zero value when we cut the transmission power of user $k$ as $p_k=E_u/(MN^2)$ and cut the transmission power of other users as $p_i = E_u/(MN), \forall i \neq k$. However, at the same time, since $\left|f_i\left(\mathbf{\Phi}\right)\right|$ is bounded, the rate of user $i$ will be zero. At this time, when $N\rightarrow\infty$ and $M\rightarrow\infty$, the dominant terms in rate expression (\ref{rate}) are those terms which are on the order of  $ MN^2$:
\begin{align}
\begin{array}{l}
\frac{E_{u}}{M N^{2}} \mathbb{E}\left\{\left\|\mathbf{g}_{k}\right\|^{4}\right\} \rightarrow E_{u}\left(\frac{\beta \alpha_{k}}{(\delta+1)\left(\varepsilon_{k}+1\right)}\right)^{2}\left(\delta \varepsilon_{k}\right)^{2} M N^{2}, \\
\frac{E_{u}}{M N} \mathbb{E}\left\{\left|\mathbf{g}_{k}^{H} \mathbf{g}_{i}\right|^{2}\right\} \rightarrow E_{u} \frac{\beta^{2} \alpha_{i} \alpha_{k}}{(\delta+1)^{2}\left(\varepsilon_{i}+1\right)\left(\varepsilon_{k}+1\right)} \delta^{2} \varepsilon_{k} M N^{2}, \\
\sigma^2\mathbb{E}\left\{\left\|\mathbf{g}_{k}\right\|^{2}\right\} \rightarrow \sigma^2\frac{\beta \alpha_{k}}{(\delta+1)\left(\varepsilon_{k}+1\right)} \delta \varepsilon_{k} M N^{2}.
\end{array}
\end{align}

Thus, after some simplification, the rate can be calculated as follows
\begin{align}\label{aligntouserk}
R_{k} &= \log _{2}\left(1+\frac{\frac{E_{u}}{M N^{2}} \mathbb{E}\left\{\left\|\mathbf{g}_{k}\right\|^{4}\right\}}  {   \sum_{i=1, i \neq k}^{K}    \frac{E_{u}}{M N} \mathbb{E}\left\{\left|\mathbf{g}_{k}^{H} \mathbf{g}_{i}\right|^{2}\right\}   +  \sigma^2\mathbb{E}\left\{\left\|\mathbf{g}_{k}\right\|^{2}\right\}     }\right) \nonumber\\
&\to\log _{2}\left(1+\frac{E_{u} \frac{\varepsilon_{k}}{\left(\varepsilon_{k}+1\right)}}{E_{u} \sum_{i=1, i \neq k}^{K} \frac{\alpha_{i}}{\left(\varepsilon_{i}+1\right) \alpha_{k}}+\left(1+\frac{1}{\delta}\right) \frac{\sigma^{2}}{\beta \alpha_{k}}}\right), \text{as }M, N\to\infty.
\end{align}

Besides, we can see that (\ref{aligntouserk}) is an increasing function with respect to $\alpha_{k}$, $\beta$ and $\delta$.

	\section{}\label{appE}
To begin with, we need to provide some necessary preliminary results. Firstly, for the ideal RIS with continuous phase shifts, we assume that the phase shift of each reflecting element $\theta_n$ is randomly and independently adjusted at each fading block following the uniform distribution of $\mathcal{U}\left[0,2\pi\right]$. Then, for $k_1$ with arbitrary values, we have
\begin{align}\label{contineous_case}
&\mathbb{E}\left\{\cos \left(k_{1}+\theta_{n}\right)\right\}=\frac{1}{2 \pi} \int_{0}^{2 \pi} \cos \left(k_{1}+\theta_{n}\right) d \theta_{n}=0,\\
&\mathbb{E}\left\{\cos ^{2}\left(k_{1}+\theta_{n}\right)\right\}=\frac{1}{2}\left(1+\mathbb{E}\left\{\cos \left(2 k_{1}+2 \theta_{n}\right)\right\}\right)=\frac{1}{2}\left(1+\frac{1}{2 \pi} \int_{0}^{2 \pi} \cos \left(2 k_{1}+2 \theta_{n}\right) d \theta_{n}\right)=\frac{1}{2}.
\end{align}

Next, for the non-ideal RIS with finite $b$ bits discrete phase shifts, we assume that  each $\theta_n$ is randomly and independently adjusted from $ \left\{0, \frac{2 \pi}{2^{b}}, 2 \times \frac{2 \pi}{2^{b}}, \ldots,\left(2^{b}-1\right) \frac{2 \pi}{2^{b}}\right\} $. When $b>1$, for $k_1$ with arbitrary values, we have
\begin{align}\label{discrete_case}
\mathbb{E}\left\{\cos \left(k_{1}+\theta_{n}\right)\right\}&=\frac{1}{2^{b}} \sum\nolimits_{t=0}^{2^{b}-1} \cos \left(k_{1}+t \frac{2 \pi}{2^{b}}\right) \nonumber\\
&=\frac{1}{2^{b}} \sum\nolimits_{t=0}^{2^{(b-1)}-1}\left(\cos \left(k_{1}+t \frac{2 \pi}{2^{b}}\right)+\cos \left(k_{1}+\left(t+2^{(b-1)}\right) \frac{2 \pi}{2^{b}}\right)\right)\nonumber\\
&\;{\mathop  = \limits^{\left( f\right)} } \;0,
\end{align}
and
\begin{align}\label{discrete_case2}
&\mathbb{E}\left\{\cos ^{2}\left(k_{1}+\theta_{n}\right)\right\}=\frac{1}{2}\left(1+\mathbb{E}\left\{\cos \left(2 k_{1}+2 \theta_{n}\right)\right\}\right)=\frac{1}{2}\left(1+\frac{1}{2^{b}} \sum\nolimits_{t=0}^{2^{b}-1} \cos \left(2 k_{1}+2 t \frac{2 \pi}{2^{b}}\right)\right) \nonumber\\
&=\frac{1}{2}  \left(1+\frac{1}{2^{b}} \left(\sum\nolimits_{t=0}^{2^{(b-1)}-1} \cos \left(2 k_{1}+t \frac{2 \pi}{2^{b-1}}\right)+ \sum\nolimits_{t=2^{(b-1)}}^{2^b-1} \cos \left(2 k_{1}+t \frac{2 \pi}{2^{b-1}}\right)\right)\right) \nonumber\\
&\;{\mathop  = \limits^{\left( g\right)} }\frac{1}{2}\left(1+\frac{2}{2^{b}} \sum\nolimits_{t=0}^{2^{(b-1)}-1} \cos \left(2 k_{1}+t \frac{2 \pi}{2^{b-1}}\right)\right)\nonumber\\
&=\frac{1}{2}\left(1+\left.\mathbb{E}\left\{\cos \left(2 k_{1}+\theta_{n}\right)\right\}\right|_{b=b-1}\right) \nonumber\\
&=\frac{1}{2},
\end{align}
where $(f)$ and $(g)$ come from $\cos (\vartheta+\pi)=-\cos (\vartheta)$ and $\cos (\vartheta+2\pi)=\cos (\vartheta)$, respectively.

(\ref{contineous_case}) $\sim$ (\ref{discrete_case2}) prove that $ \mathbb{E}\left\{\cos \left(k_{1}+\theta_{n}\right)\right\} $ and $\mathbb{E}\left\{\cos ^{2}\left(k_{1}+\theta_{n}\right)\right\}  $ have the same values for both continuous and discrete phase shifts when $b>1$. Therefore, we will not distinguish these two cases in the following derivation. Besides, since the above equations hold for arbitrary $k_1$, we can obtain the following results from trigonometric identities:
\begin{align}\label{trigonometr_icIdentitify}
&\mathbb{E}\left\{\sin \left(k_{1}+\theta_{n}\right)\right\}=\mathbb{E}\left\{\cos \left(\left(k_{1}-\frac{\pi}{2}\right)+\theta_{n}\right)\right\}=0, \\\label{trigonometr_icIdentitify2}
&\mathbb{E}\left\{\sin^2 \left(k_{1}+\theta_{n}\right)\right\}=\mathbb{E}\left\{1-\cos^2 \left(k_{1}+\theta_{n}\right)\right\}=\frac{1}{2}.
\end{align}


Then, using the above preliminary results can help us derive the asymptotic average rate with random phase shifts. Since $\bf\Phi$ is independent with the channel $\mathbf{H}_2$ and $\mathbf{h}_k$, rate $R_k$ can be calculated by substituting the terms involving $\bf\Phi$ in (\ref{rate}) with their expectation. Thus, we need to further calculate $\mathbb{E}\left\{\left|f_{k}(	{\bf\Phi})\right|^{2}\right\}$, $\mathbb{E}\left\{\left|f_{i}(	{\bf\Phi})\right|^{2}\right\}$, $\mathbb{E}\left\{\left|f_{k}(	{\bf\Phi})\right|^{4}\right\}$, $ \mathbb{E}\left\{\left|f_{k}(	{\bf\Phi})\right|^{2}\left|f_{i}(	{\bf\Phi})\right|^{2}\right\} $ and $ \mathbb{E}\left\{\operatorname{Re}\left\{f_{k}^{H}(	{\bf\Phi}) f_{i}(	{\bf\Phi})  \overline{\mathbf{h}}_{i}^{H} \overline{\mathbf{h}}_{k} \right\}\right\} $, respectively.

Recalling (\ref{fc_Phi}), we rewrite $f_{k}({\bf\Phi})$ and $ f_{i}({\bf\Phi})  $ as follows
\begin{align}
\begin{array}{l}
f_{k}({\bf\Phi})=\sum_{n=1}^{N} e^{j\left(\zeta_{n}^{k}+\theta_{n}\right)}, \\
f_{i}({\bf\Phi})=\sum_{n=1}^{N} e^{j\left(\zeta_{n}^{i}+\theta_{n}\right)}.
\end{array}
\end{align}

Then, using (\ref{contineous_case}) $\sim$ (\ref{trigonometr_icIdentitify2}) and the independence between $\theta_{n1},\theta_{n2},\forall n1\neq n2$, we have
\begin{align}\label{intermediate_results}
\begin{array}{l}
\mathbb{E}\left\{e^{j\left(k_{1}+\theta_{n1}\right)}\right\}=\mathbb{E}\left\{\cos \left(k_{1}+\theta_{n1}\right)\right\}+j \mathbb{E}\left\{\sin \left(k_{1}+\theta_{n1}\right)\right\}=0, \\
\mathbb{E}\left\{e^{j\left(k_{1}+\theta_{n 1}\right)} e^{j\left(k_{2}+\theta_{n 2}\right)}\right\}=\mathbb{E}\left\{e^{j\left(k_{1}+\theta_{n 1}\right)}\right\} \mathbb{E}\left\{e^{j\left(k_{2}+\theta_{n 2}\right)}\right\}=0, \\
\mathbb{E}\left\{\operatorname{Re}\left\{e^{j\left(k_{1}+\theta_{n 1}\right)} e^{-j\left(k_{2}+\theta_{n 2}\right)}\right\}\right\}=\mathbb{E}\left\{\cos \left(\left(k_{1}+\theta_{n 1}\right)-\left(k_{2}+\theta_{n 2}\right)\right)\right\}=0, \\
\mathbb{E}\left\{\left(\operatorname{Re}\left\{e^{j\left(k_{1}+\theta_{n 1}\right)} e^{-j\left(k_{2}+\theta_{n 2}\right)}\right\}\right)^{2}\right\}=\frac{1}{2}\left(1+\mathbb{E}\left\{\cos \left(2\left(k_{1}+\theta_{n 1}\right)-2\left(k_{2}+\theta_{n 2}\right)\right)\right\}\right)=\frac{1}{2}.
\end{array}
\end{align}

Utilizing (\ref{intermediate_results}), we will calculate the expectation of terms involving $\bf\Phi$ one by one. Firstly, the term $\mathbb{E}\left\{\left|f_{k}(	{\bf\Phi})\right|^{2}\right\}$ can be calculated as follows
\begin{align}\label{E_f_phi_k2}
\begin{array}{l}
\mathbb{E}\left\{\left|f_{k}({\bf\Phi})\right|^{2}\right\}\\
=\sum_{n=1}^{N} e^{j\left(\zeta_{n}^{k}+\theta_{n}\right)} \sum_{n=1}^{N} e^{-j\left(\zeta_{n}^{k}+\theta_{n}\right)} \\
=\sum_{n=1}^{N} 1+  \mathbb{E}\left\{\sum_{n 1=1}^{N} \sum_{n_2=1 , n_2 \neq n 1}^{N} e^{-j\left(\zeta_{n 1}^{k}+\theta_{n 1}\right)} e^{j\left(\zeta_{n 2}^{k}+\theta_{n 2}\right)} \right\}\\
=   N,
\end{array}
\end{align}

Similarly, for the term $\mathbb{E}\left\{\left|f_{i}(	{\bf\Phi})\right|^{2}\right\}$ , we also have
\begin{align}\label{E_f_phi_i2}
\mathbb{E}\left\{\left|f_{i}({\bf\Phi})\right|^{2}\right\}=   N.
\end{align}

Secondly, we focus on the term $\mathbb{E}\left\{\left|f_{k}(	{\bf\Phi})\right|^{4}\right\}$ which can be derived as:
\begin{align}\label{E_f_phi_k&i}
\begin{array}{l}
\mathbb{E}\left\{\left|f_{k}({\bf\Phi})\right|^{4}\right\}\\
=\left|\sum_{n=1}^{N} e^{j\left(\zeta_{n}^{k}+\theta_{n}\right)} \sum_{n=1}^{N} e^{-j\left(\zeta_{n}^{k}+\theta_{n}\right)}\right|^{2}\\
=\left|N+2 \sum_{n1=1}^{N-1} \sum_{n 2=n 1+1}^{N} \operatorname{Re}\left\{e^{j\left(\zeta_{n 1}^{k}+\theta_{n 1}-\zeta_{n 2}^{k}-\theta_{n 2}\right)}\right\}\right|^{2} \\
=N^{2}+4 \mathbb{E}\left\{\left|\sum_{n 1=1}^{N-1} \sum_{n2=n 1+1}^{N}  \operatorname{Re}\left\{e^{j\left(\zeta_{n 1}^{k}+\theta_{n 1}-\zeta_{n 2}^{k}-\theta_{n 2}\right)}\right\}     \right|^{2}\right\} \\
=N^{2}+4  \sum_{n1=1}^{N-1} \sum_{n2=n 1+1}^{N}    \mathbb{E}\left\{     \left( {  \operatorname{Re}\left\{e^{j\left(\zeta_{n 1}^{k}+\theta_{n 1}-\zeta_{n 2}^{k}-\theta_{n 2}\right)}\right\}     }    \right)^2    \right\}\\
= N^{2}+4 \frac{N(N-1)}{2} \frac{1}{2}\\
=2 N^{2}-N.
\end{array}
\end{align}

Thirdly, the term $ \mathbb{E}\left\{\left|f_{k}(	{\bf\Phi})\right|^{2}\left|f_{i}(	{\bf\Phi})\right|^{2}\right\} $ can be calculated as:
\begin{align}\label{E_f_phi_k2_i2}
\begin{array}{l}
\mathbb{E}\left\{\left|f_{k}({\bf\Phi})\right|^{2}\left|f_{i}({\bf\Phi})\right|^{2}\right\}\\
=\mathbb{E}\left\{\left|\sum_{n=1}^{N} e^{j\left(\zeta_{n}^{k}+\theta_{n}\right)}\right|^{2}\left|\sum_{n=1}^{N} e^{j\left(\zeta_{n}^{i}+\theta_{n}\right)}\right|^{2}\right\} \\
=\mathbb{E}\left\{\left(N+2 \sum_{n 1=1}^{N-1} \sum_{n2=n 1+1}^{N} \operatorname{Re}\left\{e^{j\left(\zeta_{n 1}^{k}+\theta_{n 1}-\zeta_{n 2}^{k}-\theta_{n 2}\right)}\right\}\right)\times\right.\\
\qquad\left.\left(N+2 \sum_{n 1=1}^{N-1} \sum_{n 2=n 1+1}^{N} \operatorname{Re}\left\{e^{j\left(\zeta_{n 1}^{i}+\theta_{n 1}-\zeta_{n 2}^{i}-\theta_{n 2}\right)}\right\}\right)\right\} \\
=N^{2}+4 \mathbb{E}\left\{\sum_{n 1=1}^{N-1} \sum_{n 2=n 1+1}^{N} \operatorname{Re}\left\{e^{j\left(\zeta_{n 1}^{k}+\theta_{n 1}-\zeta_{n 2}^{k}-\theta_{n 2}\right)}\right\} \operatorname{Re}\left\{e^{j\left(\zeta_{n 1}^{i}+\theta_{n 1}-\zeta_{n 2}^{i}-\theta_{n 2}\right)}\right\}\right\} \\
=N^{2}+4 \sum_{n 1=1}^{N-1} \sum_{n 2=n 1+1}^{N}\mathbb{E}\left\{ \cos \left(\zeta_{n 1}^{k}+\theta_{n 1}-\zeta_{n 2}^{k}-\theta_{n 2}\right) \cos \left(\zeta_{n 1}^{i}+\theta_{n 1}-\zeta_{n 2}^{i}-\theta_{n 2}\right)\right\} \\
{\mathop  = \limits^{\left( h\right)} }N^{2}+2 \sum_{n 1=1}^{N-1} \sum_{n 2=n 1+1}^{N} \cos \left(\zeta_{n 1}^{k}-\zeta_{n 2}^{k}-\zeta_{n 1}^{i}+\zeta_{n 2}^{i}\right),
\end{array}
\end{align}
where $\left(h\right)$ is obtained by using prosthaphaeresis. Since the second term in (\ref{E_f_phi_k2_i2}) is bounded, we have $ \mathbb{E}\left\{\left|f_{k}(	{\bf\Phi})\right|^{2}\left|f_{i}(	{\bf\Phi})\right|^{2}\right\} \rightarrow N^2$ when $N\rightarrow\infty$. 

The final term $ \mathbb{E}\left\{\operatorname{Re}\left\{f_{k}^{H}(	{\bf\Phi}) f_{i}(	{\bf\Phi})  \overline{\mathbf{h}}_{i}^{H} \overline{\mathbf{h}}_{k} \right\}\right\} $ is derived as:
\begin{align}\label{E_f_phi_real_k_i}
\begin{array}{l}
\mathbb{E}\left\{\operatorname{Re}\left\{f_{k}^{H}({\bf\Phi}) f_{i}({\bf\Phi}) \overline{\mathbf{h}}_{i}^{H} \overline{\mathbf{h}}_{k}\right\}\right\} \\
=\mathbb{E}\left\{\operatorname{Re}\left\{\left(\overline{\mathbf{h}}_{i}^{H} \overline{\mathbf{h}}_{k}\right) \sum_{n1=1}^{N} \sum_{n 2=1}^{N} e^{-j\left(\zeta_{n 1}^{k}+\theta_{n 1}\right)} e^{j\left(\zeta_{n 2}^{i}+\theta_{n 2}\right)}\right\}\right\} \\
=\mathbb{E}\left\{\operatorname{Re}\left\{\left(\overline{\mathbf{h}}_{i}^{H} \overline{\mathbf{h}}_{k}\right) \sum_{n=1}^{N} e^{-j\left(\zeta_{n}^{k}+\theta_{n}\right)} e^{j\left(\zeta_{n}^{i}+\theta_{n}\right)}\right\}\right\} \\
=\operatorname{Re}\left\{\left(\overline{\mathbf{h}}_{i}^{H} \overline{\mathbf{h}}_{k}\right) \sum_{n=1}^{N} e^{j\left(\zeta_{n}^{i}-\zeta_{n}^{k}\right)}\right\},
\end{array}
\end{align}
which is bounded when $N\rightarrow\infty$.

By substituting (\ref{E_f_phi_k2}) $\sim$ (\ref{E_f_phi_real_k_i}) into the corresponding terms in rate expression (\ref{rate}), we can see that when $M\rightarrow\infty$ and $N\rightarrow\infty$, the dominant terms are those which have the order of $\mathcal{O}\left(M^2N^2\right)$. Thus, when $M\rightarrow\infty$ and $N\rightarrow\infty$, we have
\begin{align}
\mathbb{E}\left\{\left\|\mathbf{g}_{k}\right\|^{4}\right\}\rightarrow &M^{2} N^{2}\left(\frac{\beta \alpha_{k}}{(\delta+1)\left(\varepsilon_{k}+1\right)}\right)^{2} \nonumber\\
&\times \left(2 \delta^{2} \varepsilon_{k}^{2}+2 \delta \varepsilon_{k}\left(2 \delta+\varepsilon_{k}+1\right)+\left(2 \delta^{2}+\varepsilon_{k}^{2}+2 \delta \varepsilon_{k}+2 \delta+2 \varepsilon_{k}+1\right)\right),
\end{align}
and
\begin{align}
\mathbb{E}\left\{\left|\mathbf{g}_{k}^{H} \mathbf{g}_{i}\right|^{2}\right\}\to M^{2} N^{2}\frac{\beta^{2} \alpha_{i} \alpha_{k}}{(\delta+1)^{2}\left(\varepsilon_{i}+1\right)\left(\varepsilon_{k}+1\right)}     \delta^{2}   \left(\varepsilon_{k} \varepsilon_{i}+\varepsilon_{k}+\varepsilon_{i}+1\right) .
\end{align}

Then, we can complete the proof after some simple algebraic simplifications:
\begin{align}\label{random_rate}
R_{k} &\rightarrow \log _{2}\left(1+\frac{  p_k \mathbb{E}\left\{\left\|\mathbf{g}_{k}\right\|^{4}\right\}}{\sum_{i=1, i \neq k}^{K} p_i \mathbb{E}\left\{\left|\mathbf{g}_{k}^{H} \mathbf{g}_{i}\right|^{2}\right\}}\right)\nonumber\\
&\rightarrow \log _{2}\left(1+\frac{p_k \alpha_{k}\left(2 \delta^{2}+2 \delta+1\right)}{\sum_{i=1, i \neq k}^{K} p_i \alpha_{i} \delta^{2}}\right), \text{as } M,N\to\infty.
\end{align}

Besides, we can find that (\ref{random_rate}) is a decrease function with respect to $\delta$.

	\section{}\label{D3}
	Firstly, by selecting the non-zero terms when all  the Rician factors grow to infinity, we can complete the derivation of (\ref{rate_los}).
	
	Secondly, we consider a conventional uplink non-RIS massive MIMO system with one BS and $K$ users. Assume that the deterministic LoS channel between the BS and user $k$ is $\sqrt {\gamma_{k}}\mathbf{\bar{h}}_k^{w/o}$, where $ \mathbf{\bar{h}}_k^{w/o} \in \mathbb{C}^{M\times1}$. To facilitate the analysis, we consider the uniform linear array (ULA), and the LoS channel $ \mathbf{\bar{h}}_k^{w/o} $ can be expressed as
	\begin{align}
	\overline{\mathbf{h}}_{k}^{w / o}=\left[1, e^{j 2 \pi \frac{d}{\lambda} \sin \vartheta_k}, \ldots, e^{j 2 \pi \frac{d}{\lambda}(M-1) \sin \vartheta_k}\right]^{T},
	\end{align}
	where $\vartheta_k$ is the AoA at the BS from user $k$. Besides, we have $\left\|\mathbf{\bar{h}}_k^{w/o}\right\|^2=M$. 

	Thus, with the MRC technique, the rate of user $k$ is given by 
	\begin{align}\label{rate_mimo}
	\bar{R}_{k}^{w / o}&=\log _{2}\left(1+\frac{p_{k} \left(   \gamma_{k}\right)^2\left\|\overline{\mathbf{h}}_{k}^{w / o}\right\|^{4}}{\sum_{i=1, i \neq k}^{K} p_{i} \gamma_{i} \gamma_{k}\left|\left(\overline{\mathbf{h}}_{k}^{w / o}\right)^{H} \overline{\mathbf{h}}_{i}^{w / o}\right|^{2}+  \sigma^{2} \gamma_{k}  \left\|\overline{\mathbf{h}}_{k}^{w / o}\right\|^{2}}\right),\nonumber\\
	&=\log _{2}\left(1+\frac{p_{k} \gamma_{k}  M}{\sum_{i=1, i \neq k}^{K} p_{i} \gamma_{i}  \frac{\left|\left(\overline{\mathbf{h}}_{k}^{w / o}\right)^{H} \overline{\mathbf{h}}_{i}^{w / o}\right|^{2}}{M}+\sigma^{2}}\right).
	\end{align}
	
	Besides, with the ULA structure, we have
	\begin{align}
\left|\left(\overline{\mathbf{h}}_{k}^{w / o}\right)^{H} \overline{\mathbf{h}}_{i}^{w / o}\right|^{2}&=\left|\sum_{i=1}^{M} e^{-j 2 \pi \frac{d}{\lambda}(i-1) \sin \vartheta_{k}} e^{j 2 \pi \frac{d}{\lambda}(i-1) \sin \vartheta_{i}}\right|^{2} \nonumber\\
&=\left|\sum_{i=1}^{M}\left(e^{j 2 \pi \frac{d}{\lambda}\left(\sin \vartheta_{i}-\sin \vartheta_{k}\right)}\right)^{(i-1)}\right|^2=\left|\frac{1-e^{j 2 \pi \frac{d}{\lambda} M\left(\sin \vartheta_{i}-\sin \vartheta_{k}\right)}}{ 1-e^{j 2 \pi \frac{d}{\lambda}\left(\sin \vartheta_{i}-\sin \vartheta_{k}\right)}}\right|^{2} \nonumber\\
&=\left|\frac{\left(e^{-j \pi \frac{d}{\lambda} M\left(\sin \vartheta_{i}-\sin \vartheta_{k}\right)}-e^{j \pi \frac{d}{\lambda} M\left(\sin \vartheta_{i}-\sin \vartheta_{k}\right)}\right) e^{j \pi \frac{d}{\lambda} M\left(\sin \vartheta_{i}-\sin \vartheta_{k}\right)}}{\left(e^{-j \pi \frac{d}{\lambda}\left(\sin \vartheta_{i}-\sin \vartheta_{k}\right)}-e^{j \pi \frac{d}{\lambda}\left(\sin \vartheta_{i}-\sin \vartheta_{k}\right)}\right) e^{j \pi \frac{d}{\lambda}\left(\sin \vartheta_{i}-\sin \vartheta_{k}\right)}}\right|^{2}\nonumber\\
&=\left|\frac{\sin \left( \pi \frac{d}{\lambda} M\left(\sin \vartheta_{i}-\sin \vartheta_{k}\right)\right)}{\sin \left(\pi \frac{d}{\lambda}\left(\sin \vartheta_{i}-\sin \vartheta_{k}\right)\right)} e^{j \pi \frac{d}{\lambda}(M-1)\left(\sin \vartheta_{i}-\sin \vartheta_{k}\right)}\right|^{2}\nonumber\\
&=\frac{\sin ^{2}\left( \pi \frac{d}{\lambda} M\left(\sin \vartheta_{i}-\sin \vartheta_{k}\right)\right)}{\sin ^{2}\left(\pi \frac{d}{\lambda}\left(\sin \vartheta_{i}-\sin \vartheta_{k}\right)\right)}.
	\end{align}
	
	Therefore, when $\vartheta_{i}\neq \vartheta_{k}$, we know $ \left|\left(\overline{\mathbf{h}}_{k}^{w / o}\right)^{H} \overline{\mathbf{h}}_{i}^{w / o}\right|^{2} $ is bounded. Thus, when $M\rightarrow\infty$, the inter-user interference terms in (\ref{rate_mimo}) becomes
	\begin{align}
\frac{\left|\left(\overline{\mathbf{h}}_{k}^{w / o}\right)^{H} \overline{\mathbf{h}}_{i}^{w / o}\right|^{2}}{M} \rightarrow 0, \text{as }M\to\infty.
	\end{align}

	\section{}\label{appF}
	 If the RIS has discrete phase shifts with $ b $ bits precision, the adjustable phase shifts $\hat{\theta}_n$ can only be selected from $ \left\{0, \frac{2 \pi}{2^{b}}, 2 \times \frac{2 \pi}{2^{b}}, \ldots,\left(2^{b}-1\right) \frac{2 \pi}{2^{b}}\right\} $. Therefore, the quantization error of RIS element $n$ can be expressed as $ \tilde{\theta}_{n}=  \theta_{n}^* - \hat{\theta}_n  \in\left[-\frac{\pi}{2^{b}}, \frac{\pi}{2^{b}}\right]$, where $\theta_n^*$ is the designed optimal phase shifts under the continuous phase shifts assumption.
	
	Assume that the phase shifts of RIS are aligned to an arbitrary user $k$, which is a simple sub-optimal solution for the maximization of sum rate $R$. In this case, when $N$ is even, the worst influence brought by phase noise can be quantified as follows
	\begin{align}
\begin{aligned}
\left|f_{k}(\mathbf{\Phi})\right|^{2}&=\left|\sum_{n=1}^{N} \exp \left(j \tilde{\theta}_{n}\right)\right|^{2} \\
&\geq\left|\frac{N}{2}\left(\exp \left(j \frac{\pi}{2^{b}}\right)+\exp \left(-j \frac{\pi}{2^{b}}\right)\right)\right|^{2} \\
&=N^{2} \cos ^{2}\left(\frac{\pi}{2^{b}}\right),
\end{aligned}
	\end{align}
	and
	\begin{align}
\left|f_{k}(\mathbf{\Phi})\right|^{4}=\left(\left|f_{k}(\mathbf{\Phi})\right|^{2}\right)^{2}\geq N^{4} \cos ^{4}\left(\frac{\pi}{2^{b}}\right).
	\end{align}
	
While for $f_{i}(\mathbf{\Phi}), \forall i \neq k$, it is still bounded when $N\rightarrow\infty$. Since  the worst rate degradation brought by RIS's phase noise is  $\cos^2 \left(\frac{\pi}{2^{b}}\right)$ which does not increase with $N$, when $N\rightarrow\infty$, user $k$'s rate still has the following orders of magnitude:
\begin{align}
&\mathbb{E}\left\{\left\|\mathbf{g}_{k}\right\|^{4}\right\}= \mathcal{O} \left(M^2N^4\right),\\
& \mathbb{E}\left\{\left|\mathbf{g}_{k}^{H} \mathbf{g}_{i}\right|^{2}\right\} =\mathcal{O} \left(M^2N^3\right),\\
&\mathbb{E}\left\{\left\|\mathbf{g}_{k}\right\|^{2}\right\}= \mathcal{O} \left(MN^2\right).
\end{align}

Therefore, the rate can still achieve a scaling law $\mathcal{O}\left(\mathrm{log_2}(N)\right)$ in the case of low-resolution phase shifts.
\end{appendices}

\bibliographystyle{IEEEtran}
\vspace{-6pt}
\bibliography{myref.bib}

\end{document}